\documentclass[fleqn,usenatbib]{mnras}

\usepackage[T1]{fontenc}

\DeclareRobustCommand{\VAN}[3]{#2}
\let\VANthebibliography\thebibliography
\def\thebibliography{\DeclareRobustCommand{\VAN}[3]{##3}\VANthebibliography}

\usepackage{graphicx}	
\usepackage{amsmath}	
\usepackage{amssymb}	
\usepackage{subcaption}
\usepackage{makecell}
\usepackage{paralist}
\usepackage[export]{adjustbox}
\usepackage{threeparttable}
\usepackage{newtxtext,newtxmath}

\DeclareMathOperator*{\argmin}{arg min}
\newcommand{\bs}{\boldsymbol}
\newcommand{\percent}{\,\mathrm{per\,cent}}
\newcommand{\logten}{\mathop{\log_{10}}}

\usepackage{xcolor}
\definecolor{mypink}{rgb}{0.858, 0.188, 0.478}

\title[Variable star classification with VVV]{Variable star classification across the Galactic bulge and disc with the VISTA Variables in the V\'ia L\'actea survey}
\author[T.~A.~Molnar, J.~L.~Sanders et al.]{
Thomas A. Molnar,$^{1}$
Jason L. Sanders,$^{2,1}$\thanks{jason.sanders@ucl.ac.uk}
Leigh C. Smith$^{1}$,
Vasily Belokurov$^1$,
\newauthor
Philip Lucas$^3$,
Dante Minniti$^{4,5}$
\\
$^{1}$Institute of Astronomy, University of Cambridge, Madingley Road, Cambridge CB3 0HA, UK\\
$^{2}$Department of Physics and Astronomy, University College London, London WC1E 6BT, UK\\
$^{3}$Centre for Astrophysics, University of Hertfordshire, College Lane, Hatfield AL10 9AB, UK\\
$^{4}$Departamento de Ciencias F\'isicas, Facultad de Ciencias Exactas, Universidad Andr\'es Bello, Fern\'andez Concha 700, Las Condes, Santiago, Chile\\
$^{5}$Vatican Observatory, Vatican City State, V-00120, Italy
}

\date{Accepted XXX. Received YYY; in original form ZZZ}

\pubyear{2020}

\begin{document}
\label{firstpage}
\pagerange{\pageref{firstpage}--\pageref{lastpage}}
\maketitle

\begin{abstract}
We present VIVACE, the VIrac VAriable Classification Ensemble, a catalogue of variable stars extracted from an automated classification pipeline for the Vista Variables in the V\'ia L\'actea (VVV) infrared survey of the Galactic bar/bulge and southern disc. Our procedure utilises a two-stage hierarchical classifier to first isolate likely variable sources using simple variability summary statistics and training sets of non-variable sources from the Gaia early third data release, and then classify candidate variables using more detailed light curve statistics and training labels primarily from OGLE and VSX. The methodology is applied to point-spread-function photometry for $\sim490$ million light curves from the VIRAC v2 astrometric and photometric catalogue resulting in a catalogue of $\sim1.4$ million likely variable stars, of which $\sim39,000$ are high-confidence (classification probability $>0.9$) RR Lyrae ab stars, $\sim8000$ RR Lyrae c/d stars, $\sim187,000$ detached/semi-detached eclipsing binaries, $\sim18,000$ contact eclipsing binaries, $\sim1400$ classical Cepheid variables and $\sim2200$ Type II Cepheid variables. Comparison with OGLE-4 suggests a completeness of around $90\,\percent$ for RRab and $\lesssim60\percent$ for RRc/d, and a misclassification rate for known RR Lyrae stars of around $1\percent$ for the high confidence sample. We close with two science demonstrations of our new VIVACE catalogue: first, a brief investigation of the spatial and kinematic properties of the RR Lyrae stars within the disc/bulge, demonstrating the spatial elongation of bar-bulge RR Lyrae stars is in the same sense as the more metal-rich red giant population whilst having a slower rotation rate of $\sim40\,\mathrm{km\,s}^{-1}\mathrm{kpc}^{-1}$; and secondly, an investigation of the Gaia EDR3 parallax zeropoint using contact eclipsing binaries across the Galactic disc plane and bulge.
\end{abstract}

\begin{keywords}
stars: variables: general -- stars: variables: RR Lyrae -- binaries: eclipsing -- catalogues -- surveys
\end{keywords}

\section{Introduction}\label{sect:intro}

Algol, the prototypical eclipsing binary system, is thought to have been discovered by the ancient Egyptians with markings in the Cairo Calendar, dated to 1271-1163 B.C., assigning luck to days with periodicity akin to the stars' fluctuations \citep{Jetsu2013,Porceddu2018}. Continued intrigue in stellar variability blossomed in the sixteenth century with a renaissance in precision astronomy and has proved essential in our progress in understanding the Universe. The periodicity of variable objects reflects underlying physical scales, from which sizes, luminosities, masses and distances can be inferred. In turn, variable stars are essential in the study both of stellar evolution and of structure within the Milky Way and the local Universe \citep[see][for a review of variable stars]{Catelan2014}.

With the advent of large-scale multi-epoch photometric surveys, the number of known variable objects has increased significantly. Over the last $25$ years, surveys such as the All-Sky Automated Survey \citep[ASAS,][]{asas}, All-Sky Automated Survey for SuperNovae \citep[ASAS-SN,][]{assassn}, Massive Compact Halo project \citep[MACHO,][]{macho}, Hipparcos Catalogue \citep{hipparcos}, Optical Gravitational Lensing Experiment \citep[OGLE,][]{OGLE}, Catalina survey \citep{Drake2014}, Asteroid Terrestrial-impact Last Alert System \citep[ATLAS,][]{atlas}, WISE survey \citep{Chen2018WISE}, EROS-II variability search \citep{EROS} and Zwicky Transient Factory \citep[ZTF,][]{Chen2020} have steadily expanded the catalogue of variable sources. Now in the era of Gaia, the number of variable sources is increasing more rapidly with approximately half a million variables in Gaia DR2 \citep{gaiadr2} and $\sim5$ million periodic variables expected in the final Gaia data release \citep{EyerCuypers}. With the imminent influx of data from the Vera Rubin Observatory \citep{lsst} providing high cadence observations of the southern sky, the possibility of extending catalogues of variables in the Milky Way (and nearby Local Group galaxies) to near completion has advanced.

With these increasingly large surveys, the demand for automated pipelines performing reliable extraction and identification of variable objects has increased. Such a problem naturally lends itself to machine-learning classification algorithms \citep{debosscher2007,richards2011,dubath2011,Kim2011,Bloom2012} where typically a set of features are constructed from photometric light curves and training labels are taken from previous variable star catalogues, such as the General Catalogue of Variable Stars \citep[GCVS,][]{gcvs} which has been continuously compiling these objects since 1946. Similar procedures have been employed recently for the ASAS-SN data \citep{jayasinghe2018, jayasinghe2019}, for Gaia DR2 \citep{Rimoldini2019}, for EROS-II light curves \citep{Kim2014} and for discovering microlensing events \citep{Husseiniova2021}. This supervised learning approach requires large sets of pre-classified variable objects which span the range of objects observed by a new survey. Any biases in the training set are naturally reflected in the outputs of the classifier, and can often be inherited from an initial human classification. Unsupervised learning algorithms, e.g. clustering, are more agnostic and are able to find natural structure within datasets as well as detect unusual objects \citep{Brett2004, Eyer2005, Sarro2009}. An iterative combination of supervised and unsupervised algorithms is potentially the only way to handle and comprehend the ever increasing datasets and to extract important variable sources.

One goal of hunting for variable objects within large photometric surveys of the Milky Way is to utilise them in the mapping of Galactic structure. The poster child for precision distance measurement from variability is the classical Cepheid variable \citep{Leavitt,Riess2019}. Although they have been used with great success to understand the Milky Way's spiral arms \citep[e.g.][]{Skowron2019}, warp \citep{Chen2019} and nuclear stellar disc \citep{Matsunaga2011}, Cepheid variables are associated with young stellar populations and relatively rare, making their use as a widespread Galactic structure tracer limited. Significantly more numerous are eclipsing binary systems and RR Lyrae stars, both of which trace different populations within the Galaxy. Traditionally, RR Lyrae stars have been used to trace the more metal-poor older populations either from in-situ early Milky Way star formation or arising from accreted substructure \citep[see chapter 6 of][]{Catelan2014}, whilst eclipsing binaries are more agnostic of specific stellar population and as such make better tracers for the entire Milky Way's evolutionary history \citep{duchene,chen2016}.

Eclipsing binaries are typically classified according to the distribution of each stellar component within a critical two-lobed figure-of-eight equipotential, known as the \emph{Roche lobes}. Following the GCVS convention, there are three main eclipsing binary types: contact (EW), semi-detached (EB) and detached (EA). In contact binaries, such as the archetypal \emph{W Ursae Majoris}, both components overflow their respective Roche lobe and share a common envelope, placing the stars in contact and thermal equilibrium. EW binaries are known to follow a tight period-luminosity relations in the infra-red on account of their common envelope evolution \citep{Chen2018}. This, and their low bias with age and metallicity of a population, makes them numerous precise tracers of local Galactic disc and bulge structure.

RR Lyrae variables are single pulsating stars, which occupy the helium-burning, horizontal branch of the Hertzprung-Russell (HR) diagram. Typically RR Lyrae stars are subdivided into types based upon their pulsation mode which is reflected in the shape of their light curves: RR Lyrae type a or b stars (RRab) oscillate in the radial fundamental period whereas RR Lyrae type c stars (RRc) oscillate at the radial overtone period. A combination of the two exist in the form of RR Lyrae type d stars (RRd), which oscillate simultaneously with both periods. Typically, RR Lyrae stars are tracers of old, metal-poor components of the Galaxy \citep{Dekany2013,Kunder2016,Du2020}. This makes them ideal for studying the very early stages of the formation of the Milky Way, and for the identification of accretion events in the Galaxy \citep{IorioBelokurov}. Recently, it has been noticed that many RR Lyrae stars appear to be on disc-like orbits \citep{Marsakov2019, IorioBelokurov2, Prudil2020} indicating that the traditional viewpoint is incomplete and alternative channels for RR Lyrae star production, possibly through binary evolution, are possible. A variable star catalogue then allows the construction of samples of different stellar populations with precise distances and low contamination, tracing different evolutionary stages in the Galaxy.

One limitation of studying the variable sky is the depth and coverage of different surveys, which can introduce biases and hinder the study of certain populations. Currently Gaia \citep{Rimoldini2019} is the only all-sky variability survey from which a relatively unbiased magnitude-limited census can be constructed (although Gaia's observing window and cadence still produces on-sky variations). However, Gaia is limited by operating in the optical so will crucially miss many important variables in the inner disc and bulge shrouded by dust. For example, \cite{Clementini2019} present catalogues of Gaia DR2 Cepheids and RR Lyrae stars of which there are $69$ and $7541$ respectively in the VVV bulge footprint. As a complement, the two decade long OGLE survey \citep{OGLE} has observed in the $I$ (and $V$) bands producing deeper catalogues of eclipsing binaries \citep{soszynskibin} and RR Lyrae stars \citep{soszynskilyr} in the inner bulge and disc, although again extinction limits the coverage within a few degrees of the Galactic mid-plane.

The deep near-IR ($0.9-2.5 \mu \mathrm{m} $) Vista Variables in the Vi\'a La\'ctea  survey \citep[VVV,][]{Minniti2010} is a perfect complement to both optical and near-IR variability surveys, such as Gaia and OGLE respectively, with \cite{soszynskilyr} reporting $44,183$ RR Lyrae stars in the VVV bulge footprint. VVV provides multi-epoch photometry in the $K_s$ band over approximately a ten year baseline for a $300\,\mathrm{deg}^2$ bulge region and a portion of the southern disc. As such it is able to reach through the dust of the inner Galaxy to previously undiscovered variables. As a result, catalogues of RR Lyrae stars along the Southern Galactic Plane \citep{Dekany2018}, inner bulge \citep{Dekany2020}, outer bulge \citep{Gran2016} and 100 arcmin from the Galactic Centre \citep{Minniti2016, Contreras2018} have been compiled for analysis. Similarly, VISTA Variables in the V\'ia L\'actea infrared variability catalogue \citep[VIVA,][]{VIVA} provides a near complete variability census for the Galactic Bulge, whilst \cite{herpich2021} have presented an analysis of previously known variable sources in VVV. These catalogues paired with preliminary results on the RR Lyrae star distribution \citep{Cabral2020}, novae \citep{vvvnovae} and the composite bulge structure of the Milky Way \citep{vvvbulgestructure} have all shown the far-reaching scientific merit of the VVV project. However, as yet, there has been no published catalogues performing a homogeneous, specialised classification of all potential periodic sources across the entire VVV footprint. This is the goal of our work, in which we forfeit absolute completeness in order to produce a catalogue with highly reliable classifications.

In this paper we carry out an automated classification of all sources in VVV using a two-stage hierarchical scheme. We first carry out an initial identification of likely variable sources through reference to a sample of non-variable sources, which are subsequently assigned a detailed classification into their variable type. In Section~\ref{sect:datadesc} we describe the dataset of VVV light curves used and the set of training data employed from the literature. In Section~\ref{sect:const} we outline the construction and performance of our hierarchical classifier, discussing the features employed for classification. In Section~\ref{sect:results} we describe the results of applying our classifier to the VVV data, before briefly touching on some basic science applications using our new VIVACE catalogue in Section~\ref{sect:science}. We close with our conclusions in Section~\ref{sect:conclusion}.

\section{Light curves and the training set} \label{sect:datadesc}

Our approach requires a set of photometric light curves taken from the VVV survey, of which a subset have pre-existing classifications or labels from other (variable) catalogues. We begin by describing the set of VVV light curves, before going on to discuss the construction of the training set from various variable catalogues. We close the section by presenting a method to establish unbiased constant source classes, using Gaia photometry, to augment the training set.

\subsection{VVV photometric dataset}\label{subsect:phot_dataset}

\begin{figure*}
    \centering
    \includegraphics[width=\linewidth]{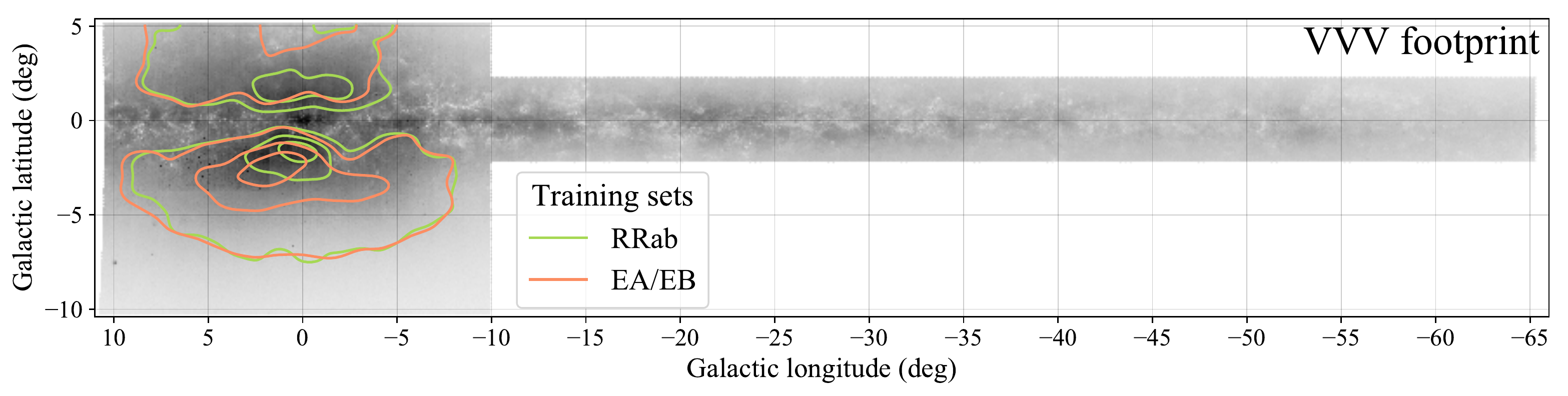}
    \caption{The on-sky density of the VVV source catalogue (with a square-root stretch) along with contours ($5,10,80\,\percent$ of peak density) showing the distribution of two training samples (RR Lyrae ab stars and EA/EB eclipsing binaries), primarily taken from OGLE.}
    \label{fig:footprint}
\end{figure*}

The VISTA Variable in the V\'ia L\'actea (VVV) public ESO (European Southern Observatory) project is an ongoing survey which has been performing near-infrared observations of the Galactic bulge and inner disk since 2010 \citep{Minniti2010,Saito2012}. The VVV photometric observations were taken from the Cerro Paranal Observatory, in Chile, using the $4\,$m ``wide-field'' VISTA telescope. The VIRCAM detector \citep{Dalton2006}, consisting of sixteen 2048$\times$2048 pixel arrays, was mounted on the telescope, allowing for a combined image of the sky covering a $1.5 \times 1.1\,\mathrm{deg}^2$ tile. The detector was equipped with 5 broad-band filters $Z Y J H K_s$ and two narrow-band filters centred at 0.98 and 1.18$\,\mu$m. The VVV survey used the 5 broad band filters spanning $0.84$ to $2.5\, \mu$m. The full on-sky area spanned by the survey is 562 deg$^2$ discretised into 348 tiles, 196 tiles in the bulge ($-10.0^{\circ}\leq l \leq +10.4^{\circ}$ and $-10.3^{\circ} \leq b \leq +5.1^{\circ}$) and 152 in the disk ($294.7^{\circ}\leq l \leq 350.0^{\circ}$ and $-2.25^{\circ} \leq b \leq +2.25^{\circ}$). A single observation of a tile is composed of six VIRCAM pawprints resulting in between one and six detections per source per tile observation. Here we work with the individual pawprint detections. In Fig.~\ref{fig:footprint} we display the footprint and source density of the survey. The original VVV survey concluded in 2015, but from 2016 the survey was granted a five-year extension (VVVX) to extend both the baseline of observations of the original VVV footprint and the spatial coverage of the survey. Here we consider observations (both VVV and VVVX) only within the original VVV footprint.

The principal data product of VVV is multi-epoch $K_s$-band photometry taken across a $\sim10$ year baseline from which variable objects can be identified and classified. The number of $K_s$-band observations is typically $130-230$ in the disc region and $170-320$ in the bulge region, except in high-cadence bulge tiles ($+2.417^{\circ}\leq l \leq +6.807 ^{\circ}$ and $-3.138^{\circ} \leq b \leq -2.043^{\circ}$) where there are $660-1250$ observations. The high-cadence region is visible as a darker rectangular overdensity in Fig.~\ref{fig:footprint} as a result of the increased visibility of sources. The $K_s$ band photometry is complemented by several $Z Y J H$ observations allowing for sparsely sampled time-series analysis in these bands.

The initial VVV data reduction and pre-processing was executed via the VISTA Data Flow System \citep{vista}, by a collaboration of the Cambridge Astronomy Survey Unit (CASU) and UK Wide-Field Astronomy Unit (WFAU). The catalogue employed here consists of the unpublished 2nd version of the VVV Infrared Astrometric Catalogue (VIRAC, Smith et al., in prep), an upgrade to the original VIRAC catalogue \citep{virac}. The primary purpose of the VIRAC project was to provide astrometry (parallaxes and proper motions) from the multi-epoch VVV observations. VIRAC v2 provides improved astrometry and a new photometric reduction, which utilises point-spread-function photometry using a version of DoPhot \citep{dophot} modified by \cite{dophot2}. In total, $1.14\times10^{11}$ sources were detected by DoPhot across all images. The initial photometry was then calibrated using the CASU calibration \citep{GonzalezFernandez2018}. As part of VIRAC v2, an SDSS-like ubercal has been carried out on the photometry, where a time and pixel-dependent zeropoint for each chip was fitted using a pool of 2MASS reference sources. A pool of astrometric reference sources observed by Gaia (DR2) were used to calibrate the astrometry of each image. A source catalogue was then formed from first grouping nearby detections, performing astrometric fits, and then iteratively reassessing the source matching using the astrometry. The final VIRAC v2 catalogue we utilise contains all non-duplicate sources (groups of sources within $0.339"$ are labelled duplicates if they don't have the most detections of the group or if more than $20\,\percent$ of detections are shared by another grouped source) detected in more than $20\,\percent$ of observations and with a five-parameter astrometric solution. This results in a set of $539,845,118$ sources equipped with $Z Y J H K_s$ photometric `light curves', with the majority of the detections in $K_s$.

For each $K_s$ light curve, there is a set of quality indicators for individual detections which are useful for removing low quality or spurious data. Each detection is assigned a non-zero \texttt{ambiguous\_match} flag if it is shared by more than one non-duplicate source. The astrometric residual chi-squared (\texttt{ast\_res\_chisq}) gives a measure of whether a detection is likely spurious based on the corresponding source's astrometric fit. All of our light curves are `cleaned' by ensuring \texttt{ambiguous\_match}$=0$, \texttt{ast\_res\_chisq}$<11.829$ (approximately equivalent to a $3\sigma$ boundary)  and photometric point spread function fit chi-squared \texttt{chi}$<5$ for bright detections ($K_s<13.2$). We further only retain photometric light curves with number of $K_s$ epochs $>20$ and $11.5\leq \mathrm{mean}(K_s) \leq 17$. The magnitude boundaries were included to avoid saturation errors for bright stars and high observational errors associated with faint stars. The restriction on number of epochs was put in place in order to ensure reliable period recovery and limit the appearance of extremely unevenly sampled sources. This further set of cuts reduces our considered dataset to approximately $490$ million sources. Of these, $Z Y J H$ light curve analysis was only able to be performed for sources with more than one detection in each auxiliary band

\subsection{Variable star training sets} \label{subsect:variable trainset}

To construct our hierarchical classifier, we require a comprehensive set of accurately classified variable stars with corresponding VVV light curves. We considered all periodic variable types with sufficiently high-amplitude variations and with significant representation in our dataset: detached/semi-detached (EA/EB), contact (EW) and ellipsoidal (Ell) eclipsing binary stars, fundamental mode (RRab), overtone (RRc) and double-mode (RRd) pulsating RR Lyrae stars, classical (CEP) and Type II (T2CEP) Cepheid variables and long period variables (LPV) formed of Mira, semi-regular and OGLE small-amplitude variables. $\delta$-Scuti stars (DSCT) were included in early tests but the quality of the VVV data did not allow for their confident separation from non-variable sources. Furthermore, we don't include young stellar objects (YSOs) in our training set. These are typically intrinsically red objects and can show a range of stochastic and quasi-periodic variability. Although our algorithm is designed to target periodic variables, we antipicate some level of contamination from YSOs in the LPV class as seen by \cite{Mowlavi2018}.

The primary source of our classification labels is the \emph{Optical Gravitational Lensing Experiment} \citep[][OGLE]{OGLE}. OGLE has monitored the brightness of roughly 2 billion stars in both the $V$ and $I$ bands over a 2750 deg$^2$ footprint covering the Galactic plane and bulge. The VVV footprint has excellent overlap with the bulge/disc component of OGLE as shown in Fig.~\ref{fig:footprint}. The lack of representation of the Galactic mid-plane in OGLE when compared to VVV is also apparent. OGLE has provided $\sim450,000$ ellipsoidal and eclipsing binary star classifications towards the Galactic bulge \citep{soszynskibin}, of which $80.0\,\percent$ were positionally cross-matched to our previously defined VVV dataset within $1\,\mathrm{arcsec}$. Similarly, \cite{soszynskilyr} have provided a catalogue of $\sim78,000$ RR Lyrae stars in the Galactic bulge and disc using OGLE data. A previous version of the RR Lyrae star catalogue containing $\sim38,000$ RR Lyrae stars was used for this work \citep{Soszynski2014}, of which $90.8\,\percent$ had matches to VVV within $1\,\mathrm{arcsec}$. It was found that a sufficient number of RR Lyrae stars were included to fully represent the narrow clustering of the class in feature space, such that the diminished subset of RR Lyrae stars considered did not hamper the accuracy of the classification. We chose to combine overtone and double-mode RR Lyrae stars into a single class (RRcd), as their light curves are for the most part indistinguishable. Long-period variable and Cepheid variable training samples were taken from corresponding OGLE variable catalogues \citep{Soszynski2013,Udalski2018}. This yielded 291 CEP, 673 T2CEP and 837 LPV variables cross-matched within $1\,\mathrm{arcsec}$ to our VVV dataset.
We supplemented this set using the AAVSO International Variable Star Index\footnote{https://www.aavso.org/vsx/} \citep[VSX, accessed October 2020,][]{VSX} catalogue with DSCT, additional long period variables labelled as M, SRV, SRA, SRB or SRS and classical Cepheids labelled as CEP, DCEP or DCEP(B). We found 160 DSCT, 52 LPV and 41 CEP variables in our light curve set, removing any duplicates found in OGLE.

Due to many known LPVs being too bright and saturated in VVV, we found that representation of the class within the training set was inadequate. We therefore supplemented the training set with $\sim1700$ Mira variables in the nuclear stellar disc ($|\ell|<1.5\,\mathrm{deg}, |b|<1.5\,\mathrm{deg}$) discovered from VIRAC v2 data (Sanders et al. , in prep.), of which 673 are non-duplicates found in our photometric dataset.

In Table~\ref{tab:trainset} we give the number of sources in each class used for the different classification stages. In total there are $405, 686$ known variable sources with VIRAC-2 light curves satisfying our quality cuts (primarily on magnitude and number of detections).

\newlength{\storetabcolsep}
\setlength{\storetabcolsep}{\tabcolsep}
\setlength{\tabcolsep}{5pt}
    \begin{table}
        \caption{Training set of variable sources in VVV (predominantly from the OGLE and VSX variable catalogues) as used in both stage 1 ($N_1$) and stage 2 ($N_2$) of the hierarchical classifier. Period mismatch rate gives the percentage of sources with Lomb-Scargle false alarm probability $<1\times10^{-10}$ which don't have VVV periods matching the corresponding literature period within $10\,\percent$.} \label{tab:trainset}
        \centering
        \begin{tabular}{lcccc}
            \hline
            Class & $N_1$ & $N_2$ & \makecell{Period \\mismatch \\($\percent$)}\\ \hline
            Contact Ecl. (EW) & 67 926 & 40 116 & 1.71 \\
            Detached Ecl. (EA/EB) & 281 480 & 109 744 & 2.74\\
            Ellipsoidal (Ell) & 18 055 & 11 263 & 1.61\\
            RR Lyrae ab (RRab) & 24 837 & 22 419 & 0.56\\
            RR Lyrae c/d (RRcd) & 10 661 & 6 679 & 3.64\\
            Classical Cep. (CEP)& 332 & 306 & 0.66\\
            Type II Cep. (T2CEP) & 673 & 644 & 1.09\\
            Long period (LPV) & 1 562 & 753 & 32.00\\
            $\delta$-Scuti (DSCT) & 160 & n/a & n/a\\
            \hline
            Total & 405 686 & 191 924\\ \hline
        \end{tabular}
    \end{table}
\setlength{\tabcolsep}{\storetabcolsep}

\subsection{Gaia constant source reference set} \label{subsect:gaia const}

\begin{figure}
\centering
\includegraphics[width=\columnwidth]{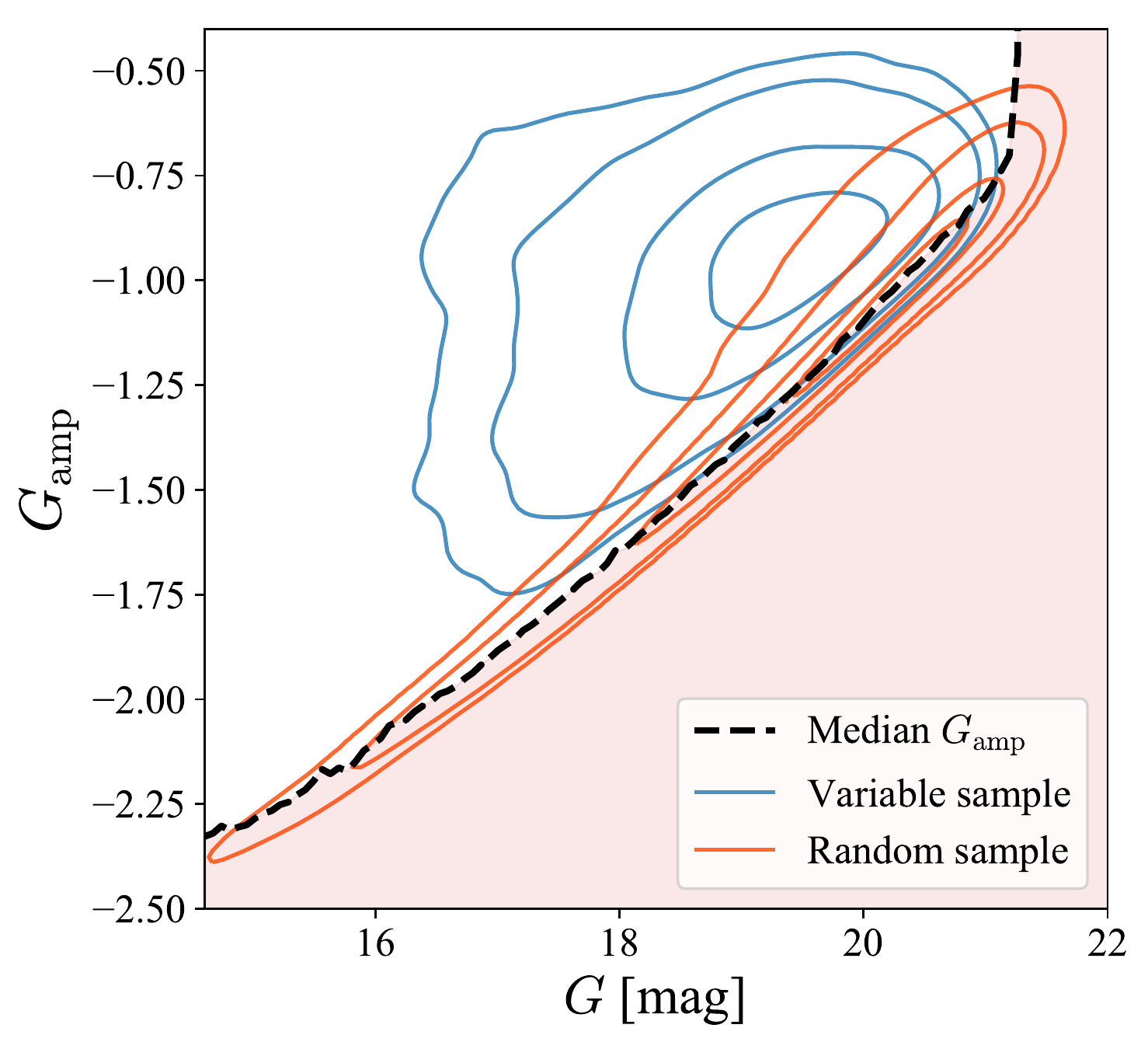}
\caption{Distribution of $G_\mathrm{amp}$ against $G$ mean magnitude for samples of known variable stars and random Gaia EDR3 sources (contours correspond to $10, 20, 50$ and $80\,\percent$ of the peak density). The variable sample consists of our compiled variable training set cross-matched to Gaia within $0.4\,\mathrm{arcsec}$ whilst the random sample depicts a representative subset of all Gaia EDR3 data. The black dashed line highlights the median $G_\mathrm{amp}$ trend for the random sample, below which constant sources are selected from the red shaded area.}
\label{fig:gaia_extract}
\end{figure}

To initially classify a source into variable vs. non-variable, we define a training set of likely non-variable sources (labelled CONST). Although constant sources should in principle be easily identified, there are cases where systematics (e.g. blending) give rise to spurious variability (occasionally periodic variability due to seasonal seeing variations). Furthermore, selecting non-variable sources from VVV itself, using photometry statistics or by other means, would introduce unwanted selection biases into our modelling, as the initial classification would be based on these same statistics. Therefore, to construct an unbiased sample we utilise photometry from the Gaia Early Data Release 3 \citep[EDR3,][]{Gaia,GaiaEDR3}. Gaia has a smaller point-spread function than VVV allowing for cleaner separation of sources in crowded regions. For each Gaia source, we compute the scatter in $G$ \citep{Belokurov2017} as
\begin{equation}
    G_{\text{amp}} = \logten{\left(\sqrt{N_\mathrm{obs}} \frac{\sigma_{\overline{I_G}}}{\overline{I_G}}\right)},
\end{equation}
where $N_\mathrm{obs}$ is the number of observations, $\overline{I_G}$ is the mean $G$-band flux and $\sigma_{\overline{I_G}}$ is the error in the mean flux estimate. Due to the continually changing brightness of variable stars, it is expected that, in the absence of systematic issues, stars at a given magnitude with larger $G_{\mathrm{amp}}$ are likely intrinsically variable. Consequently, we use this statistic as a magnitude-dependent signal-to-noise proxy for analysing variability in source light curves.
For a set of Gaia sources cross-matched to VVV within an angular separation of $0.4\,\mathrm{arcsec}$ (accounting for the Gaia EDR3 and VVV epoch difference using the Gaia proper motions) and satisfying the previously-mentioned VVV epoch and magnitude selection criteria, we characterise the typical $G$-band magnitude scatter as a function of $G$ and identify non-variable sources as those that lie beneath the median trend. This method assumes that at a given magnitude the variable stars are in the minority. Testing the procedure yielded clear placement of known variables, taken as our VVV variable training set cross-matched to Gaia in the same way, above the median trend of Gaia sources as shown in Fig.~\ref{fig:gaia_extract}. Using this procedure, sets of constant VVV sources could be extracted for any Galactic region of interest.

\section{Two-stage hierarchical classifier} \label{sect:const}

With $490$ million sources selected from VVV, the task of performing detailed analysis on each light curve is highly computationally intensive. We therefore adopt an initial selection procedure to identify likely variable objects using computationally-cheap variability indicators. This step makes up \emph{stage 1} of the hierarchy. The greatly reduced sample of candidate variables is then further classified into detailed variable types, \emph{stage 2} of the method, based on more comprehensive analysis of the light curves. In both stages we utilise supervised machine learning algorithms. In particular, we view stage 1 as a binary and stage 2 as a multi-class classification problem. The aim in either exercise is to determine sources which cluster, in some feature space, equivalently to representative samples of general variable objects for stage 1 and distinct variability classes for stage 2.

Ensemble tree classifiers rank among the most efficient binary classification methods \citep{Pashchenko2018} and have been used to great effect in initial variable detection for the Gaia Variability Analysis pipeline \citep{Eyer2017,Rimoldini2019}. Similar conclusions are noted for detailed variable classification \citep{debosscher2007,dubath2011,richards2011}, which lead us to adopt these methods for our purposes. Ensemble methods group large sets of weak classifiers with predictive ability slightly above uniform chance. This grouping enables robust discrimination between parent classes without overfitting. To construct such models we require \begin{inparaenum}
\item training set classes of stellar sources with known (broad or specific) variability labels,
\item a list of features, to be extracted for each source, representing the differences between classes.
\end{inparaenum}
We give a brief description of the algorithms used in each step before discussing the training set and feature spaces employed in detail.

The initial variable/non-variable binary classification stage uses a Random Forest \citep[RF]{Breiman2001}. This algorithm is well suited to the problem in hand due to \begin{inparaenum}
\item the ability to model hidden underlying relations in large feature spaces,
\item the ease with which multiple class partitions are incorporated and
\item the low computational cost \citep{debosscher2007,richards2011}.
\end{inparaenum}
Given a training set, a decision tree is `grown' through sequential splits of the feature space at nodes, in order to minimize the statistical entropy of the resulting divided feature space. The tree expands until splits are no longer beneficial, at which point each path, along nodes, terminates in a leaf. The class attributed to a leaf subspace is then defined in terms of the fraction of particular class training examples populating the subspace. Decision trees of this kind are grouped into ensembles forming Random Forests. RFs reduce overfitting by utilising \emph{bootstrap aggregation} where training subsets, with randomly replaced features, are used in each tree and \emph{random feature subspaces} where only a random subset of features are considered at each potential node.

The second stage detailed variable classification uses the gradient boosting tree algorithm as implemented in XGBoost \citep[eXtreme Gradient Boosting,][]{XGBoost}. This algorithm sequentially constructs a series of decision trees, minimising the loss function at each stage with the inclusion of a regularization to reduce overfitting. The choice of XGBoost for our second stage classifier was based on the increased speed of computation with the greater number of features employed and also a small improvement in performance on our test set.

We first describe the training classes and features for the variable/non-variable RF classifier in Section~\ref{subsect:stage1} and the detailed variable XGBoost classifier in Section~\ref{subsect:stage2}. This discussion is followed by an analysis of the performance of each classifier based on k-fold cross validation of their training samples in Section~\ref{subsect:train}.

\subsection{Stage 1: variable/non-variable classification}\label{subsect:stage1}

The first stage of our classifier consists of selecting a list of candidate variables from the full VVV survey. Here we describe the training sets and feature spaces utilised to construct the binary classifiers necessary for candidate extraction.

\subsubsection{VVV tile training sets}
Training set examples are given broad labels of either variable (VAR) or non-variable (CONST). The VAR class consists of all the variable star sets from Section~\ref{subsect:variable trainset} grouped together. As the majority of collected variables are taken from OGLE, the variable class footprint mirrors the OGLE footprint with a peak in source density around $1<|b|<3\,\mathrm{deg}$ in the Galactic bulge (see Fig.~\ref{fig:footprint}). Though morphological differences of light curves exist among variability classes, continuous periodic magnitude variation is a common feature. We therefore employ variability summary statistics reflecting this commonality to separate them from non-variable objects.

Observational cadence and intrinsic physical differences such as stellar source density vary widely over the VVV footprint. This can lead to irregular photometric data for distinct on-sky regions. In particular, blending of multiple light sources increases dramatically towards the crowded regions of the Galactic plane. As a consequence, observations of a constant magnitude source within a crowded region may differ from those of an isolated constant source found at higher latitudes. If considered equivalently, the relative loss in accuracy of the former could be interpreted as a signal for variability of the source.  For this reason, we decide to construct region specific binary classifiers. The natural tiling provided by the VVV observational procedure is chosen to be of adequate size to reflect near-homogeneous conditions for constant source extraction \citep[see][for a discussion of feature variation across the VVV footprint]{Cabral2021}. We therefore construct 348 binary classifiers, one for every VVV tile, each with a unique constant (CONST) class which we attempt to distinguish from the shared variable source class. Splitting the variable class into tile-specific subsets is not found to be vital as the intrinsic variability of these sources overwhelms the effect of location dependent spurious detections. The distinct CONST classes consist of Gaia-sampled constant sources, extraction described in Section~\ref{subsect:gaia const}, limited to the respective tile on-sky regions. We down-sample the constant training set in each tile in order to have at least as many sources as the variable training set ($\sim400,000$) with many tiles having fewer sources ($50,000$ at a minimum but $50\,\percent$ of tiles with more than $300,000$).

\begin{figure*}
    \centering
    \includegraphics[width=\textwidth]{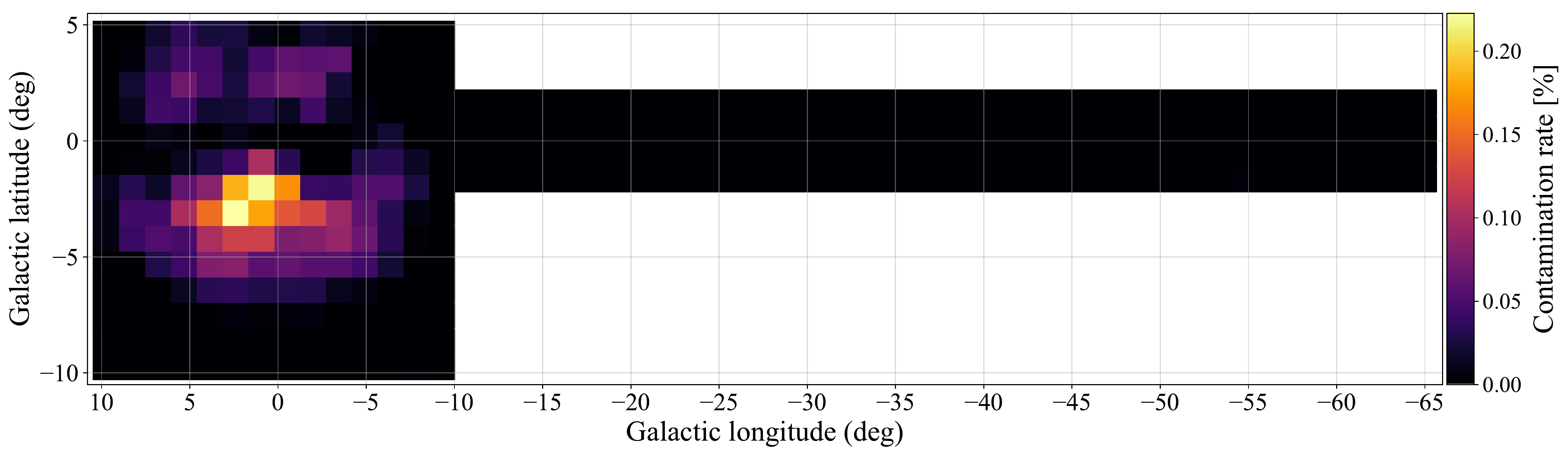}
    \caption{Grid of 348 binary classifiers coloured by the contamination ratio of known variable sources within each tile's non-variable training set. The distribution mirrors the OGLE variable catalogue density as contamination correlates with known variable source density. Maximum contamination (bright yellow) is noted in the OGLE complete region, where the majority of variable sources are identified by OGLE.}
    \label{fig:const_contam}
\end{figure*}

Certain low scatter variables, predominantly detached binary systems, exhibit minimal time-averaged flux variation from the mean. This is due to their light curves showing near constant magnitude, reflecting the brightness of the dominant source in the pair, with brief sharp drops due to eclipses. This led to a small number of known variable star contaminants within the non-variable samples. The mean contamination rate as a fraction of total constant class size for the 348 binary classifiers is 0.05$\,\percent$ with variation over tiles shown in Fig.~\ref{fig:const_contam}. Lack of completeness of the known variable training set suggests these numbers are underestimates, as many variable contaminants could be present which we have not collected. However, contamination peaks at 0.23$\percent$ for tiles in the OGLE complete region, where the majority of variables which are not extinction limited are accounted for and hence should be viewed as a minimal upper bound. We therefore remove any identified contaminants and construct each tile's constant class from the remaining sources.

\subsubsection{Variability features}
Having established the respective training samples to adopt, the PSF photometry available from the VIRAC v2 catalogue is exploited to determine a set of easily computable \emph{variability indices}. These features are chosen to reflect the degree of variability present in the magnitude observations of individual sources, typically the scatter about the mean or correlation between time-ordered detections. Crucially for our purposes, their estimation must be computationally scalable \citep[for a review of possible variability indices see][]{sokolovsky}. We define the variability indices employed in the upper part of Table~\ref{tab:var_indices} in Appendix~\ref{appendix::feature_table}. These include the median absolute deviation, standard deviation, higher order moments such as the skewness and kurtosis, the von Neumann ratio $\eta$ \citep{vonneumann1941}, the Stetson $I$, $J$ and $K$ indices \citep{stetson_i} as well as percentile ranges illustrating the scatter of points in light curves. The Stetson $I$ index computation considered pairs of observations separated by less than $1$ hour. Following \cite{Shin2009}, the Stetson $J$ and $K$ indices used all pairs of observations \citep[note in the definition of Stetson $J$ from][there is no normalising factor for the number of pairs]{Shin2009}. We supplement this feature list with error-weighted versions of the median absolute deviation, standard deviation and percentile ranges. These features were computed on the full VIRAC-2 dataset using all detections with no \texttt{ambiguous\_match} flag, a looser quality cut than that described in Section~\ref{subsect:phot_dataset}.

\begin{figure*}
\centering
\includegraphics[width=.4\linewidth]{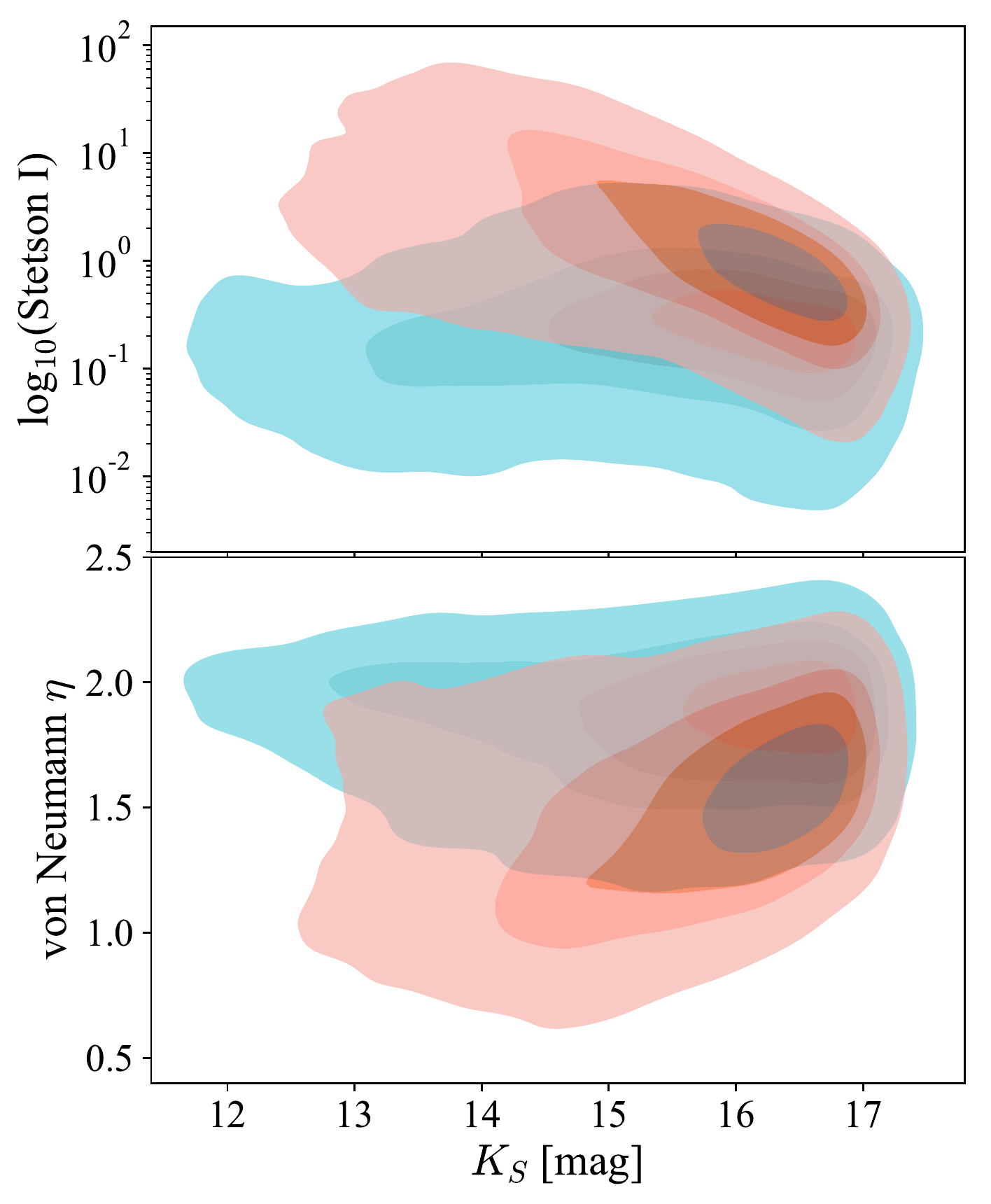}
\includegraphics[width=.4\linewidth]{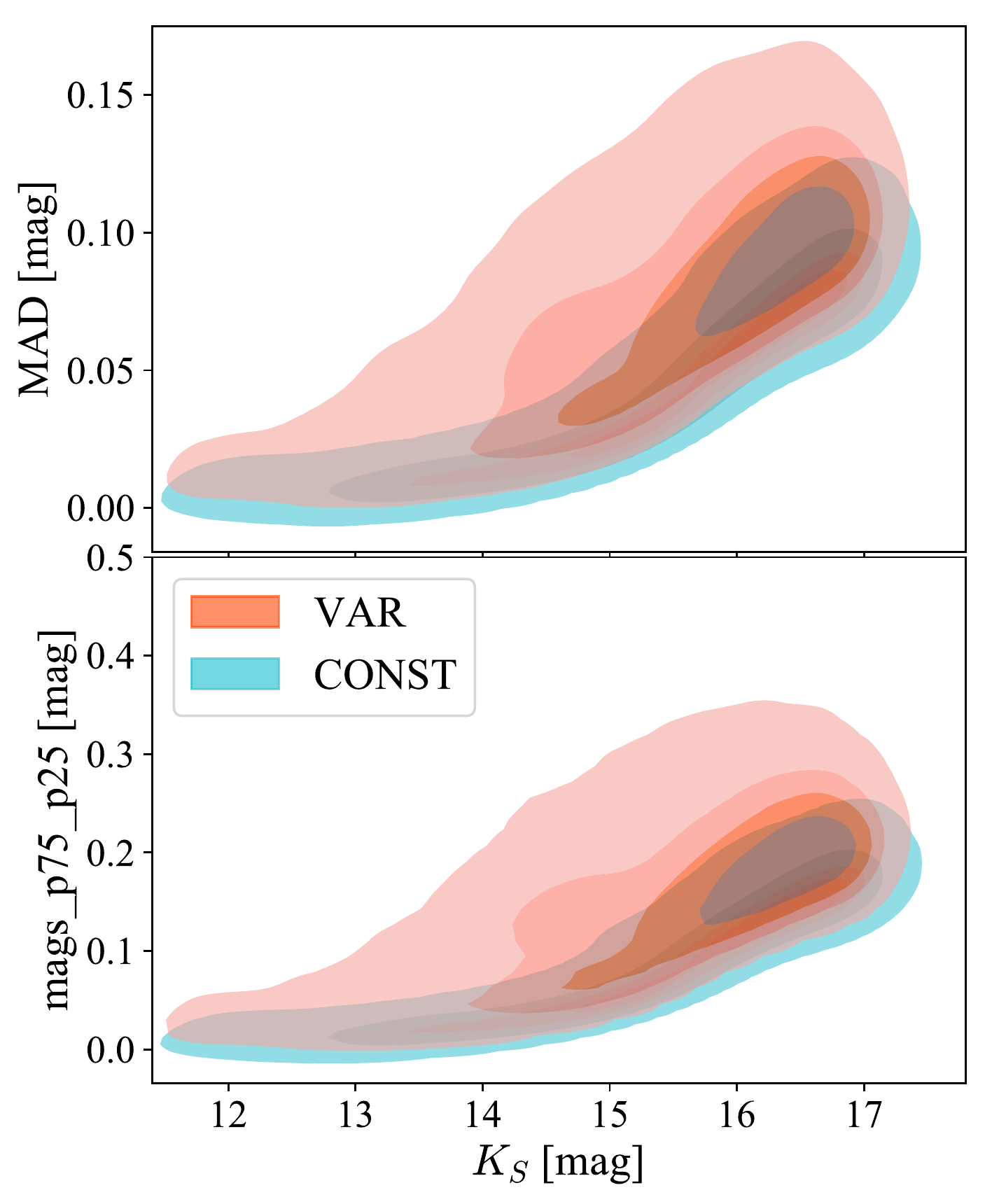}
\caption{Distribution of variability indices, used for initial variable/non-variable classification, for the full variable training set (VAR) and a constant class sampled from the full VVV footprint (CONST). The contours correspond to $10, 20, 50$ and $80\,\percent$ of the peak density. Separation is significantly noticed for $\log_{10}\left(\mathrm{Stetson}\,I\right)$ and von Neumann $\eta$ (left panel) which quantify time-correlation of magnitudes and to a lesser extent for scatter features (right panel).}
\label{fig:varindices_dist}
\end{figure*}

In Fig.~\ref{fig:varindices_dist}, we show distributions of a subset of these indices against magnitude for the VAR class and an example CONST class sampled from the full VVV footprint. Separation between the samples is greatest for the time-correlated features Stetson $I$ and von Neumann $\eta$ and is present to a lesser extent in the scatter features. This follows expectations as the former describes light curve smoothness present in variables with underlying periodicity and absent in constant sources with uncorrelated datapoints on short timescales.

\subsection{Stage 2: detailed variable classification}\label{subsect:stage2}

The second stage in the hierarchical classifier involves detailed classification of identified candidate variable objects using a set of additional features. The variable training set from Section~\ref{subsect:variable trainset} is taken with the following separate labels: EA/EB, EW, Ell, RRab, RRcd, CEP, T2CEP and LPV. In addition, a constant class, again labelled CONST, is included as a comparative baseline. The creation of this class follows the procedure outlined in the previous section with sources chosen spanning the full VVV footprint, rather than being chosen from a particular Galactic region. The size of the class was chosen to be a tenth of the full variable dataset to match the approximate average size of the variable classes.

As with other large-scale variable source classification projects (see Section~\ref{sect:intro}), we carry out a periodic decomposition of our $K_s$-band light curves, yielding Fourier components and period estimates. The period of a source is an important feature for distinguishing different variable types, whilst light curve shape and asymmetry allows separation of RR Lyrae stars and eclipsing binaries, for example. Similarly, lack of periodicity of CONST sources entails arbitrary periodic feature distributions which diverge from the constrained distributions for variable stars.

We first describe our method for periodic feature extraction before highlighting additional non-periodic features which have proved useful for this stage. We then examine our period estimates with respect to values quoted in OGLE and VSX. A comprehensive list of the novel classifier features for stage 2 is shown in the lower part of Table~\ref{tab:var_indices}. Note these are employed in addition to the previously considered stage 1 variability indices.

\subsubsection{Periodic features}\label{subsubsect:periodic_features}
Higher-order Fourier models are necessary to accurately fit the complex qualities of light curves under consideration, such as sharp and uneven minima produced by detached binary sources. However, computational expense increases linearly with the number of Fourier terms. Hence, we take advantage of the approximate sinusoidal nature of all variable sources by first employing a floating-mean Lomb-Scargle periodogram \citep{Lomb1976,Scargle1982,Zechmeister2009,Vanderplas2018} to identify a set of candidate frequencies which we can further process. The Lomb-Scargle periodogram is equivalent to fitting the function
\begin{equation}
    m(t|\omega, \boldsymbol{\theta}) = c + a\sin{(\omega t)} +b\cos{(\omega t)},
     \label{eq:model1}
\end{equation}
for a grid of trial frequencies $\omega$ and $\boldsymbol{\theta}=\left(a, b, c\right)$, where $a$ and $b$ are Fourier coefficients and $c$ is an offset. Utilising the `fast' implementation in \textsc{Astropy} \citep{astropy:2018}, we compute the Lomb-Scargle periodogram for light curves detrended using a cubic polynomial (which accounts for variability on timescales longer than the maximum considered period). A regular grid of frequencies is trialled between $6.67\times10^{-4}\,\mathrm{day}^{-1}$, corresponding to a period of roughly half the baseline of the survey, and $20\,\mathrm{day}^{-1}$, falling below the shortest period expected. The frequency spacing for the trial grid is $0.2/(\mathrm{max}(t)-\mathrm{min}(t))$ ensuring sufficient samples around each peak \citep{Vanderplas2018}. To form a list of candidate frequencies, we select the top $30$ local maxima, or `peaks', in the periodogram. Frequencies are then removed from this list if the Lomb-Scargle power is less than the corresponding power in the window function periodogram \citep[where the magnitudes are replaced with a constant and $c=0$,][giving an indication of the power as result of observational cadence]{Vanderplas2018}. Furthermore, we have found that, if present, short frequencies $f_\mathrm{s}$ can couple with the typical observational cadence (at some frequency $f_\mathrm{d}$) giving rise to beat alias peaks at $(f_\mathrm{d}\pm f_\mathrm{s})$. This produces narrow peaks in the output period distributions of the training set around multiples of a sidereal day. In theory, for a single light curve, it is impossible to distinguish between $f_\mathrm{s}$ and its beat alias. However, it is much more probable for a source to vary on long timescales than timescales very close to the observational alias. We therefore choose to remove a frequency $f$ from the list if the window power at $f\pm f_\mathrm{s}$ is greater than the Lomb-Scargle power at $f$. We define short frequencies as any peak in our original list with $f<f_\mathrm{tol}=0.005\,\mathrm{day}^{-1}$. This procedure has the effect of removing candidate frequencies within $f_\mathrm{tol}$ of likely observational aliases (identified by the window function), but only if there is evidence of an accompanying significant long-term trend. However, the procedure still allows the possibility of true periods very close to a multiple of a sidereal day (which can be important for RR Lyrae stars around $0.5$ days). We supplement the final list of candidate frequencies with half frequencies as binary sources tend to have orbital frequencies half that of the sinusoidal fit frequency.

For each candidate frequency, $f=\omega/2\pi$, we fit a multi-term Fourier series with an extra quadratic component,
\begin{equation}
    m(t|\omega, \boldsymbol{\theta}) = \theta_{0} + \theta_{1}t + \theta_{2}t^2 + \sum_{n=1}^{N_f} a_n\sin{(n\omega t)} +b_n\cos{(n\omega t)},
     \label{eq:model2}
\end{equation}
where $\omega$ and $\boldsymbol{\theta}=(\theta_0, \theta_1, \theta_2, a_1, b_1, a_2, b_2, \cdots)$. The quadratic terms account for any genuine or systematic long term trend.
At each candidate frequency, $\omega$, the best-fitting coefficients are obtained by minimising $\chi^2$ using linear least-squares \citep{Palmer2009}. We also adopt a regularization scheme described in Appendix~\ref{appendix::optimal_regularization} to penalize highly oscillatory model fits. For each light curve we then vary $N_f$ between $4$ and $10$ and select the $(\omega, \boldsymbol{\theta}, N_f)$ combination that minimises the Akaike information criterion, AIC,  $\chi^2+2(2N_f+3)$.  $N_f$ = 4 was found to be adequate for the majority of the light curves, although a greater $N_f$ is favoured for light curves of EA/EB type which can exhibit deep narrow minima.

Once the AIC-minimised results are obtained, a subroutine is carried out for sources whose greatest amplitude is not associated with the first Fourier terms. This is expected to occur when a harmonic of the true physical frequency is taken as the fundamental frequency and once again is mainly specific to binary systems. In this case, harmonic multiples of the best fitting period are passed through the spectral analysis again and taken as correct if the resulting $\chi^2$ is further minimised. This can occur due to the discrete frequency grid employed.

From the set of coefficients $\{a_n,b_n\}$, we calculate amplitudes $A_i=\sqrt{a_i^2+b_i^2}$, phases $\Phi_i=\arctan(-b_n/a_n)$, amplitude ratios, $R_{ij}=A_j/A_i$, and phase differences, $\Phi_{ij}=j\Phi_i-i\Phi_j$, following the conventions of \citet{Petersen1987}. We compute the model amplitude $A_\mathrm{model}$ as the maximum minus minimum magnitude of the best-fitting Fourier model (ignoring the polynomial term) and the data amplitude $A_\mathrm{data}$ as the maximum minus minimum magnitude of the data (similar to mags\_p100\_p0, defined in Table~\ref{tab:var_indices}, but using slightly different quality cuts). Furthermore, we compute the difference in log likelihood per datapoint between a constant source and the best fitting Fourier model, based on residuals when compared to the observed photometry, and the false alarm probability (FAP) associated with the highest peak in the Lomb-Scargle periodogram \citep{Baluev2008}. Both quantities provide good measures of the signal-to-noise ratio in the light curve and accuracy of the model. We detail these in the lower part of Table~\ref{tab:var_indices}.

\begin{figure}
    \centering
    \includegraphics[width=\columnwidth]{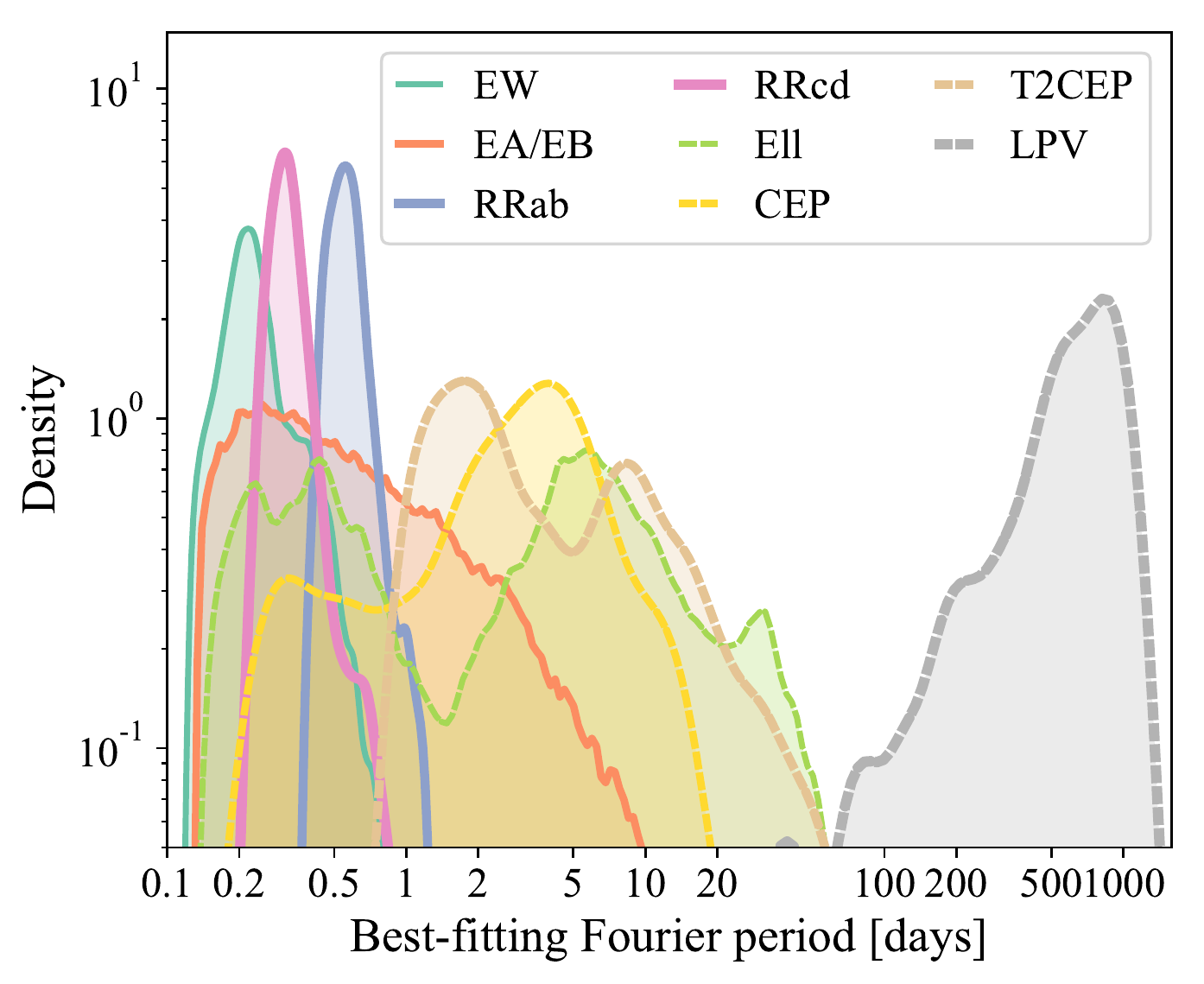}
    \caption{Distribution of best-fitting Fourier model periods for the variable training set.
    }
    \label{fig:period_dist}
\end{figure}

\begin{figure*}
\includegraphics[width=\linewidth]{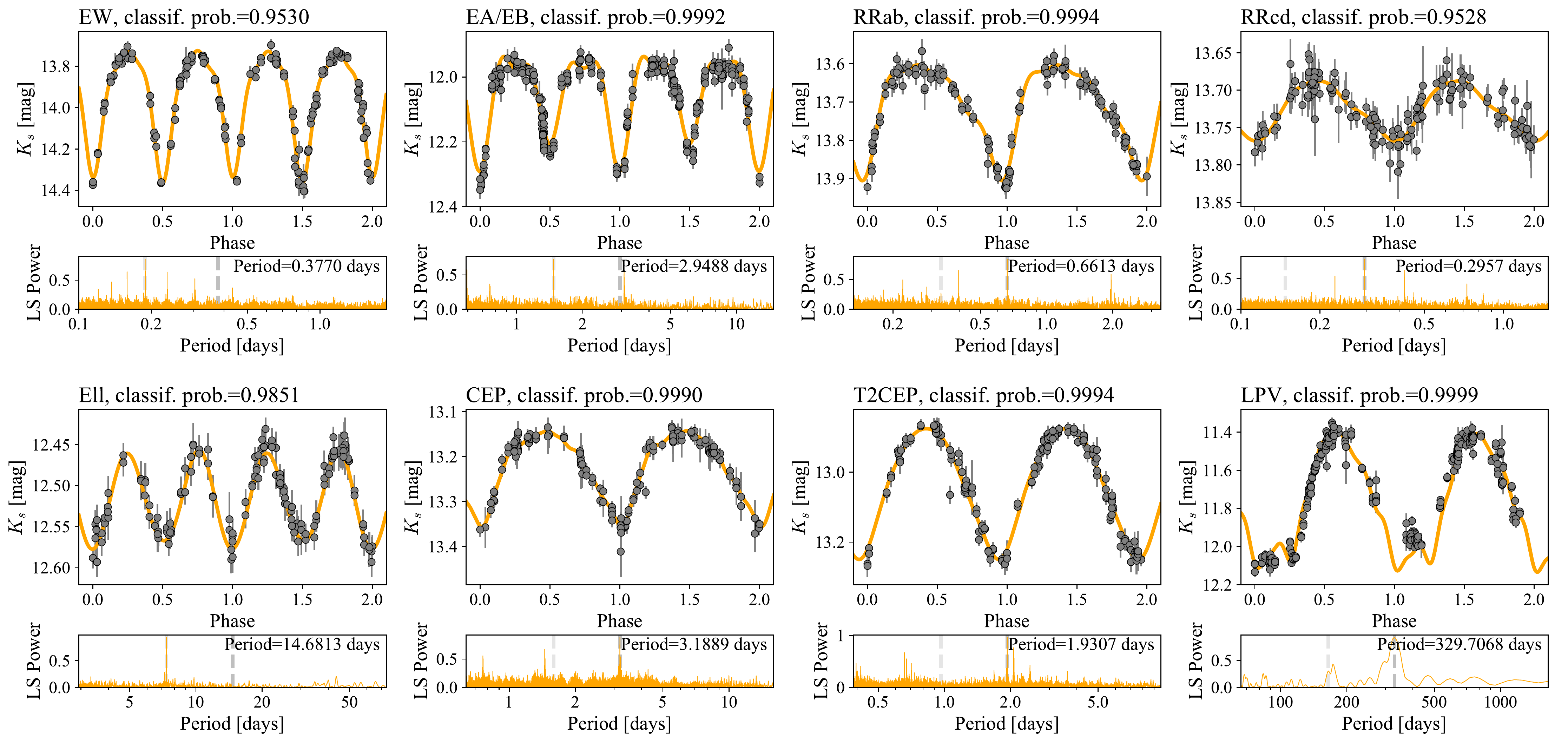}
\caption{Example training set light curves for the eight variable classes. Each set of two panels shows the light curve points in grey with a Fourier model fit in orange above a Lomb-Scargle periodogram with the period and half-period marked with grey lines. Above each set of panels, we give the classification probability from cross-validation.}
\label{fig::example_lc}
\end{figure*}

\begin{figure*}
    \centering
    \includegraphics[width=\textwidth]{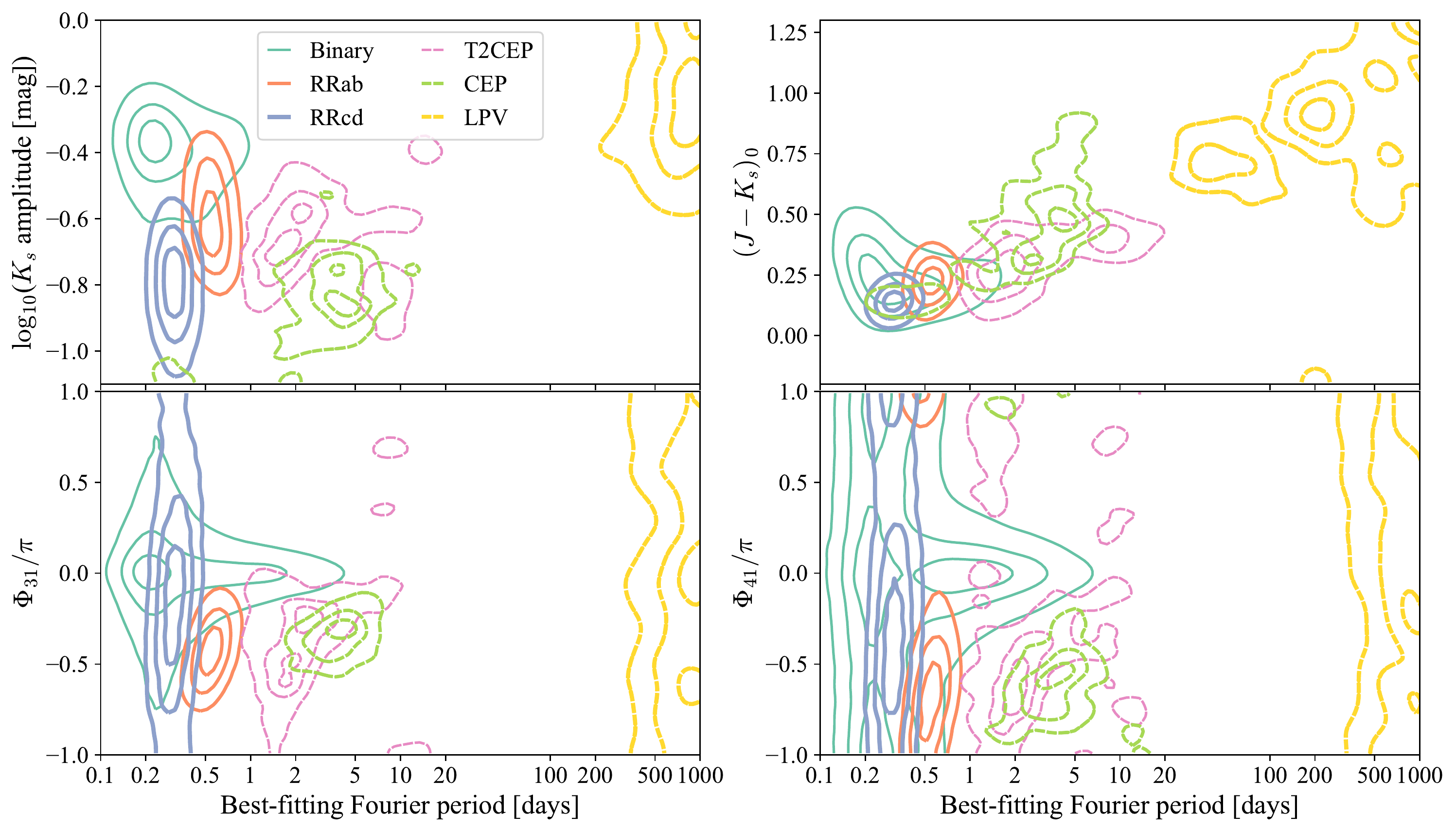}
    \caption{Feature distributions for the full training set (contours correspond to $20, 50$ and $80\percent$ of the peak density). We have combined the EA/EB, EW and ellipsoidal (Ell) classes into a single `Binary' class.
    The top row shows the amplitude and unextincted $(J-K_s)$ colour against period, whilst the bottom row shows the differences in Fourier phase $\Phi_{ij}=j\Phi_i-i\Phi_j$. The LPV distribution has been truncated at periods of $3.1$ days for clarity of presentation.
    }
    \label{fig:period_amp_feature}
\end{figure*}

The best-fitting Fourier model periods need some modification to match the definitions and conventions of a period for each source. For all sources other than EA/EB and Ell, if $A_n>A_1$ for $n\neq1$ we divide the period by $n$. For EA/EB and Ell, we double the periods if $A_1>A_2$. Finally, for EW we double the resultant period, as a period for a contact binary corresponds to a complete orbit. These corrections cannot be performed for unclassified sources. To maintain equivalence between our training set and general sources, we use the best-fitting Fourier periods in our classification pipeline and apply the required modifications when classifications are finalised. All periods quoted in upcoming feature distributions consist of unmodified periods unless comparison to literature periods is performed, for which we use modified periods. We show the period ranges for the eight variable classes in Fig.~\ref{fig:period_dist}.

Example phase folded light curves for each variable class, with the best fitting Fourier model overlaid, are illustrated in Fig,~\ref{fig::example_lc}. Two full periods are depicted in each case. Below every light curve, we show the Lomb-Scargle periodogram used for candidate frequency extraction with the adopted period annotated. This period is equivalent to the highest power period for intrinsic variables and is twice that for eclipsing systems. The notable asymmetry of the RRab, CEP, T2CEP light curves is quantified by the Fourier phase differences, for example $\Phi_{31}$ and $\Phi_{41} $, which we show differ in value from symmetric variables (e.g. in the bottom row of Fig.~\ref{fig:period_amp_feature}). Class differences are also clearly surmised in period-amplitude space (see the upper left panel of Fig.~\ref{fig:period_amp_feature}) where narrow regions are populated by each class.

\subsubsection{Period output comparison with source catalogues}\label{subsubsect:periodmatch}

\begin{figure*}
\includegraphics[width=.33\linewidth]{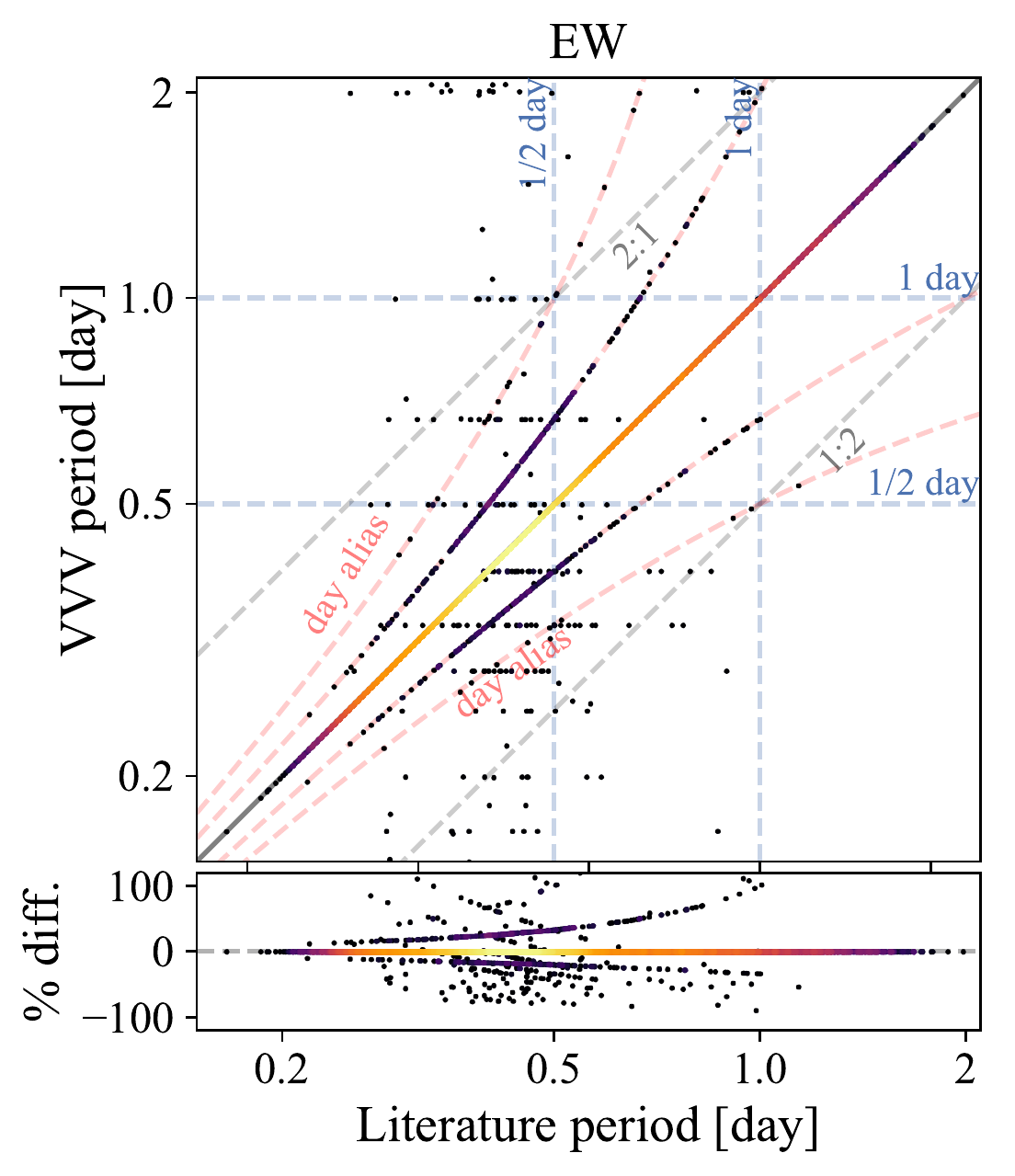}
\includegraphics[width=.33\linewidth]{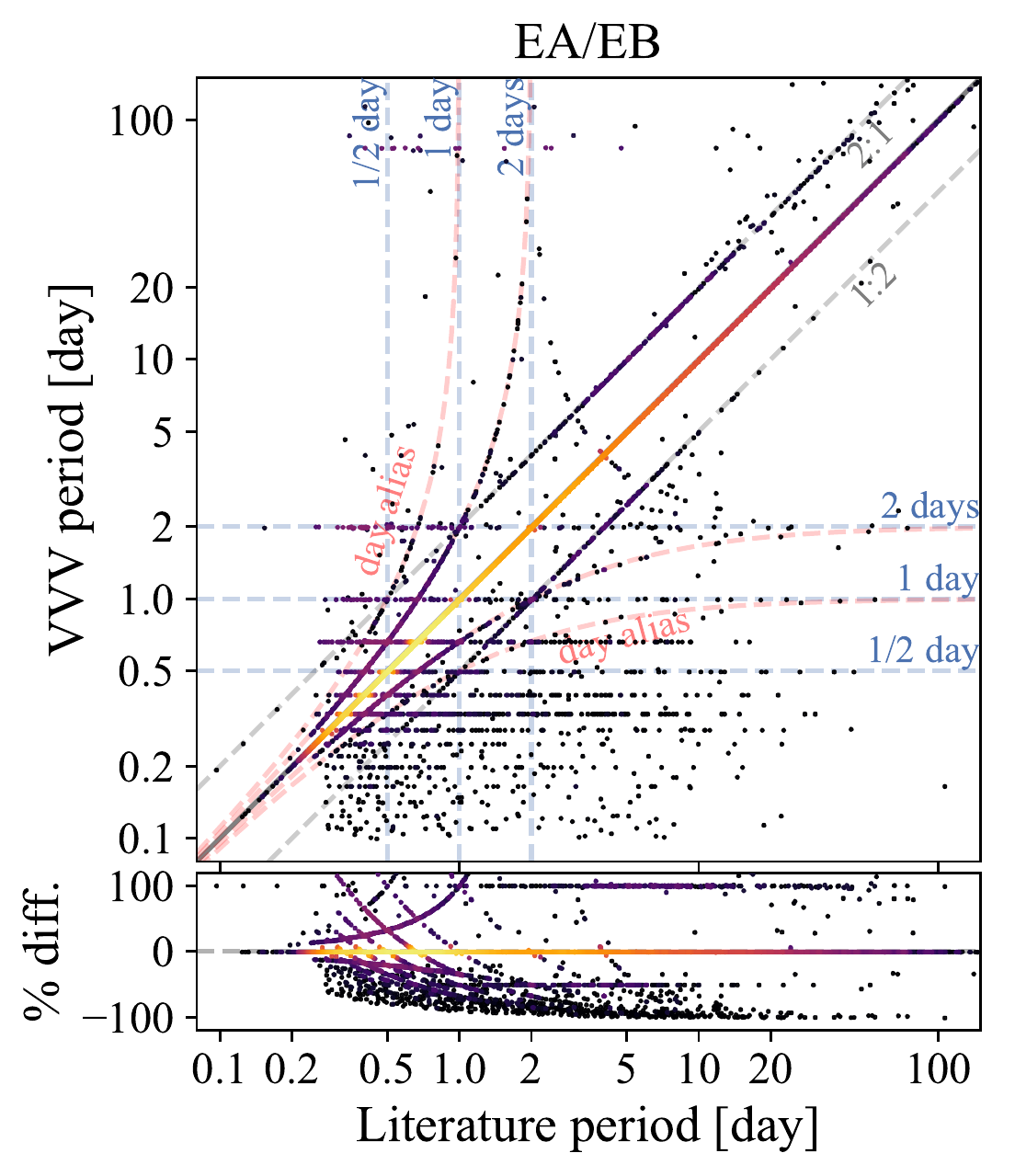}
\includegraphics[width=.33\linewidth]{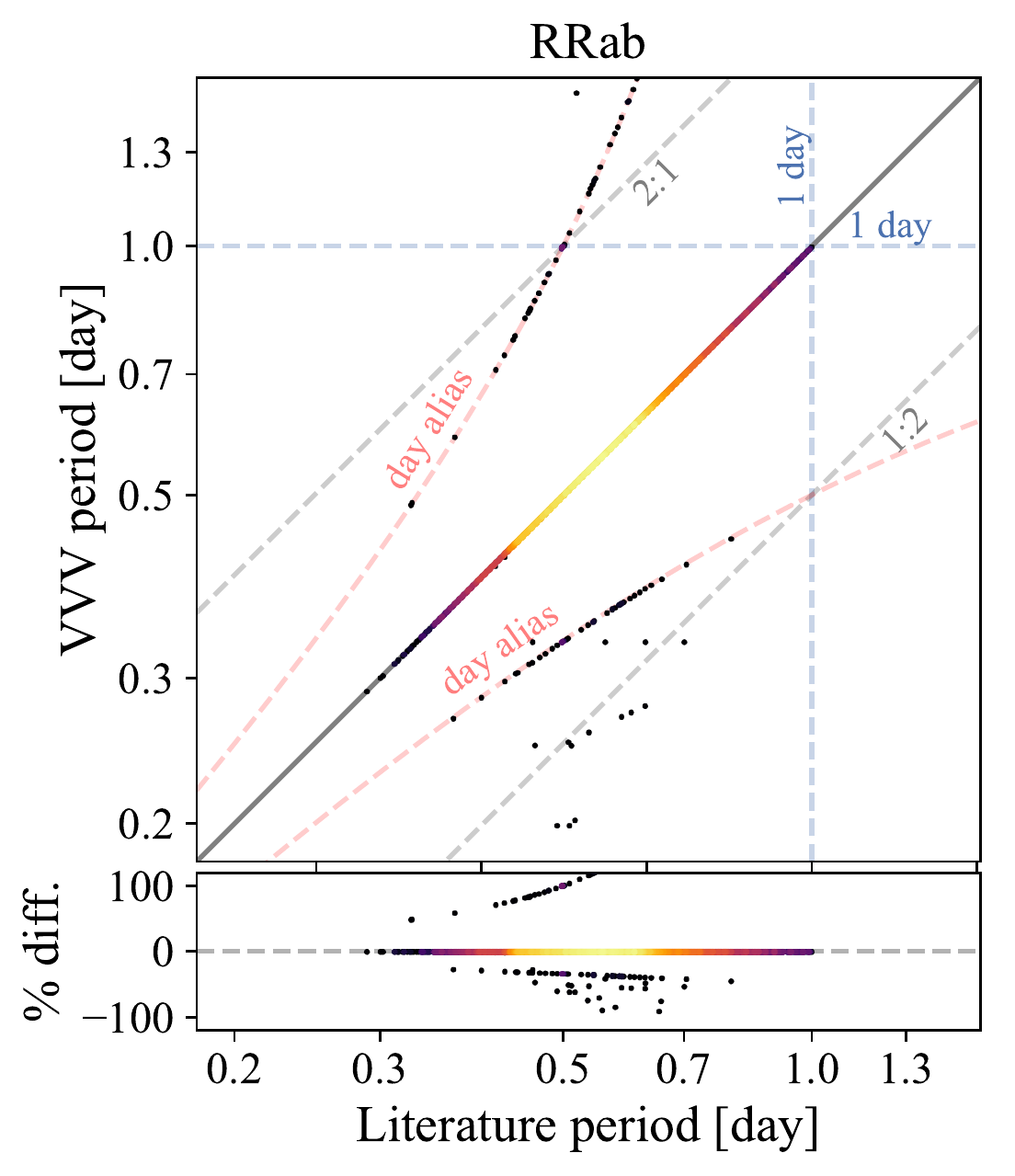}
\includegraphics[width=.33\linewidth]{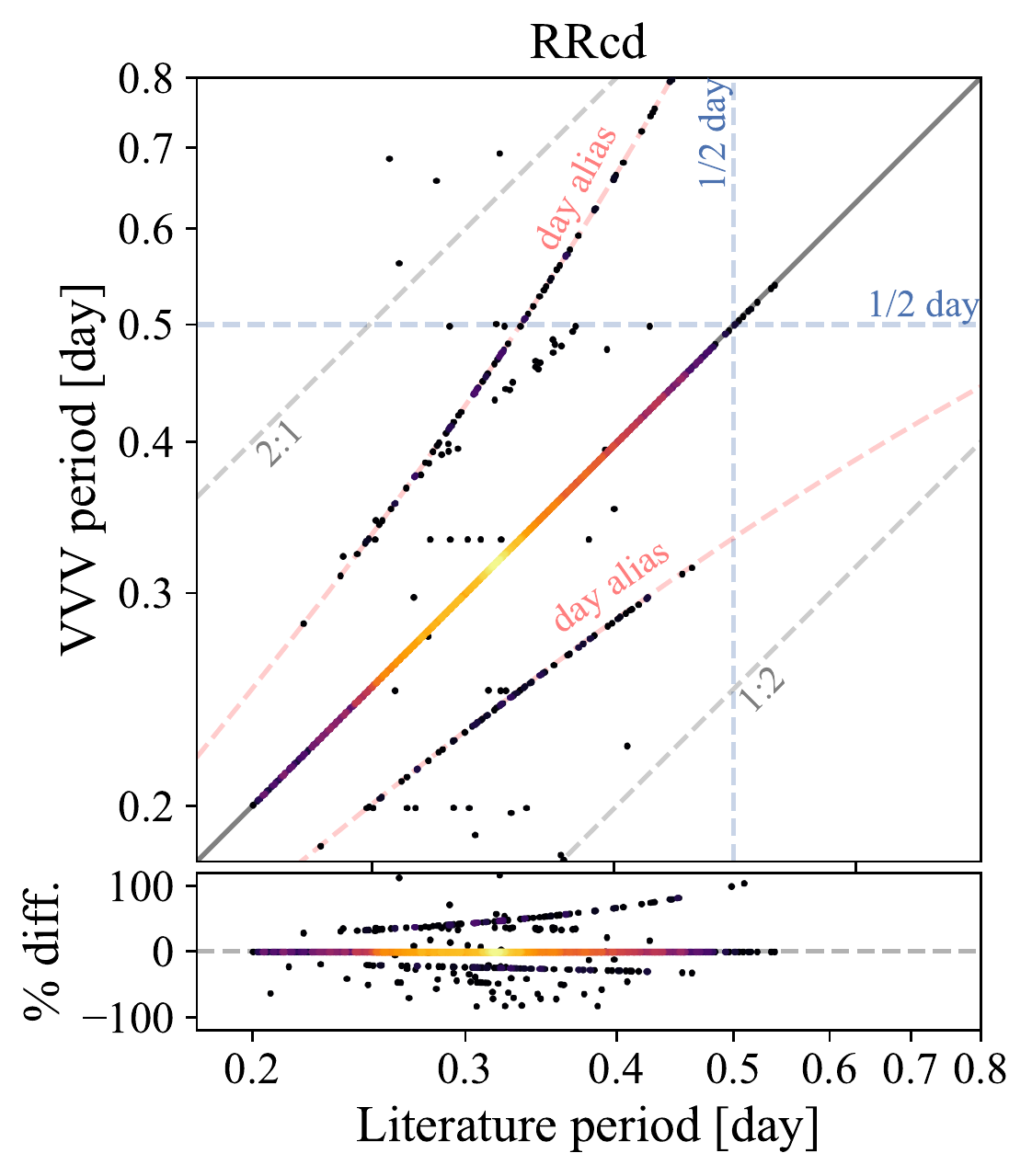}
\includegraphics[width=.33\linewidth]{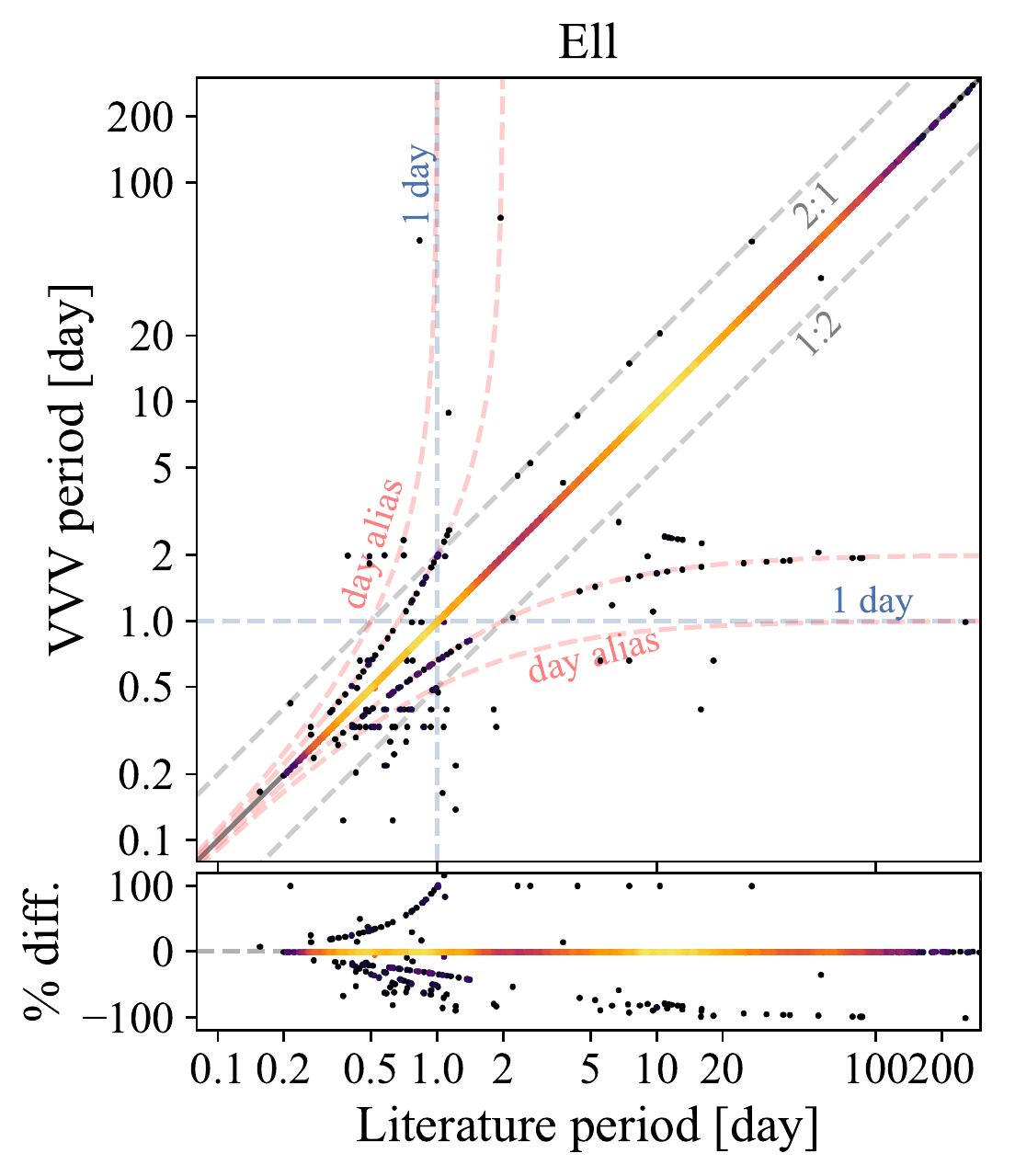}
\includegraphics[width=.33\linewidth]{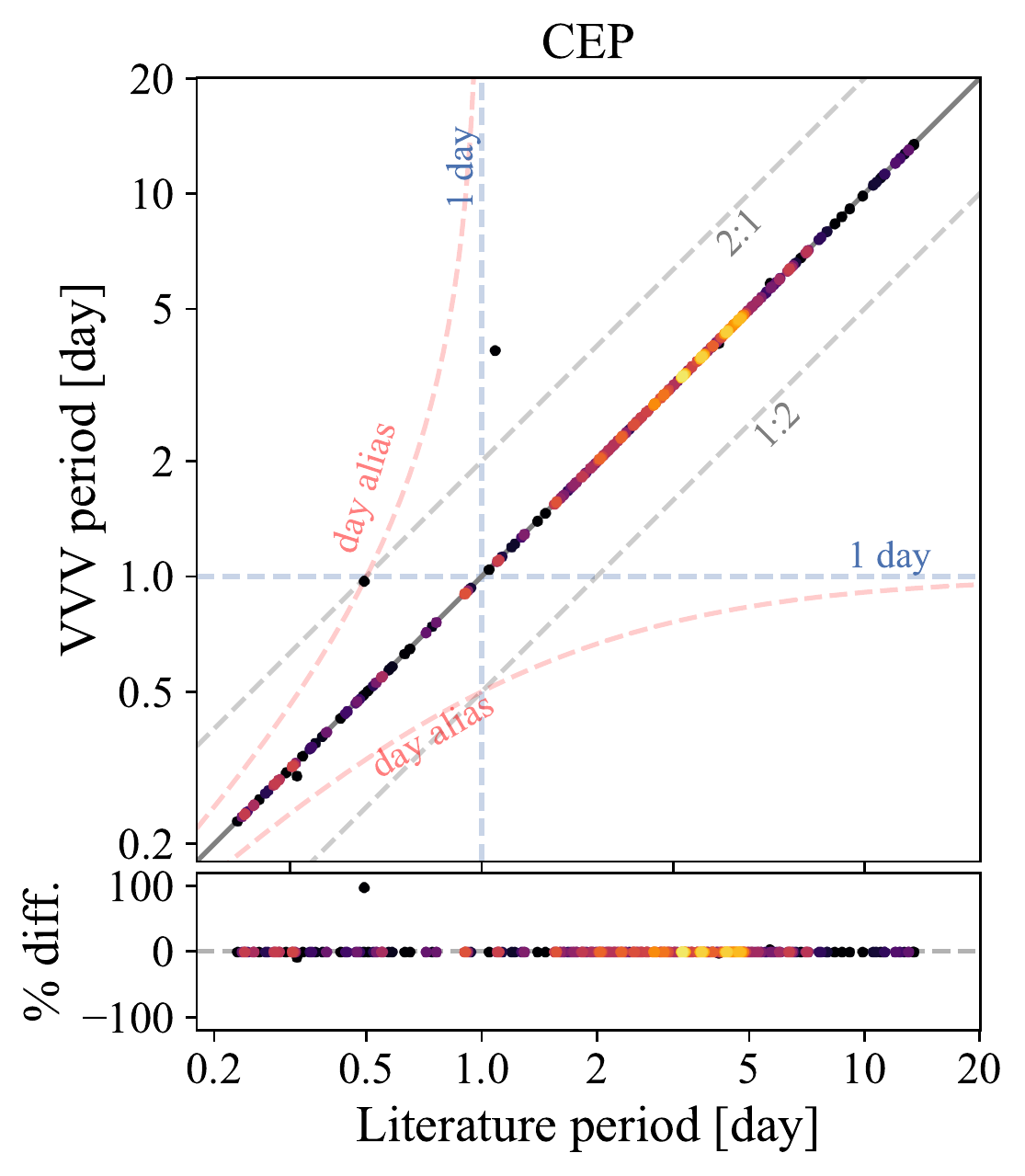}
\includegraphics[width=.33\linewidth]{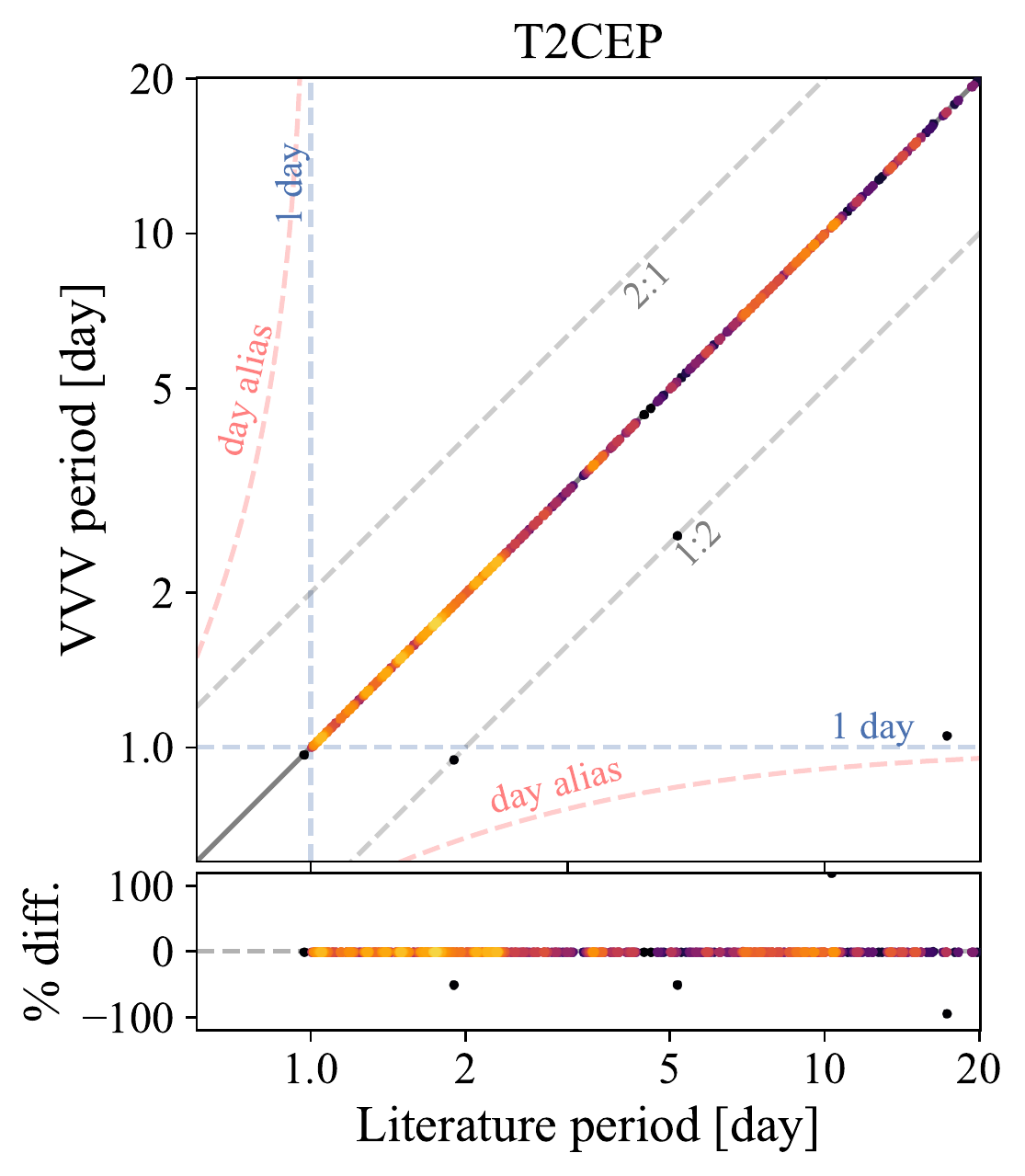}
\includegraphics[width=.33\linewidth]{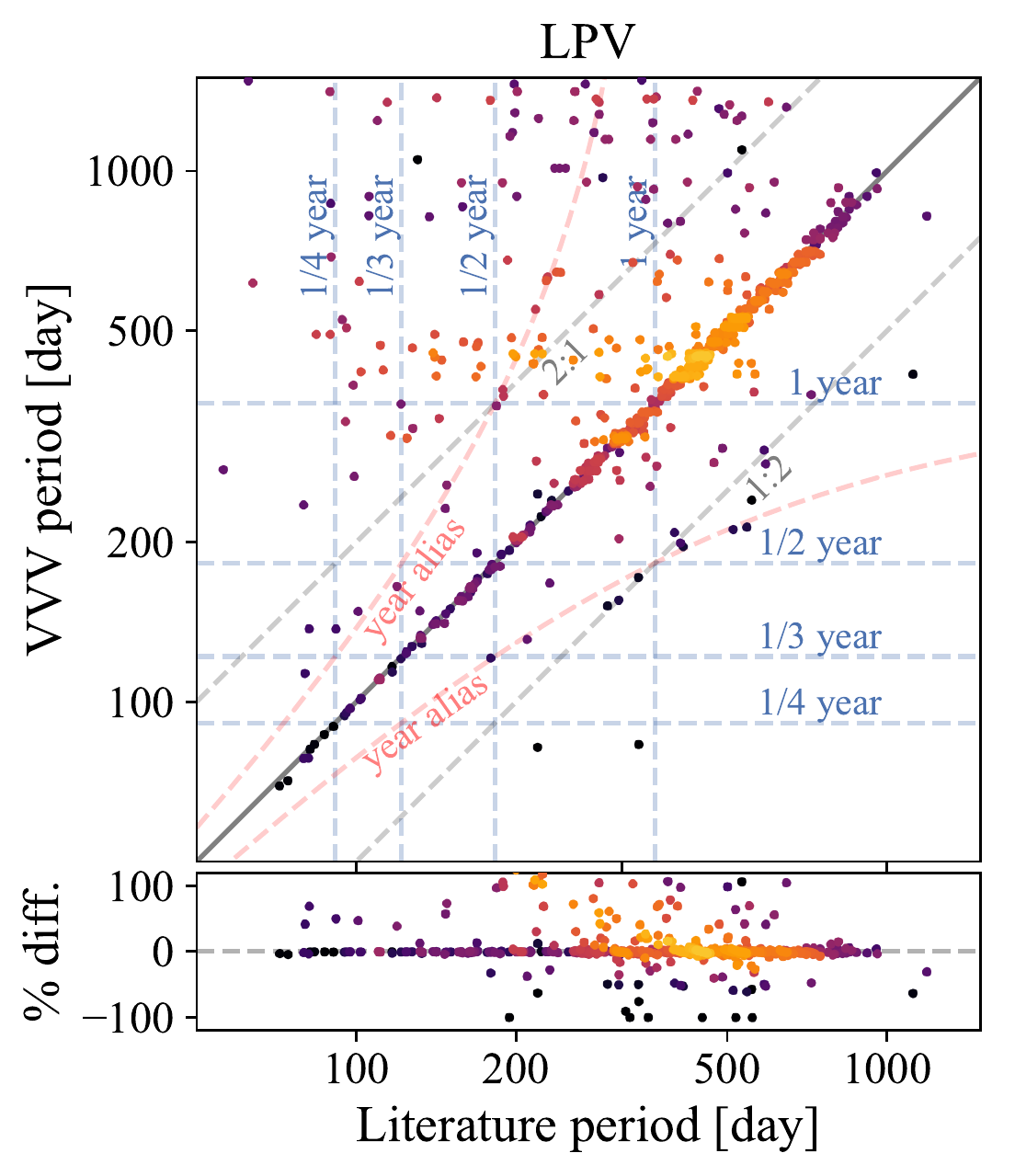}

\caption{Comparison of the periods extracted from the VVV light curves with the literature periods for our variable star training sample (with $\log_{10}\mathrm{FAP}<-10$). Each panel shows a different variable class. Points are coloured by the logarithm of the density of points (yellow being high density). Common aliases and period multiples are marked, and we display the percentage difference between the two periods in the lower panels. Note the different period ranges for each plot.}
\label{fig:periodcomp}
\end{figure*}

We evaluate our period extraction procedure through comparison with literature periods for our variable training set. Fig.~\ref{fig:periodcomp} shows the modified best-fitting Fourier periods plotted against literature estimates for sources from each of the eight variable classes and with $\log_{10}(\mathrm{FAP})<-10$. Table~\ref{tab:trainset} gives the percentage of sources with periods matching within $10\,\percent$. Although a large fraction of sources lie along the expected 1-to-1 relation, a few per cent are more discrepant, primarily due to aliasing effects. In particular, we note \begin{inparaenum}
\item \emph{observational aliases} resulting from regular cadence of ground-based observations appearing as horizontal lines at multiples of a day or a year,
\item\emph{beat aliases} where the physical periodicity of a source and observational aliases are combined, taking the form $P_{\mathrm{obs},n} = \left(\frac{1}{P_{\mathrm{true}}}+\frac{n}{P_{\mathrm{alias}}}\right)^{-1}$ where $P_{\mathrm{true}}$ and $P_{\mathrm{obs}}$ are the correct and observed periods,  $P_{\mathrm
{alias}}$ is an observational alias and $n$ is an integer \citep{Vanderplas2018},
\item \emph{harmonic aliases} for which multiples of the anticipated period are found as shown by clustering along the 2:1 and 1:2 trends, and most apparent for the EA/EB class.
\end{inparaenum}
We aimed to limit inclusion of observational aliases in our procedure through the means described in Section~\ref{subsubsect:periodic_features}. However, complete removal of alias and artificial periods is unfeasible.

\begin{figure}
    \centering
    \includegraphics[width=\columnwidth]{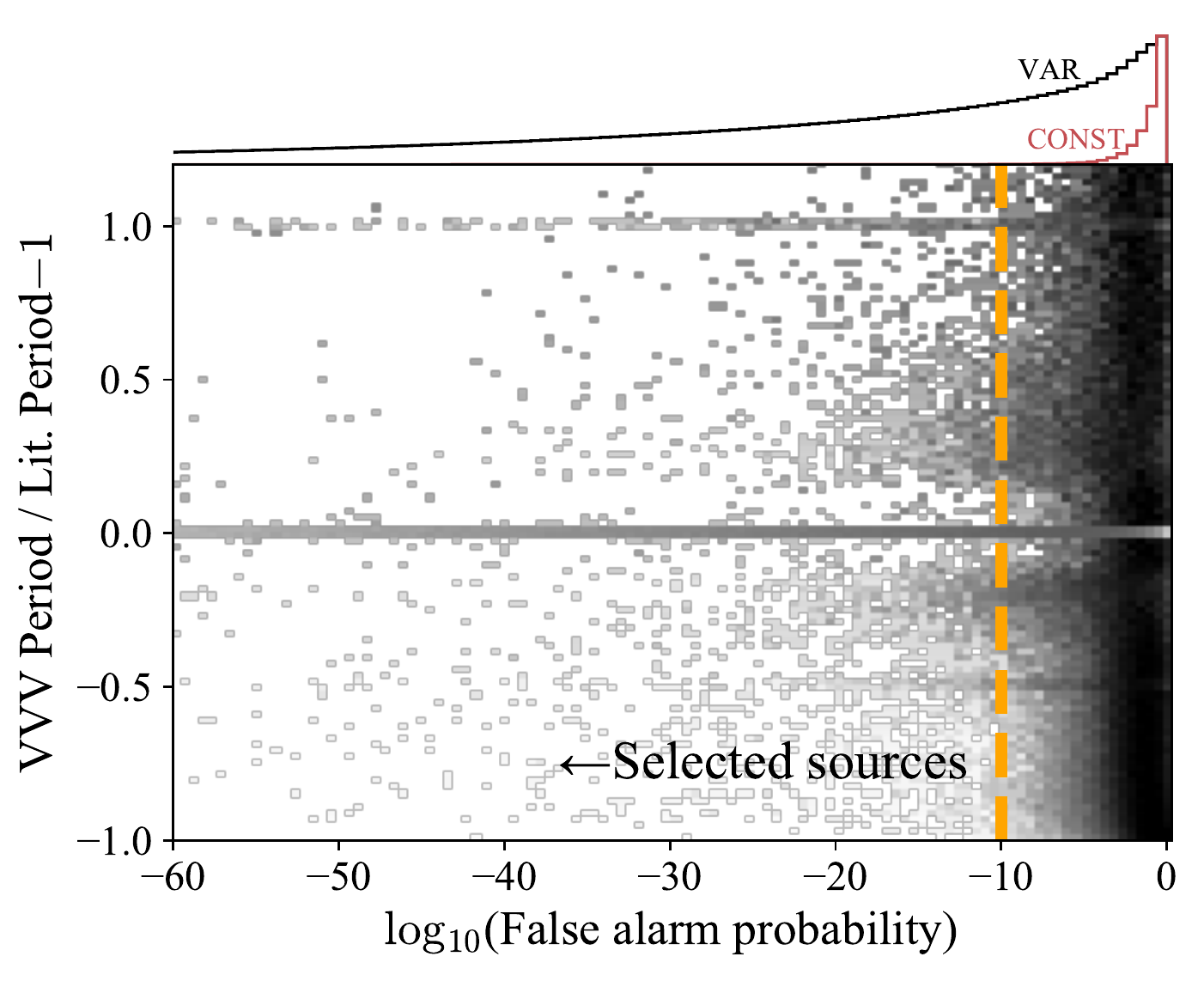}
    \caption{Row-normalised distribution of the logarithm of Lomb-Scargle false-alarm probability against the relative difference of our periods and the literature periods. The greyscale is logarithmic in the density. The false-alarm cumulative probability distributions of the variable (VAR) and constant (CONST) training sets are shown above. We only consider sources with false alarm probability smaller than $1\times10^{-10}$.}
    \label{fig:log10fap}
\end{figure}

Fig.~\ref{fig:log10fap} displays the joint distribution of the ratio of our periods to the literature periods and the Lomb-Scargle false alarm probability for \emph{all} stars in our variable set. We see that for high false alarm probability the quality of period recovery significantly deteriorates. This motivates us to only use sources in the variable training set with $\log_{10}\left(\mathrm{FAP}\right)<-10$ as only these will have trustworthy Fourier parameters allowing for reliable classification. Whilst $\sim90\,\percent$ of the RRab and CEP training set are retained by this cut, only $40(60)\,\percent$ of EA/EB (EW/Ell/RRcd) are retained and in the extreme only $25\,\percent$ of LPVs. Nearly all sources in our constant class fall short of this cut as depicted in the top section of Fig.~\ref{fig:log10fap}.
A downside of the method is the artificially created bias towards low FAP when applying the classifier in general. This suggests a significant number of truly variable sources will be rejected. We therefore forego completeness of our final catalogue in order to reliably extract the clearest periodic stars with minimal contamination.

LPV periods range from 100 to 1000 days with $35\percent$ of the $\log_{10}\mathrm{FAP}$ set having discrepant periods relative to the literature values (at the $10\,\percent$ level). The majority of the mismatch periods have VVV periods greater than the literature periods suggesting a bias in our procedure towards long-term trends and an inability to find shorter fluctuations in the light curves. This is particularly the case for semi-regular variables which exhibit strong variation of light curve amplitude from period to period, which can be interpreted as long-term periodicity. Whilst the mismatch may be related to our procedure, there is the possibility the literature periods are in error (possibly due to aliasing). We find that only $30\,\percent$ of the mismatching LPVs have one of the top $30$ peaks in the Lomb-Scargle periodogram within $20\,\percent$ of the literature period suggesting erroneous literature periods or incorrect cross-matches/blending issues.

Phase folding light curves with improper periods (long-term or observational aliases) leads to abnormal gaps in the phase space coverage. We therefore compute both the maximum phase separation as well as the ratio of the difference between the maximum and mean phase separation to the standard deviation of these separations. These quality features are included in the second stage classifier and used for selection of well sampled light curves.

\subsubsection{Eclipsing binary features}\label{subsect:eclipingbinary}
\begin{figure*}
    \centering
    \includegraphics[width=\textwidth]{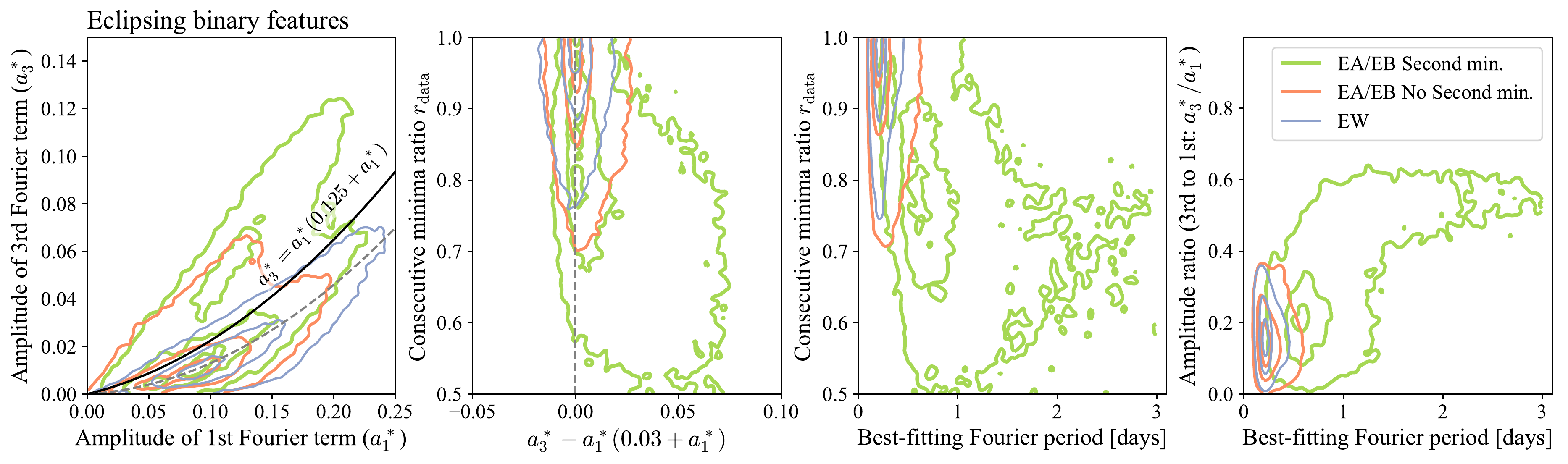}
    \caption{Feature distributions for the eclipsing binary training set (contours correspond to $10, 20$ and $80\percent$ of the peak density). We split the sample into three: detached EA/EB with a significant secondary minimum detected (green),
    detached EA/EB without a significant secondary minimum detected (orange),
    and contact eclipsing binaries (EW, purple).
    The amplitude of the 3rd Fourier term, $a^*_3$, relative to the 1st Fourier term, $a^*_1$, (using the Fourier fits at double the period where appropriate, see Section~\ref{sec::non-periodic_feats}) allows separation between clearly detached (uneven minima) and contact. \protect\cite{rucinski1993} propose the separating line $a^*_3=a^*_1(0.125+a^*_1)$. The samples also separate in $r_{\text{data}}$, the ratio between consecutive minima depths, and the period. However, in general, there is significant confusion between these classes and high-quality data is required to determine if the system is genuinely contact.}
    \label{fig:eclipsing_binary}
\end{figure*}
Detached eclipsing binary sources can differ visually in their light curves due to a lack of symmetry between the two minima per orbital revolution. In the case of other binary systems, contact (EW) and ellipsoidal (Ell), these minima are of nearly equal depth, although detached eclipsing binary systems can also have equal depth minima. We capture this potential difference through the ratio of consecutive light curve minima depths. First we identify any secondary minimum in the best-fitting Fourier model. To do this, we shift the phases of observations to place the primary model minimum at zero and search for any secondary model minima in the phase range $0.35-0.65$ with depth with respect to the neighbouring maxima of greater than seven times the model uncertainty. This phase range corresponds to eccentricities $\lesssim0.25$. In the Galactic bulge eclipsing binary sample of \cite{Devor2005} only $3\percent$ of systems had more extreme eccentricities whilst in the second Kepler Eclipsing Binary catalogue of \cite{Slawson2011} $\sim6\percent$ do. Although this procedure will miss this minority of high eccentricity systems, we anticipate such systems will be well classified by their Fourier parameters. With a secondary minimum identified, we compute $r_{\text{model}}$ as the ratio of the secondary to primary model minimum relative to the maximum of the model light curve. If no second minimum is found, this ratio is set to unity. As a complementary feature, we also compute the minimum ratio more directly from the data, $r_{\text{data}}$. For this, we instead measure the depth of the primary and secondary minima using the inverse-variance-weighted mean of measurements within $0.025$ phase of the previously identified model minima (the bin is broadened until at least $3$ datapoints are included in the mean computation) relative to the $1$st percentile of the light curve. If no secondary model minimum is found, every other primary minima is treated as a secondary minimum. Orbital eccentricity and inclination can produce eclipsing binaries with only one minimum per orbit which will be confused with equal depth binaries with two observed minima per orbit. The majority of our EA/EB class with noted secondary minima have $r_{\text{data}}$ vastly below unity, as featured in the second and third panels of Fig.~\ref{fig:eclipsing_binary}. This exemplifies the uneven light curve minima associated with detached binary systems, which also tend to have greater periods than contact binaries. In contrast, even minima EW and EA/EB without a secondary minimum cluster towards unit $r_{\text{data}}$.

For sources with no secondary minimum, we compute an additional set of amplitudes $A^*_i$, phases $\Phi^*_i$, amplitude ratios, $R^*_{ij}$, and phase differences, $\Phi^*_{ij}$ called \emph{double-period features} using a Fourier model with period twice that of the previously best-fitting Fourier period. For sources with a second minimum, these features are taken as duplicates of the prior quantities. This puts all eclipsing binary features on an equal footing, irrespective of whether our Fourier model has successfully distinguished between minima of differing depths or not.
In Fig.~\ref{fig:eclipsing_binary}, we display the training set of eclipsing binaries using these double-period features. As noted by \cite{rucinski1993}, EW binaries predominantly lie below the line $a_3^*=a_1^*(0.125+a_1^*)$. However, we find there is also significant contamination in this region from EA/EB type. In fact, the subset of EA/EB for which we fail to identify a secondary minimum show strong overlap with the EW class in \emph{all} feature sub-spaces. For example, Fig.~\ref{fig:eclipsing_binary} shows the similarity of the distributions in the consecutive minimum ratios ($r_\mathrm{model}$ and $r_\mathrm{data}$), period and amplitude ratio, highlighted by \cite{jayasinghe2019} as useful for distinguishing binary types. This suggests that for many cases assessing whether stars in a binary system are in contact or not is not possible with our data. We therefore accept a grey zone where EW and EA/EB are practically indistinguishable in feature space. Note that when faced with a similar issue, \cite{jayasinghe2019} performed a visual re-classification of 15,000 eclipsing binaries using ASAS-SN data resulting in a much cleaner separation of the different classes in the feature sub-spaces of Fig.~\ref{fig:eclipsing_binary}. Here we accept that the classification of near-contact eclipsing binaries is poorer for our sample than in OGLE, and suggest that any users interested in forming more complete samples of contact eclipsing binaries could utilise cuts in the fuller feature space illustrated in Fig.~\ref{fig:eclipsing_binary}.

\subsubsection{Non-periodic features}\label{sec::non-periodic_feats}

\begin{figure}
    \centering
    \includegraphics[width=\columnwidth]{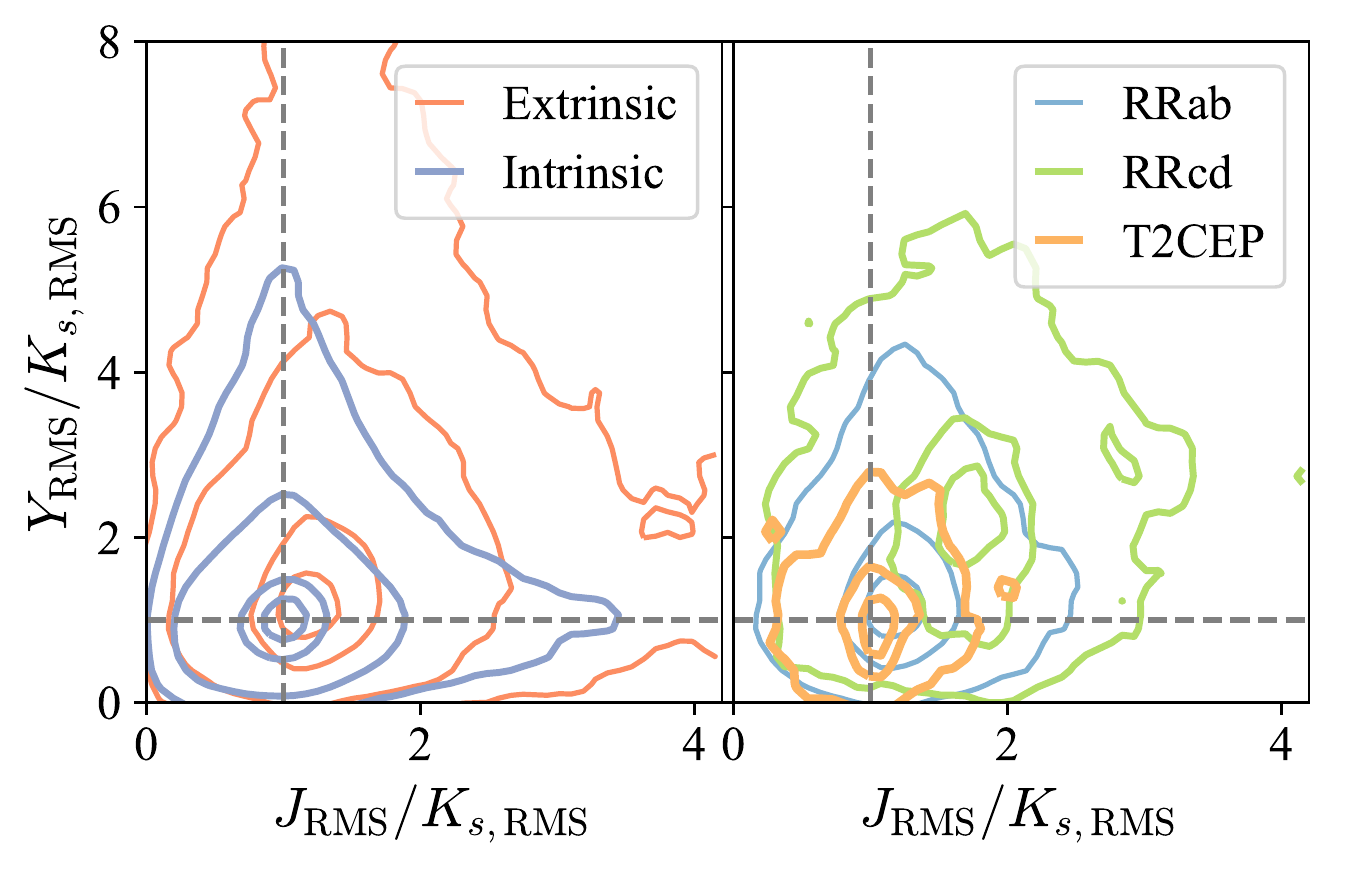}
    \caption{Distribution of the training set in $X_\mathrm{RMS}/K_{s,\mathrm{RMS}}$, the ratio of the RMS of the $X$-band observations compared to the RMS of the Fourier model evaluated at the epochs of the $X$ observations (contours show $10,50$ and $80\,\percent$ of peak density). Left panel shows the training set grouped by `extrinsic' variables (EA/EB, EW, Ell) and `intrinsic' (RRab, RRcd, CEP, T2CEP, LPV). The right panel shows a breakdown of three intrinsic classes: RRab, RRcd and T2CEP. Note how extrinsic variables typically concentrate around unity whilst intrinsic variables have a tendency for more variability in shorter wavelength bands. This is particularly pronounced for the RRc class.}
    \label{fig:intrinsic_extrinsic}
\end{figure}

Following work illustrating varying amplitudes of NIR filter band light curves for RR Lyrae stars \citep{Braga2018} and Cepheids \citep{Inno2015}, we utilise variability in the VVV $Z Y J H$ bands with respect to the $K_s$-band observations. We only perform this analysis if more than one detection is available in any of the bands. We compute $X_\mathrm{RMS}/K_{s,\mathrm{RMS}}$ as the ratio of the inverse-variance-weighted root-mean-squared (RMS) of measurements in band $X=\{Z,Y,J,H\}$ to the RMS of the $K_s$ Fourier model magnitudes at the epochs of the $X$ observations. Similarly, we find the factors, $X_\mathrm{scale}$, by which the $K_s$ Fourier models need to be scaled to fit the $X$ band measurements. We show in Fig.~\ref{fig:intrinsic_extrinsic} how $X_\mathrm{RMS}/K_{s,\mathrm{RMS}}$ is of use in separating intrinsic variables, primarily RR Lyrae stars and Cepheid variables, from extrinsic eclipsing binary classes. In general, intrinsic variables, particularly RRcd, show increased variability in the shorter wavelength bands.

The classification feature space was further supplemented by extinction-corrected colour information, in particular $(H-K_s)$ and $(J-K_s)$ colours. Averaged values of $J$, $H$ and $K_s$ apparent magnitudes were determined from VVV observations. Colours were dereddened using colour excesses, here $E(H-K_s)_\mathrm{RC}$ and $E(J-K_s)_\mathrm{RC}$, calculated from red clump stars \citep{Gonzalez2012} over an adaptive nested Healpix grid in Galactic longitude and latitude with around $25$ red clump stars per pixel. Red clump giant stars exhibit a clear peak in colour-magnitude space so are easily identifiable and have a small range of intrinsic colours ($(J-K_s)_0=0.62$, $(H-K_s)_0=0.09$). The colour excesses are only reliable for stars located behind the same dust screen as the red clump tracers. This assumption is strictly only valid at higher latitudes where the intervening dust is primarily due to the nearby dust in the foreground Galactic disc. However, we have found that the inclusion of colour information is particularly important for separating the relatively intrinsically blue RR Lyrae stars and red LPVs. Segregation in colour space by class is seen in the upper right panel of Fig.~\ref{fig:period_amp_feature}. For this reason, despite the shortcomings, the colour information was considered in the final analysis.

\subsection{Training the Classifier}\label{subsect:train}

We now describe the implementation and training of the two stage classifier. To deal with inherent infinities and outliers in feature distributions, due to erroneous computations, we undertake a pre-processing routine for all training sets. We first take sine and cosine components of the phase differences, $\Phi_{ij}$ and $\Phi^*_{ij}$, in stage 2. This maps the values in the range $[-1, 1]$ and removes the artificial difference between $0$ and $2\pi$ values. The data in all features with a large dynamic range is made to look more Gaussian-like through the application of a feature-wise Yeo-Johnson power transformation \citep{Yeo2000}. Both positive and negative infinities are then replaced by \textsc{NaN} values, to be ignored during the computation of distribution statistics. This is followed by a clipping program, where a filter labels sources as outliers if one or more of their features are located beyond 10 sigma of the median feature distribution value. As an exception, we do not identify as outliers larger values of Stetson $I$, $(J-K_s)_0$ and $(H-K_s)_0$ as this would remove a large fraction of LPVs. Outliers are removed from the training set decreasing the size of the set by roughly $2\,\percent$. We replace the \textsc{NaN} entries with the mean value of the 5 nearest neighbours found in the feature space through k-Nearest Neighbour imputation \citep[implemented in \textsc{scikit-learn},][]{scikit}. These steps ensured reliability and robustness of the classifiers removing any irregular bias towards our selected samples.

For stage 1, each binary classifier is generated using the \textsc{scikit-learn} implementation of the RF ensemble method \citep{scikit}. Furthermore, the classifiers are initialised with \verb|n_estimators|=100 defining the number of decorrelated decision trees in the ensemble, \verb|max_depth|=8 pruning the size of each tree, \verb|min_samples_split,min_samples_leaf|=5 restricting the number of samples required for a node split and leaf respectively, \verb|max_features|=\verb|sqrt| defining the number of features considered when deciding to split and \verb|class_weight|=\verb|balanced_subsample| giving equal weight to each classification class. The best classifier hyper-parameters were chosen based on a grid search including a large array of possible parameters, with restriction placed on certain parameters to avoid overfitting to the training set.

For stage 2, we create a gradient-boosting classifier using the \textsc{scikit-learn} interface to \textsc{XGBoost} \citep{XGBoost} instantiated with \verb|n_estimators|=100, \verb|learning_rate|=0.15 which is the factor by which weights are updated after each learning epoch, \verb|gamma|=2 as minimum reduction in loss function required to split a node, \verb|max_depth|=8 as in the RF case, \verb|min_child_weight|=3 defining the minimum sum of weights of all observations required in a child (equivalent effect to \verb|min_samples_leaf| in RF), \verb|subsample|=0.8 outlining the fraction of observations considered in each tree, \verb|colsample_bytree|=0.9 denoting the fraction of features to be randomly sampled for each tree (equivalent to \verb|max_features| in RF) and tree generator algorithm chosen as \verb|tree_method|='hist'. We then guarantee uniform class balancing through inversely scaling example weights by their parent class size using the \verb|compute_sample_weight| functionality.

We now discuss the accuracy of each stage of classification and finalise the training phase of the hierarchical classifier.

\subsubsection{Performance of classifier} \label{sub:performance}

\begin{figure*}
    \centering
    \includegraphics[width=0.4\textwidth,valign=t]{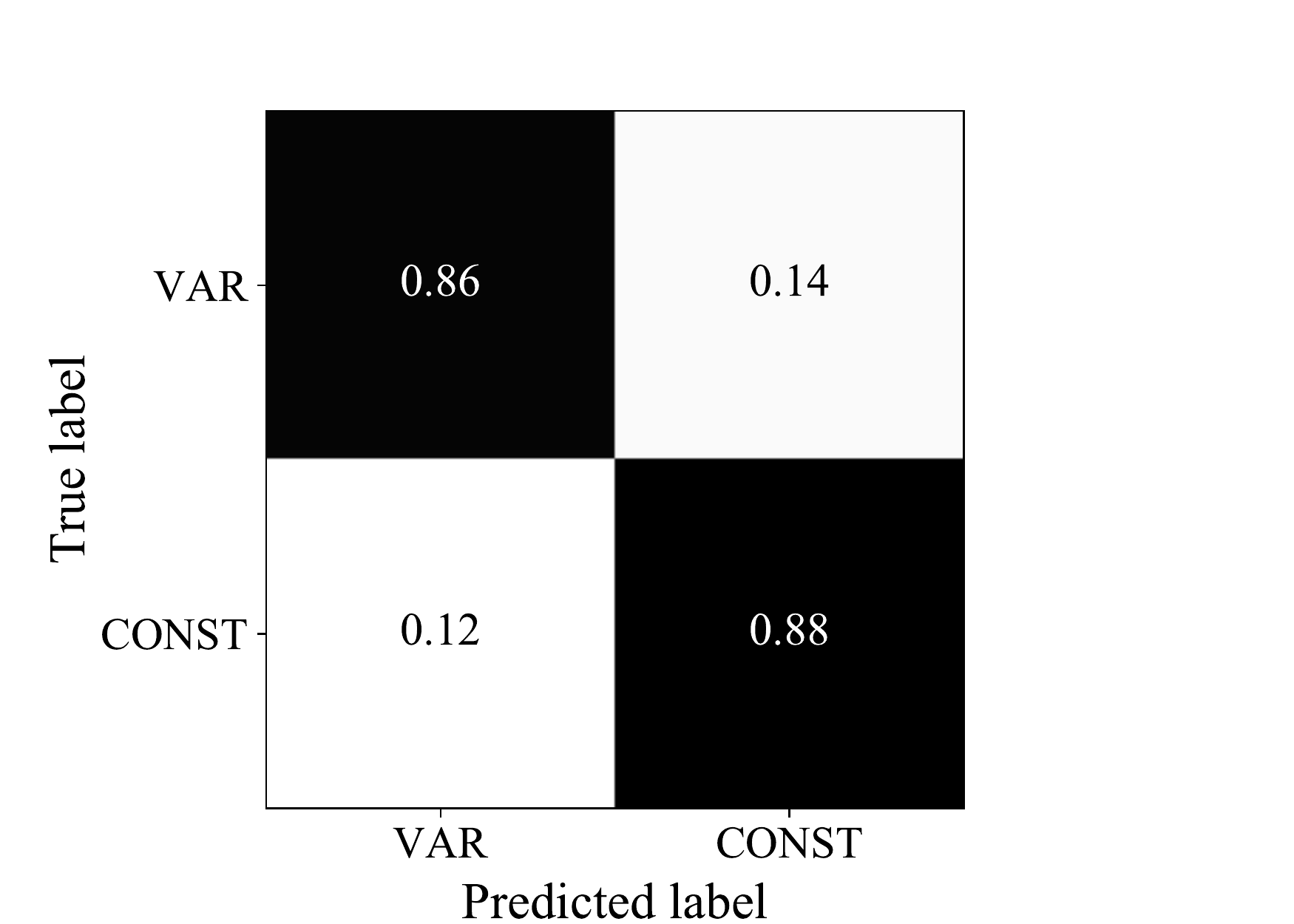}
    \includegraphics[width=.59\textwidth,valign=t]{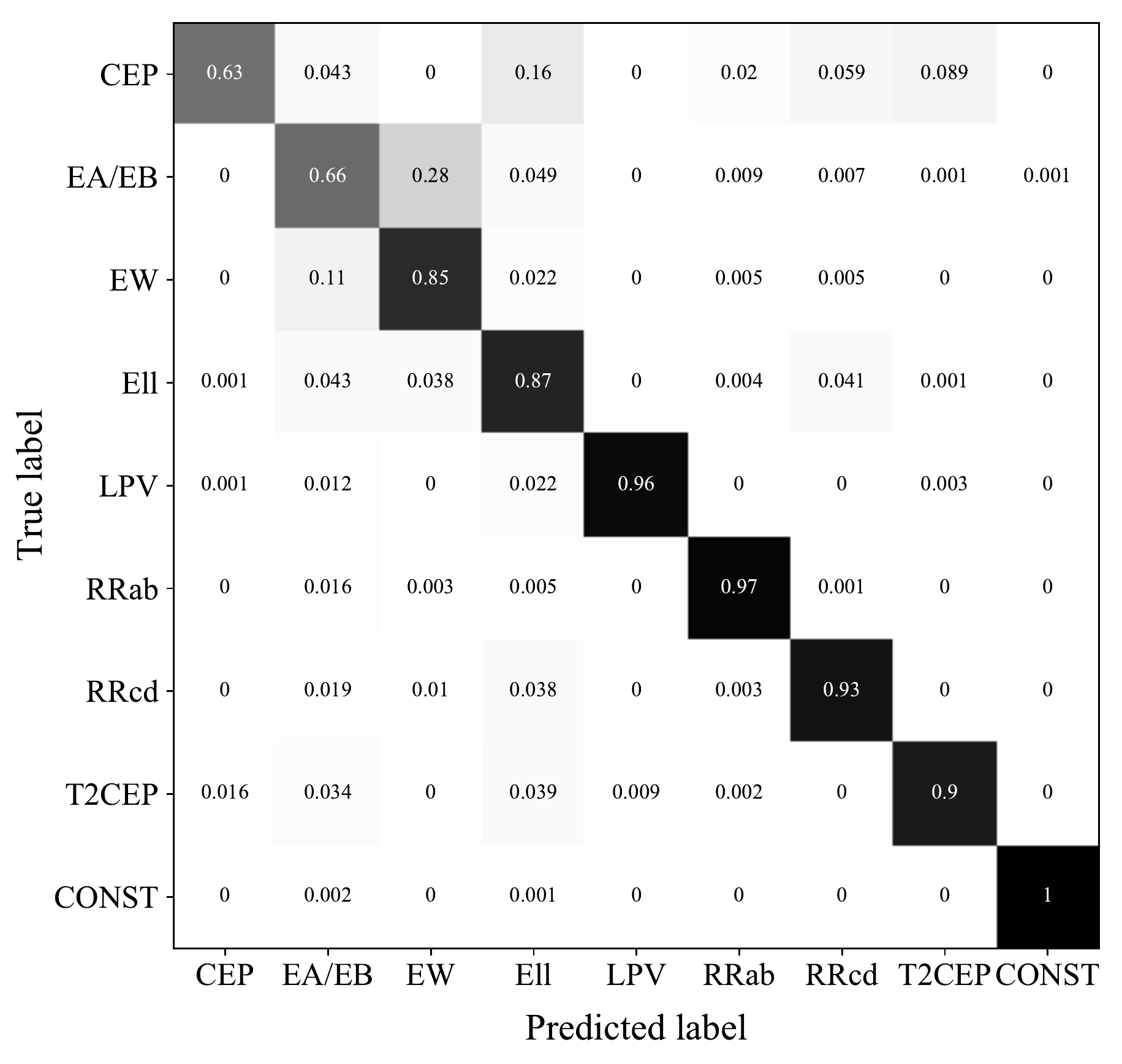}
    \caption{Normalised confusion matrix for 10-fold cross-validated variable/non-variable star classifier (left) and variable star classifier (right). The x-axis represents the classification predicted by the model for the set of known labels shown on the y-axis.}
    \label{fig:confusion_matrices}
\end{figure*}

\begin{table*}
	\caption{Aggregated 10-fold cross-validation results for the 348 binary classifiers. The arithmetic mean of each performance score and class size is shown, with the standard deviation of the scores quoted as well.}
    \centering
    \begin{tabular}{lcccc}
    	\hline
        Broad Class & precision & recall & $F_1$ score & class size\\ \hline
        VAR & 0.915 $\pm$ 0.048 & 0.862 $\pm$ 0.018 & 0.887 $\pm$ 0.031 & 401 363\\
        CONST & 0.774 $\pm$ 0.081 & 0.877 $\pm$ 0.030 & 0.820 $\pm$ 0.048 & 255 060\\ \hline
        Macro average & 0.844 $\pm$ 0.027 & 0.870 $\pm$ 0.021 & 0.854 $\pm$ 0.021 & 656 423\\
        Weighted average & 0.877 $\pm$ 0.025 & 0.868 $\pm$ 0.019 & 0.870 $\pm$ 0.020 & 656 423\\ \hline
    \end{tabular}
    \label{tab:crossval_binary}
\end{table*}

\begin{table}
        \caption{$10$-fold cross-validation results for second stage classifier divided by individual classes. Macro and weighted performance averages are also given where scores are aggregated with equal weighting or weights proportional to class size respectively.}\label{tab:crossval_variable}
        \centering
        \begin{tabular}{lcccc}
            \hline
            Variability Class & precision & recall & $F_1$ score & class size\\ \hline
            EW & 0.522 & 0.852 & 0.648 & 39 874\\
            EA/EB & 0.927 & 0.655 & 0.768 & 109 127\\
            Ell &  0.592 & 0.874 & 0.706 & 11 156\\
            RRab & 0.945 & 0.975 & 0.961 & 22 387\\
            RRcd & 0.814 & 0.928 & 0.867 & 6 663\\
            CEP & 0.802 & 0.647 & 0.716 & 300 \\
            T2CEP & 0.805 & 0.870 & 0.836 & 545\\
            LPV & 0.879 & 0.918 & 0.898 & 364\\
            CONST & 0.997 & 0.996 & 0.997 & 34 449\\\hline
            Macro average & 0.809 & 0.857 & 0.822 & 224 865 \\
            Weighted average & 0.847 & 0.794 & 0.801 & 224 865\\ \hline
        \end{tabular}
        \end{table}

\begin{figure*}
\includegraphics[width=\linewidth]{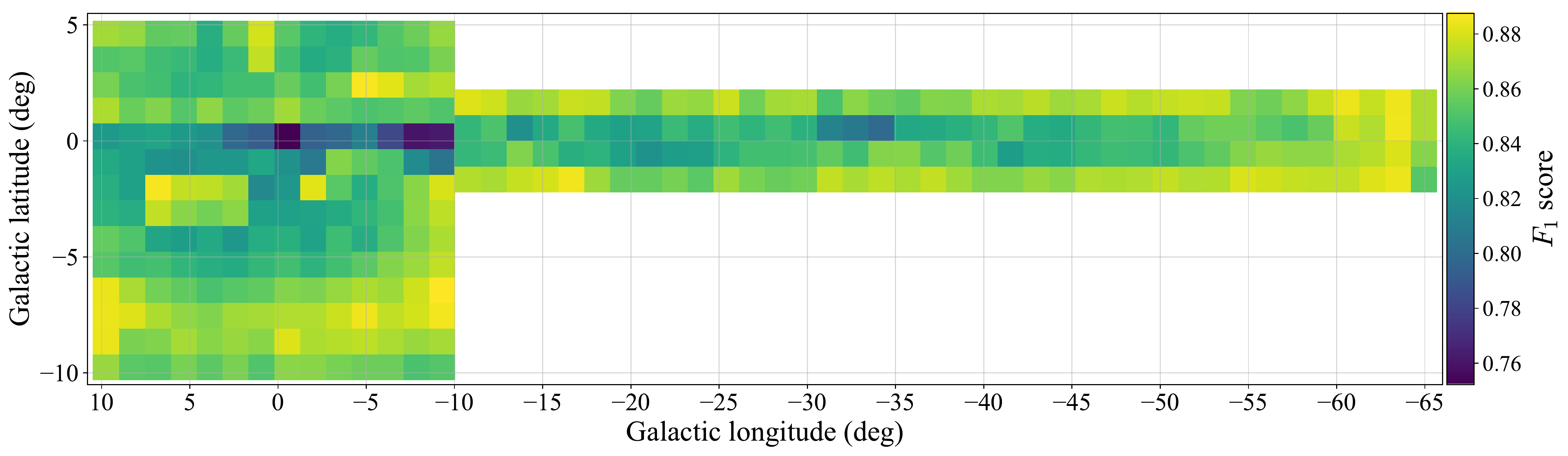}
\caption{Grid of 348 VVV tile binary classifiers colour mapped based on $F_1$ score of $10$-fold cross-validation. Performance is greatest for high latitude tiles and those found in the high cadence region ($+2.417^{\circ}\leq \ell \leq +6.807 ^{\circ}$ and $-3.138^{\circ} \leq b \leq -2.043^{\circ}$) and decreases towards the crowded regions of the Galactic midplane ($|b|<2^{\circ}$).}
\label{fig:f1_footprint}
\end{figure*}

\begin{figure}
    \centering
    \includegraphics[width=\columnwidth]{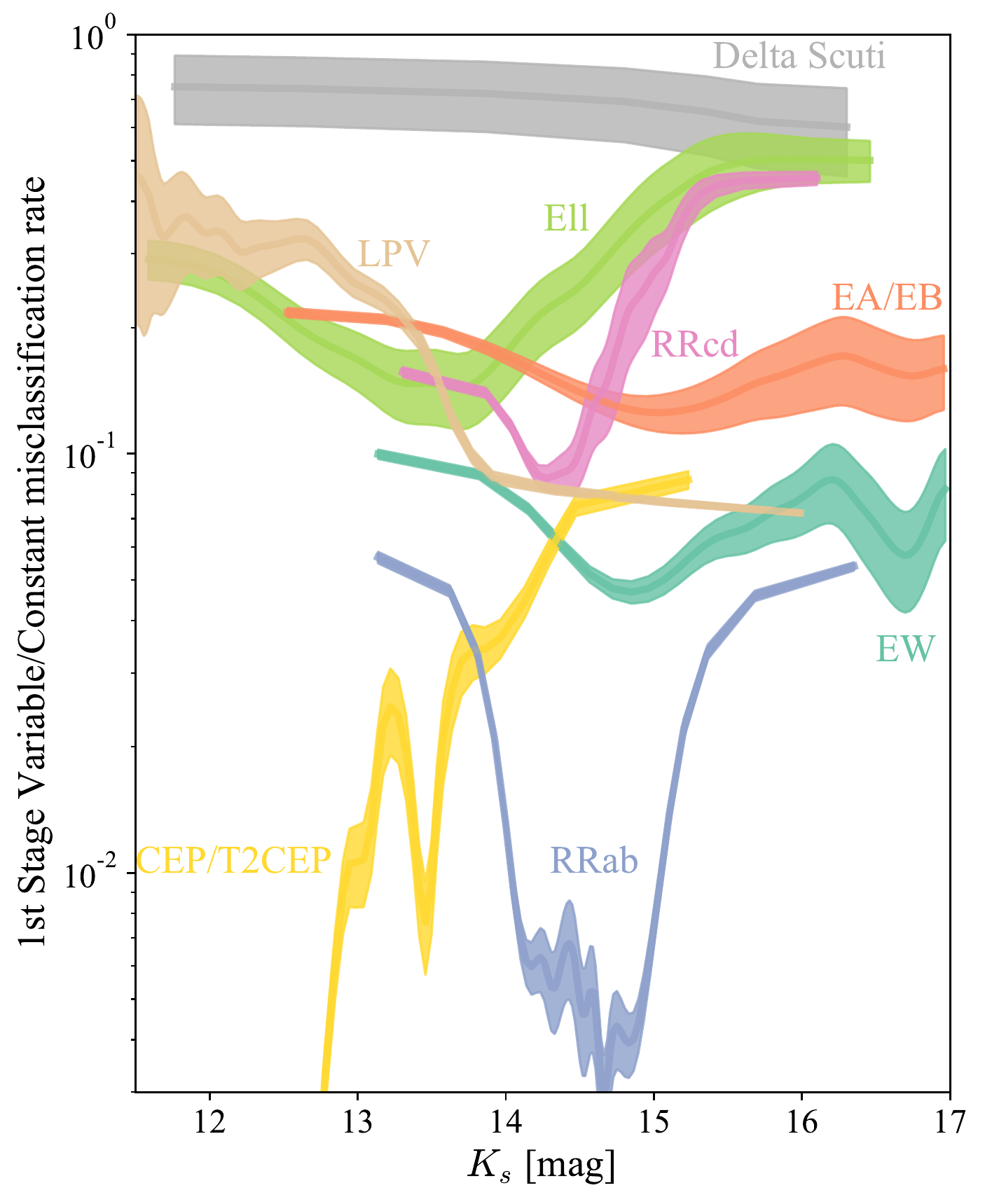}
    \caption{Misclassification rate for the 1st stage binary classification. We show the results for the variable star training set as a function of $K_s$ magnitude averaged over VVV tiles. The line width reflects the training set density with a broader band corresponding to higher density. Note Delta Scuti are not included in the second classification stage due to the high misclassification rate.}
    \label{fig:misclass}
\end{figure}

Performance of the classifiers is examined by dividing each full training set into a training and validation subset. This allows for the construction of a classifier with the training portion, whose accuracy can subsequently be tested through the classification of the validation set. The assessment procedure can be enhanced by increasing the number of initial divisions to $k$ ``folds''. The process follows the same steps with every combination of $k-1$ folds compiled to form the training subset, leaving the remaining fold in each iteration for validation. The overall effectiveness of the classifier is then obtained as the average performance over each fold. This process is referred to as $k$-fold cross-validation. We choose $k=10$.

Performance metrics for the validation tasks typically rely on combinations of the number of true positive (TP), true negative (TN), false positive (FP) and false negative (FN) classifications. The following metrics were considered:
\begin{equation}
    \text{recall}=\frac{\text{TP}}{\text{TP+FP}}\qquad  \text{precision}=\frac{\text{TP}}{\text{TP+FN}}\\
\end{equation}
\begin{equation}
    F_1=2\left( \frac{\text{recall}\times \text{precision}}{\text{recall}+\text{precision}}\right)\\
\end{equation}
The recall (more often called completeness in astrophysics applications) indicates the fraction of input example sources, for a given class, that are correctly labelled, whilst the precision (purity) shows the fraction of output predictions which are truly part of the class. The harmonic mean of these metrics constitutes the $F_1$ score, giving an overall measure of accuracy for the classification. Confusion matrices also serve as useful visual tools to inspect the performance of a classification process. The recall for each class is shown along the leading diagonal with the fraction of false negatives and false positives, with respect to other classes, shown in adjacent cells along each row and column respectively.

A quantitative summary of the averaged $10$-fold cross-validation scores obtained for the $348$ initial binary classifiers is shown in Table~\ref{tab:crossval_binary}. We quote the mean and standard deviation of the metrics over the classifiers. The corresponding confusion matrix is given in the left panel of Fig.~\ref{fig:confusion_matrices}. The macro $F_1$ average, where macro signifies normal averaging such that all predictions are given equal weight, is $0.854 \pm 0.021$ showing a notable level of discrimination between the variable and constant training classes. Classifier accuracy depends highly on the Galactic positioning of the tile in question, as shown in the distribution of $F_1$ scores over the footprint in Fig.~\ref{fig:f1_footprint}. Performance peaks in the high cadence region of VVV ($+2.417^{\circ}\leq \ell \leq +6.807 ^{\circ}$ and $-3.138^{\circ} \leq b \leq -2.043^{\circ}$ where there are $\sim200-600$ detections per source) and more sparsely populated parts of the bulge and disk, where constant sources are sufficiently sampled or isolated to limit the effect of spurious magnitude measurements. In contrast, the crowded tiles towards the Galactic midplane exhibit lower $F_1$ scores of typically 0.77 due to the decrease in observational accuracy.

We show in Fig.~\ref{fig:misclass} the average misclassification rate averaged over all first-stage classifiers, equivalent to $1-\text{recall}$, as a function of $K_s$ magnitude and split by detailed class. RRab, the majority of which are within $14 \leq K_s \leq 15$, are efficiently identified with misclassification rates strictly below $1\,\percent$ within said boundaries. EW and the joint Cepheid class CEP/T2CEP display good recovery with rates of $5$ and $2\,\percent$ at peak source density respectively. Lower amplitude variable stars such as EA/EB, Ell and RRcd are harder to discern from constant sources and are misclassified as constant sources in $10$ to $30 \percent$ of predictions. Around $30\,\percent$ of LPV sources are misclassified towards the bright magnitude end where saturation and bleeding are more commonplace leading to erroneous variability indices. Though significant misclassification is apparent for these classes, the sources affected do not have satisfactory VVV light curves for accurate period extraction and further classification. Our sample of DSCT sources, also observed to have insignificant amplitudes, are rarely classified as variables with a standard recall of $0.3$. We are therefore unable to determine these sources in the survey and ignore the class for stage 2. This analysis leads us to believe that the first stage selects the majority of worthwhile variables across the whole footprint as intended.

Similar analysis is given in Table~\ref{tab:crossval_variable} for the cross-validation performance of the second-stage classifier. We report scores computed for each class and give the corresponding confusion matrix in the right panel of Fig.~\ref{fig:confusion_matrices}. The macro $F_1$ for the classifier is $0.822$. It is clear that RR Lyrae variables are most accurately classified, with $97\percent$ true positive rate for RRab and $93\percent$ for RRcd. This is a product of the narrow and distinctive regions in feature space populated by these sources as illustrated previously (see Fig.~\ref{fig:period_amp_feature}). The LPV class also achieves high accuracy with only an $8\percent$ misclassification rate. This is expected as the majority of the class consists of high amplitude red Mira variables, easily discernible from other entities in period, amplitude and colour. In contrast to this, detached and contact eclipsing binaries are significantly mixed, with $28\percent$ of the training set EA/EB identified as EW and $12\percent$ of EW sources labelled as EA/EB. This is due to the inherent similarities between both variable types and no decisive physical class boundary as discussed in Section~\ref{subsect:eclipingbinary}. The ellipsoidal binary class (Ell) suffers a smaller loss in accuracy due to resemblance with its eclipsing counterparts, and is misclassified $8\percent$ of the time as either EW or EA/EB classes. Relatively low levels of EA/EB, EW and Ell contamination in other classes are noted. However, as the EA/EB represents the largest training class, any degree of misclassification brings about a large number of contaminants. This entails low precision scores for the severely contaminated classes such as 0.522 and 0.592 for EW and Ell classes respectively.  Classical Cepheid variables are recovered in $65\percent$ of predictions as a result of the limited training set available and similarity to Ell examples. We also note a complete separation between all variables classes and the CONST class as a result of the differing log$_{10}(\mathrm{FAP})$ distributions and periodic features. Note that our quoted statistics reflect the success with which we recover the labels in the training set. Any misclassifications in the training set likely reduce the genuine success rate with which we classify true class members.

With high accuracy, specifically for variable classes of primary interest, we evaluate the second stage of the classifier as successful. The efficacy of the whole method can be examined by compounding the performance of each stage. We believe this results in a competent workflow to extract and identify variable sources. Furthermore, we bias towards sources with high degree of periodicity producing low contamination from constant sources and between variable classes. This in turn leads to an incomplete but somewhat reliably classified output catalogue. We proceed by retraining all classifiers in both stages on their full respective training sets and store them for application on unknown sources.

\subsubsection{Feature importance}

\begin{figure*}
    \centering
    \includegraphics[width=0.8\linewidth, valign=c]{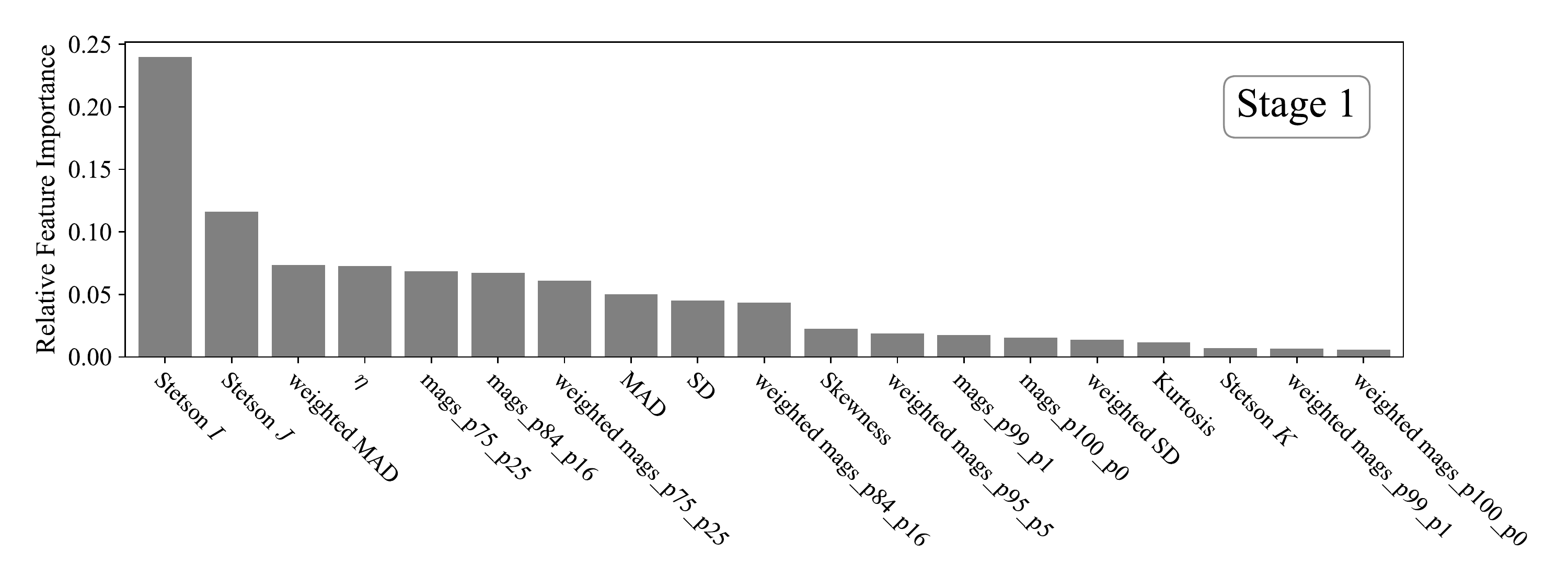}
    \includegraphics[width=\linewidth, valign=c]{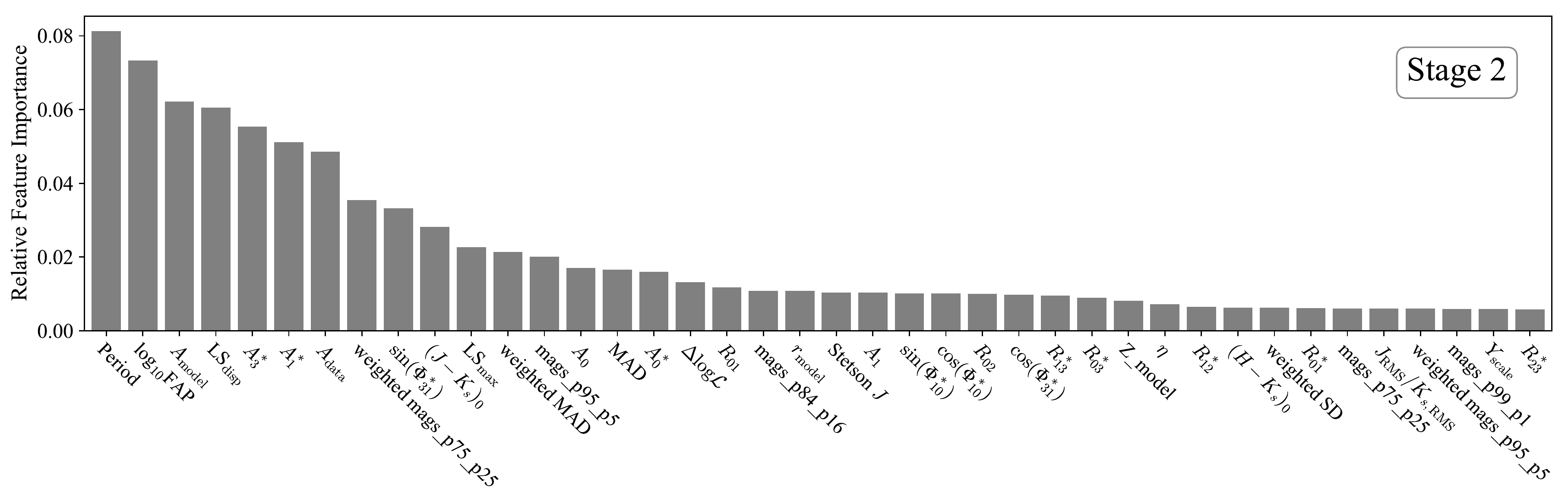}
    \caption{Relative feature importances for all binary classifier features (top) and top 40 variable star classifier features (bottom). All importances for features employed in either case sum to unity.}
    \label{fig:feature_imp}
\end{figure*}

A key advantage of ensemble classification is the ability to assess individual feature importances. These measure the normalised mean decrease in decision tree impurity as a result of feature inclusion for RF classifiers \citep{Breiman2001} and the overall feature contribution to loss reduction along tree branches for XGBoost ensembles \citep{XGBoost}. In either case, all incorporated feature importances sum to unity. Furthermore, each feature importance is computed independently and does not account for correlation between attributes. Hence, quantities outlining the same physical properties may have equally large independent importances. Consequently, feature combinations of these quantities would lead to meager increases in importance.

Table~\ref{tab:var_indices} in Appendix~\ref{appendix::feature_table} gives the full feature ranking for the first and second stage classifiers. Fig.~\ref{fig:feature_imp} displays this information visually, limiting to the $40$ most important features for the secondary classifier. The greatest distinguishing factor for the initial classification is the level of correlation between consecutive time-series points included as the Stetson $I$ and $J$ indices. Other light curve spread statistics, such as percentile ranges and median absolute deviation, are also influential in stage 1 classifications. Similarly, the degree and cadence of periodicity quantified as $\log_{10}(\mathrm{FAP})$, the period and the amplitude are used to greatest effect when determining the specific variability type of the sources in stage 2.

It must also be noted that these quantities establish the general importance of a feature in \emph{all} classifications. We therefore expect distinction between specific class boundaries to rely on certain features, possibly quantified as relatively unimportant overall. For example, phase differences are particularly useful differentiating between asymmetric and symmetric light curves such as RRab and EW, respectively, and colour amplitude ratios in separating extrinsic and intrinsic variables. However, these features are not instrumental in all classifications.

\section{Classification of the entire VVV survey}\label{sect:results}

We now apply the constructed hierarchical classifier to first extract candidate variable stars from the full VVV survey and then categorise their variability type.

The 490 million unclassified VVV source $Z Y J H K_s$ photometric light curves and variability indices, described in Section~\ref{subsect:phot_dataset}, are stored in a spatially arranged grid of 22 585 files created through \verb|HEALpix| pixelization \citep{HEALPix} of the on-sky survey footprint. We begin by splitting the file group into 128 `chunks' and parallelise the computation over individual pixels in each chunk and over the chunks themselves. This severely reduces the compute time.

The end-to-end procedure carried out for each pixel is the same and begins by locating the VVV tiles within which the pixel sources are found. The corresponding tile binary classifiers are then loaded from disk and used to classify the input sources based on their computed variability indices. Note that we pre-process and normalise the feature entries before classification in the same way as in Section~\ref{subsect:train}. Candidate variables are selected as outputs labelled variable with first stage classification probability greater than 0.65, where probabilities denote the fraction of trees in the Random Forest which reach majority consensus. Stage 2 periodic and non-periodic features, outlined in Section~\ref{subsect:stage2}, are then computed for each candidate in order to predict the variability type using the trained detailed variable classifier loaded from disk.

On average this resulted in $24.5\percent$ of sources labelled as candidate variables to be classified further. The vast majority of these were determined to be constant sources by the secondary classifier due to ambiguous periodic or non-periodic features. Inspection of the output set of roughly 1.9 million variable sources showed narrow overdensities in source count around observational aliases, a signature of spurious period extraction. Sources with output frequency (1/Period) within a narrow $\pm 2.5 \times 10^{-4}\,\mathrm{day}^{-1}$ window of a reciprocal sidereal day, and its integer multiples $f_n = n*\mathrm{\left(sidereal\,day\right)}^{-1}\,\mathrm{day}^{-1}$ for $n=2,3,...,9$, were removed. Additionally, we included the above sidereal frequencies $\pm 1/365.25\,\mathrm{day}^{-1}$ to the removed range in order to account for coupling with year aliases. Following this procedure, a remaining set of $1,364,732$ sources were characterised as periodic variables with no clear alias signal. We display the classification results by class as well as their grouping based on classification probability in Table~\ref{tab:results} and denote the number of sources recovered from our initial training set. $198,476$ predictions detail sources found in our initial variable set, of which $155,676$ have matching class labels. We attribute the loss of $51.1\percent$ of initially collected variables primarily to the strict $\log_{10}(\mathrm{FAP})$ cut performed when constructing the second stage training set. This places a selection bias in favour of entries with a high degree of periodicity recognised in their Lomb-Scargle periodograms. In contrast, we recover $90.6\percent$ of the second stage training set which satisfy the false alarm probability constraint. The $42,800$ predictions of known variables with incorrect labelling ($21.6\percent$ of all recovered sources) agrees roughly with  $20\percent$ average contamination rate between variable classes attained from cross-validation of the training set (see Section~\ref{sub:performance}).

\begin{figure}
	\centering
    \includegraphics[width=\columnwidth]{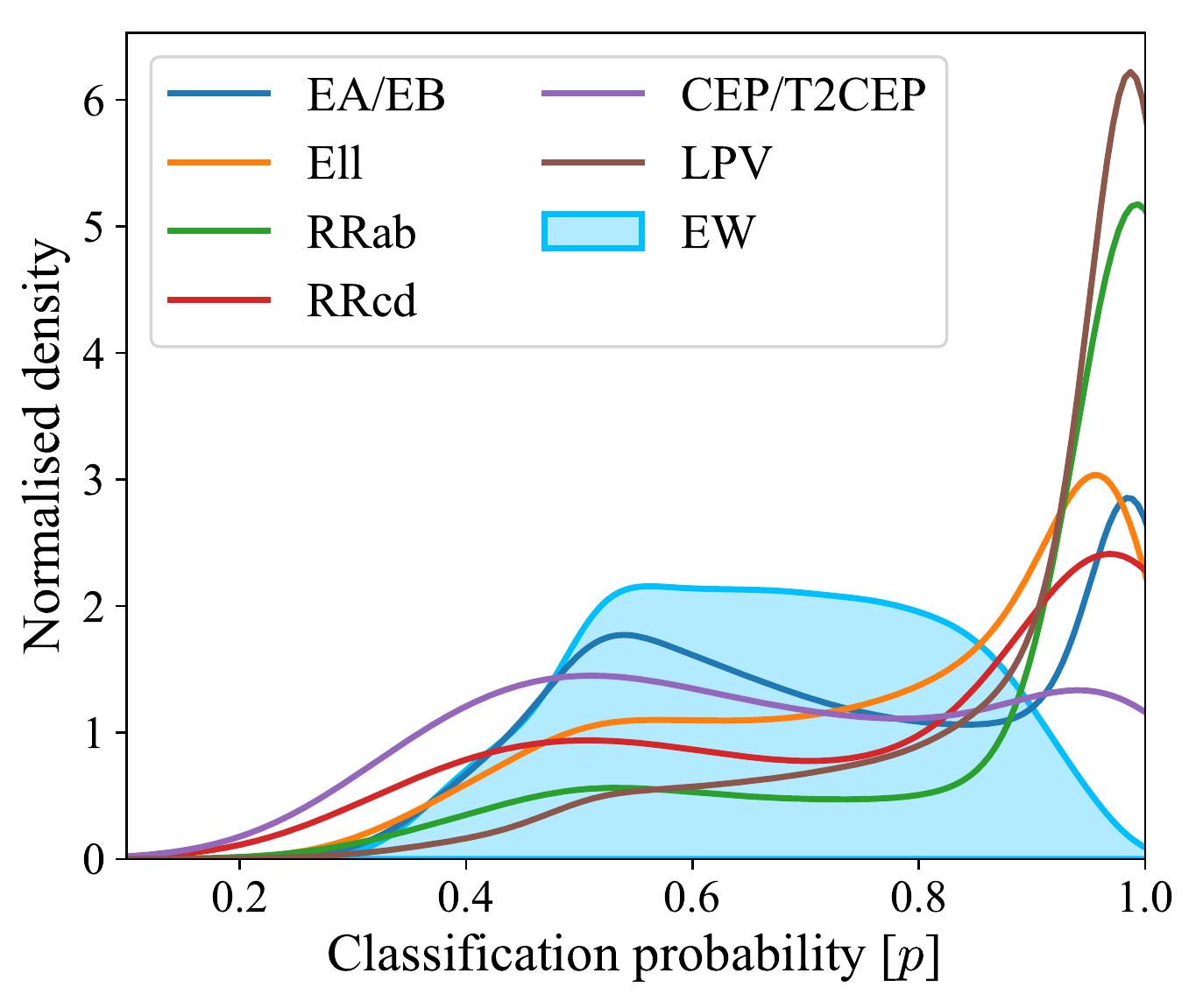}
    \caption{Distribution of classification probability for catalogued variable sources. Area is normalised to one for each class. Bimodality with a high probability ($p>0.9$) peak is present for each class apart from the shaded light blue EW distribution peaking around $p=0.55$.}
    \label{fig:prob_dist}
\end{figure}

\begin{figure*}
    \centering
    \includegraphics[width=\textwidth]{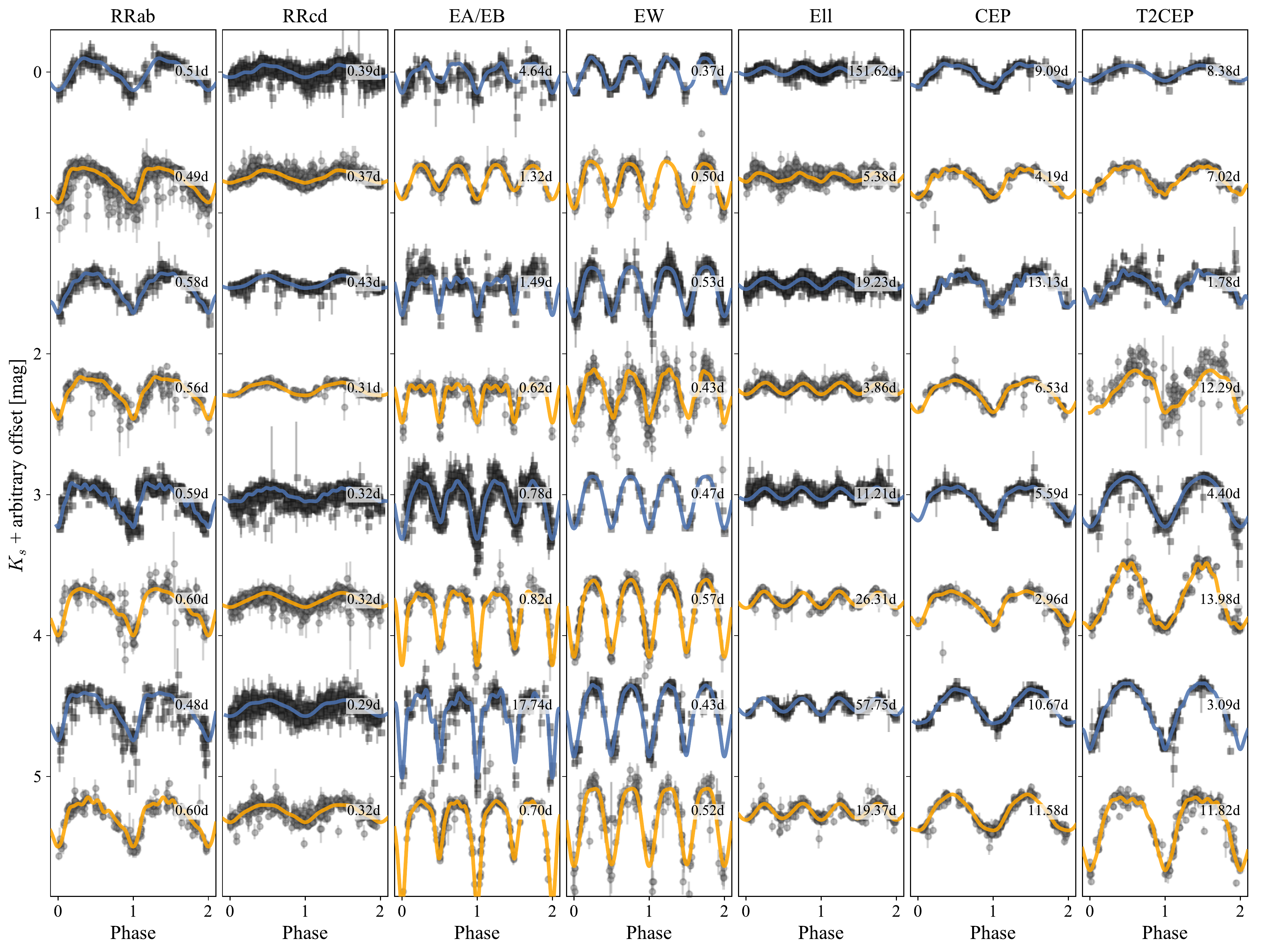}
    \caption{Example light curves classified by our algorithm. Each column corresponds to a different class. All sources were selected to be within $2\,\mathrm{deg}$ of the Galactic plane, to have high classification probability ($p>0.9$) and to have log-likelihood difference between the periodic and constant model $>300$.}
    \label{fig:lc_gallery}
\end{figure*}

XGBoost classification probability, $p_C$, for a class, $C \in$ Classes $=\{$EA/EB, EW, Ell, RRab, RRcd, CEP, T2CEP, LPV$\}$, is given by
\begin{equation}
p_C(x) = \frac{\mathrm{e}^{F_C(x)}}{\sum\limits_{i\in \text{Classes}}{\mathrm{e}^{F_i(x)}}}
\end{equation}
where $x$ is the example feature entry and $\{F_i(x)$ for $i \in$ Classes$\}$ is the set of class prediction scores output by the classifier. From here on, we discard class notation and annotate the probability of classification as $p$ with $C$ taken as the predicted class. In Fig.~\ref{fig:prob_dist} we show the distribution of second-stage classification probability for predictions in each class. Bimodal distributions are noticed for nearly all of the classes, with a sharp peak at high probability and a lower peak in the range 0.4 to 0.6. The former peak corresponds to sources with features in complete agreement with their predicted training class, whilst the latter incorporates sources with similarity to multiple classes. Contact binaries (EW) stand out with a distinct lack of high probability predictions, with only $1.0\percent$ of predictions having $p\geq0.95$. We believe this is a result of a broad EA/EB training set incorporating a portion of sources near identical to EW in all features (see Section~\ref{subsubsect:periodic_features}). This entails significant classifier scores associated with the EA/EB class for EW predictions, limiting high probability outputs. Examples of high confidence, $p>0.9$ and $\Delta\log\mathcal{L}>300$, extracted variable light curves are shown in Fig.~\ref{fig:lc_gallery}. These sources are found within 2 deg of the Galactic plane, showing the ability to extract and determine periodic stars in challenging observational regions with the NIR survey. The similarity between the first EA/EB light curve and the EW light curve column emphasises the difficulty in differentiating between eclipsing binary subtypes. The confusion between eclipsing binary subtypes highlights a limitation of the supervised learning approach. Unsupervised learning algorithms such as locally linear embedding or t-distributed stochastic neighbor embedding may allow for reclassifications of the spectrum of eclipsing binaries in our dataset \citep{matijevic2012,kirk2016,Bodi2021} and a cleaner separation of classes. However, our reported statistics and the classification probabilities give an honest reflection of the ambiguity in classification given the set of well-motivated classes we employ and the quality of our data.

\begin{table*}
        \caption{Classification summary of the entire VVV survey. We give the number of classifications for each class and the percentage with classification probability, $p$, greater than $0.8$, $0.9$ and $0.95$. The total number of sources recovered from the variable set constructed in Section~\ref{subsect:variable trainset} is also quoted as well as the number of these for which our classifications agree with the literature.}
        \label{tab:results}
        \begin{center}
        \begin{tabular}{lc c c cc c}
            \hline
            Class & \# of classifications  & $p\geq 0.8$ (\%) & $p\geq 0.9$(\%) & $p\geq 0.95$ (\%)& \makecell{ \# recovered \\ sources} & \makecell{\# with matching \\ class}\\ \hline
            EW & 338 647 & 22.5 & 5.3 & 1.0 & 69 085 & 37 361\\
            EA/EB & 599 529 & 41.7 & 31.2 & 25.2 & 85 673 & 81 318 \\
            Ell &  155 822 & 53.6 & 37.8 & 25.1 & 12 722 & 8 266 \\
            RRab & 53 728 & 77.3 & 71.9 & 67.4 & 23 200 & 22 031 \\
            RRcd & 16 369 & 57.5 & 48.3 & 40.9 & 5 993 & 5 052 \\
            CEP & 5 900 & 33.0 & 23.4 & 17.3 & 320 & 283\\
            T2CEP & 8 830 & 34.9 & 24.5 & 18.6 & 708 & 629\\
            LPV & 185 907 & 77.1 & 65.7 & 55.6 & 775 & 736\\\hline
            Total & 1 364 732 & 44.6 & 31.9 & 25.1 & 198 476 & 155 676\\ \hline
        \end{tabular}
        \end{center}
\end{table*}

\subsection{Contamination of the catalogue}
\begin{figure}
    \centering
    \includegraphics[width=\columnwidth]{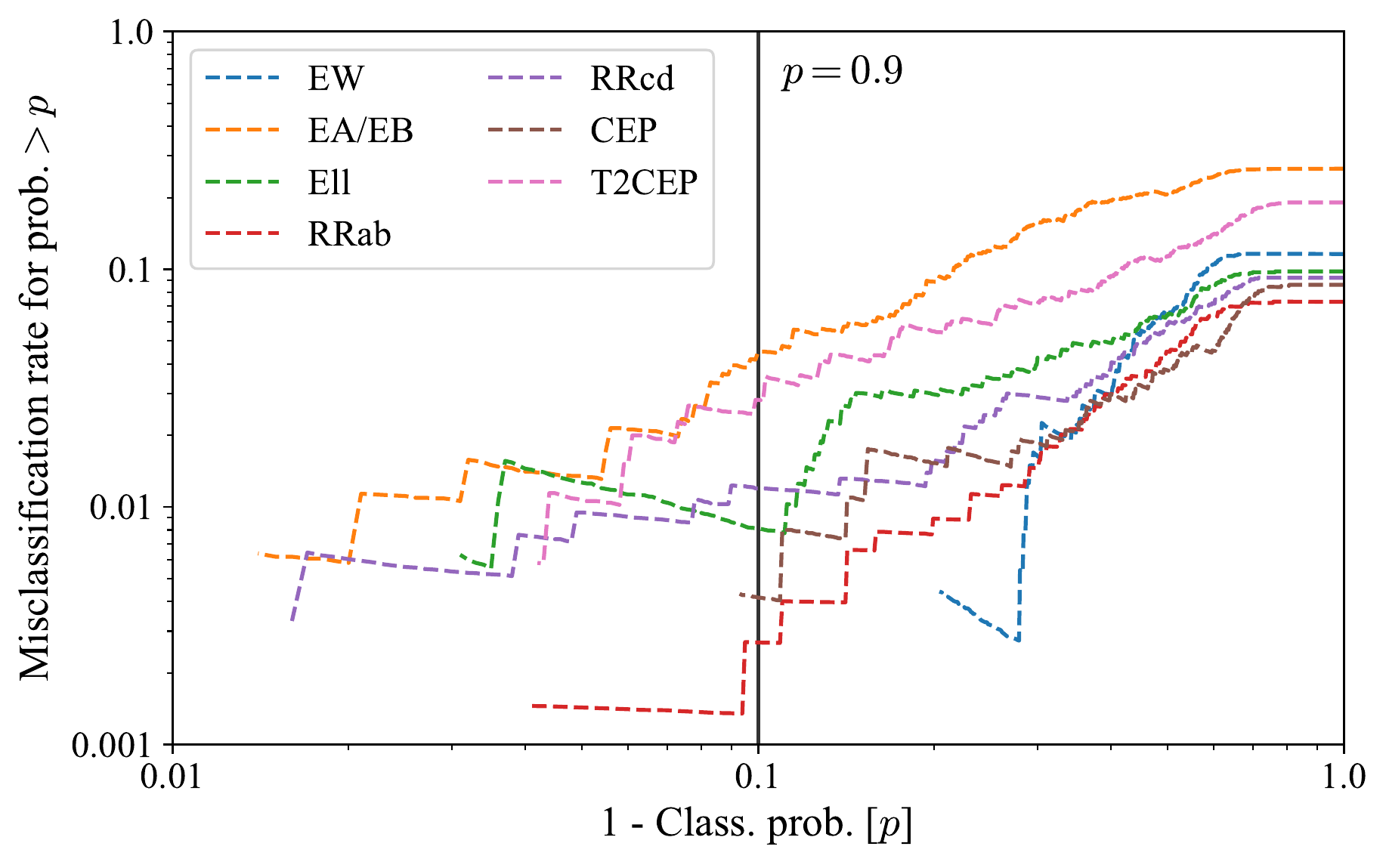}
    \caption{Visually determined cumulative misclassification rate against probability of classification. Rates for high probability ($p>0.9$) sources are shown by intersections with the vertical black line. Distributions end when no more contaminants are found in the probability range of the samples.}
    \label{fig:eyeball_misclass}
\end{figure}

Complete visual inspection of the output catalogue requires large amounts of time and countless pairs of eyes. We therefore limit the exercise to a thousand sources in each class to gauge misclassification estimates in various probability ranges. This is carried out through reviewing the shape of phase-folded light curves in comparison to known variable light curves. Misclassifications are noted if no periodicity is apparent or light curve shape resembles that of a class other than the one predicted. We find the former much more prevalent and due primarily to a faulty period extraction yielding observational aliases. We outline in Fig.~\ref{fig:eyeball_misclass} the results of the visual inspection. All classes have below $5\percent$ misclassifications for predictions with $p>0.9$. The RRab sample is particularly impressive with lower than $1\percent$ misclassified sources for $p>0.8$, where the majority of the output class lie. The same is seen for the EW class, with $p=0.8$ marking the high end of its shifted probability distribution. We neglected consideration of the LPV class due to concerns of largely erroneous period extraction, mentioned in Section~\ref{subsubsect:periodmatch}, which would entail severe misclassification rates.

We provide the catalogue in its entirety and advise the use of known class boundaries for physical features, such as period or amplitude ranges, paired with probability cuts in order to provide minimally contaminated samples, primed for scientific use-cases. These representative samples can be further refined by restricting features outlining the quality of the model fit or periodicity (e.g. $\Delta\log\mathcal{L}$, FAP, LS$_\mathrm{max}$ or LS$_\mathrm{disp}$).

\begin{figure*}
\includegraphics[width=0.93\linewidth]{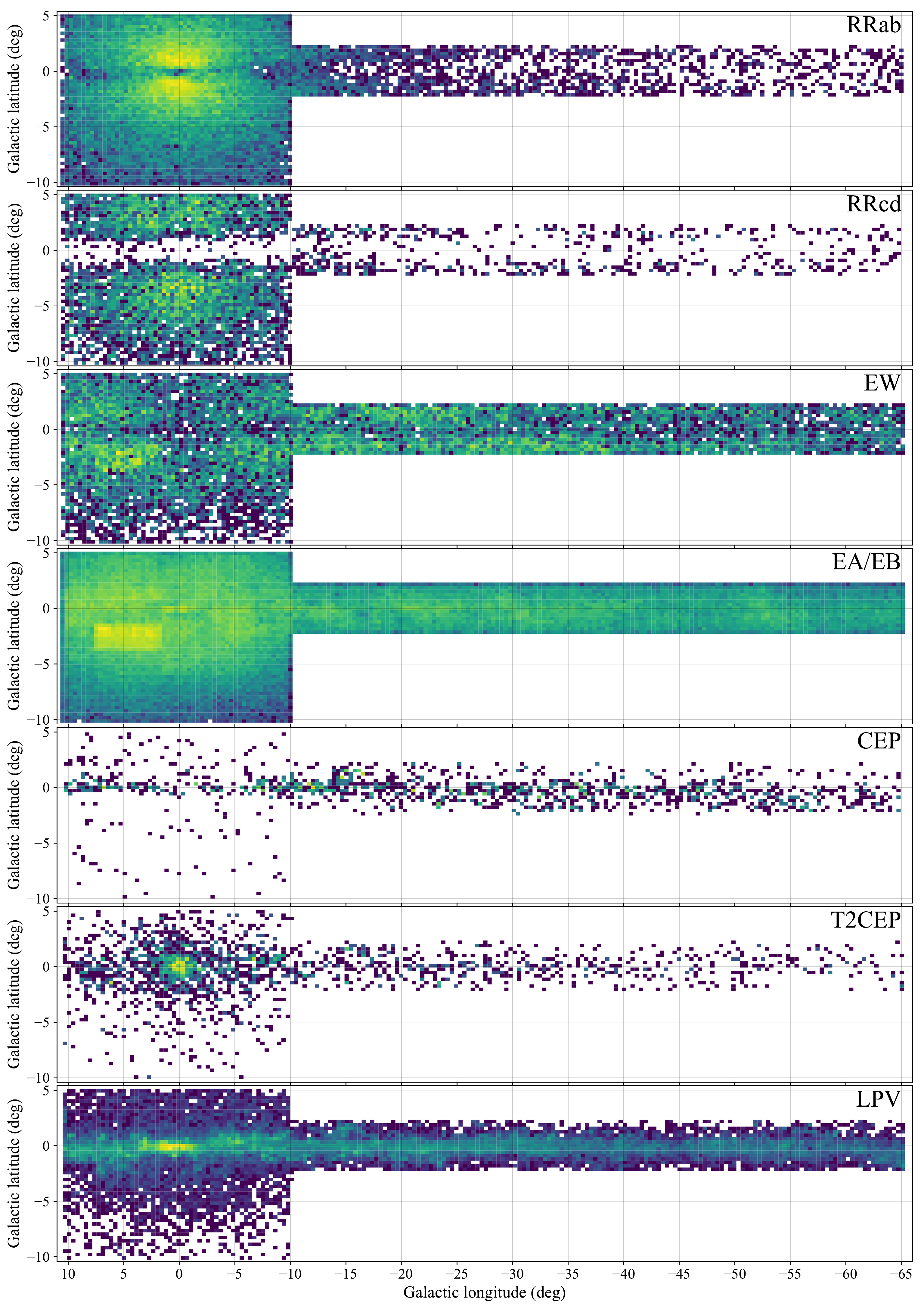}
\caption{On-sky distributions of high-probability ($p>0.9$) variable sources of each class. The colour-scale is logarithmic.}
\label{fig:result_footprints}
\end{figure*}

The normalised on-sky density distribution for sources with $p\geq0.9$ is illustrated in Fig.~\ref{fig:result_footprints} for all variability classes. The EW and RRcd maps show lower density in the Galactic midplane where quality of light curves is more limited. This effect is less pronounced for the RRab sample which shows reasonable coverage in the mid-plane and dense sampling across the rest of the footprint. In contrast we see classical and Type II Cepheid variables peaking in the central bulge regions, possibly tracing in part a young inner disk population \citep{Dekany2015}. We also note the flattened LPV distribution which appears to follow the extinction. This suggests that colour is a primary factor in classifications for this class, with long period redder sources predominantly labelled as such. However, it should also be noted that typically these sources would be intrinsically very bright in VVV so significant extinction is required to avoid saturation. The LPV overdensity towards the Galactic centre region is likely related to the nuclear stellar disc \citep{Launhardt2002}.

\subsection{Catalogue completeness}
The design of our algorithm has been chosen to minimise contamination at the expense of lowering the completeness. In general, the completeness of our catalogue is difficult to assess as any calculation is biased by the properties of our training set. However, here we investigate the completeness of the RR Lyrae star sample with respect to the latest OGLE results \citep{soszynskilyr}.

\begin{figure*}
\includegraphics[width=\linewidth]{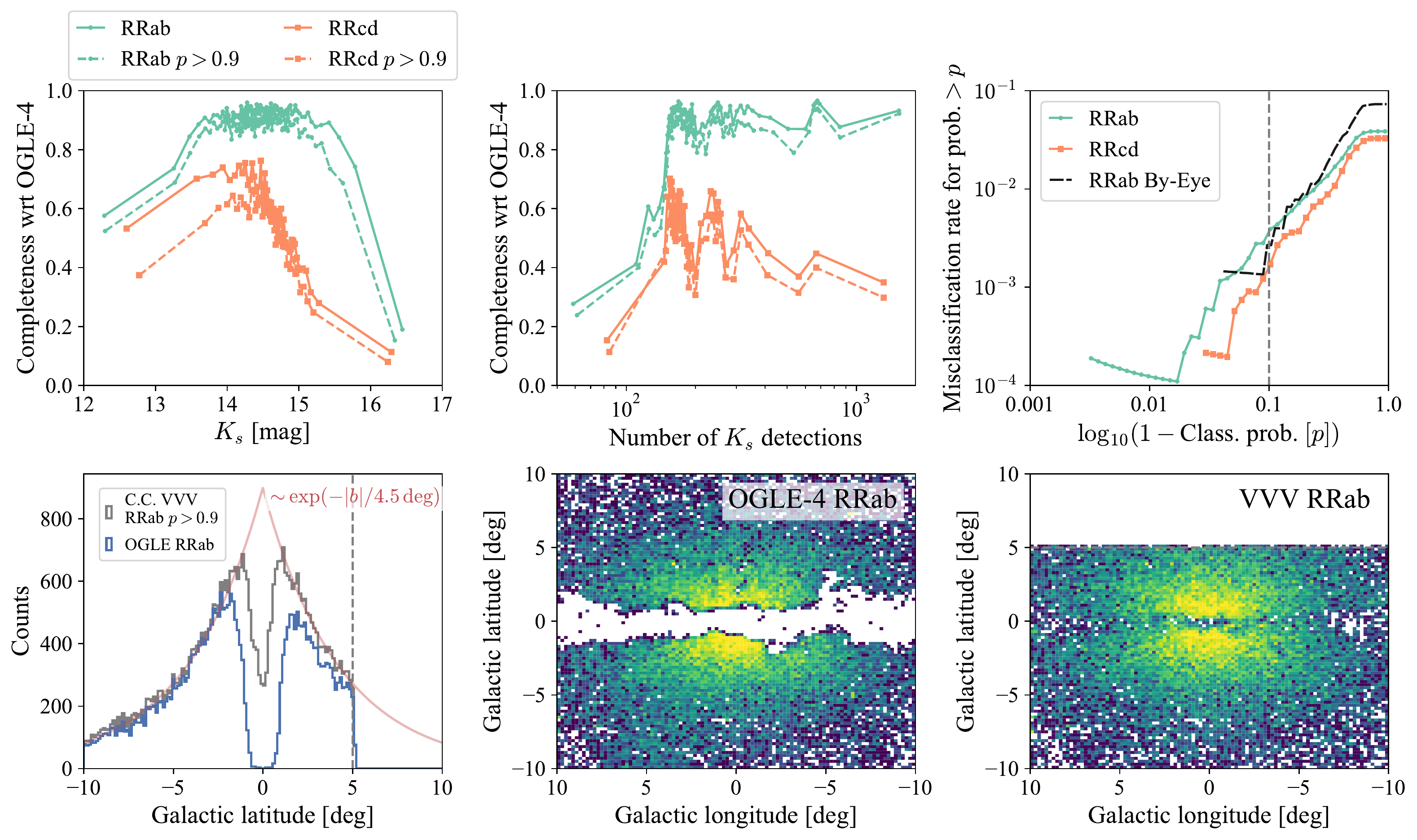}
\caption{Properties of the RR Lyrae stars in comparison to the OGLE-4 catalogue of \protect\cite{soszynskilyr}. The top row shows the completeness (number of OGLE sources in VIRAC-2 recovered by our procedure) against $K_s$ magnitude (left) and number of detections (middle). The right top panel shows the cumulative rate of RR Lyrae stars in OGLE classified differently by us. We also display the misclassification rate from visual inspection of the RRab set (from Fig.~\protect\ref{fig:eyeball_misclass}) in black dashed, which agrees well with the OGLE comparison. The lower set of panels show the density of RRab in OGLE-4 and VVV. The VVV density is completeness-corrected with respect to OGLE using the dashed green curve in the upper left panel. The left panel shows the density in Galactic latitude for the bulge region ($|\ell|<10\,\mathrm{deg}$,$|b|<10\,\mathrm{deg}$) using stars with $K_s<16$ where we also show a by-eye exponential fit with scaleheight $4.5\,\mathrm{deg}$. The right two panels show the on-sky number counts for the full OGLE RRab sample and the completeness-corrected VVV sample with $K_s<16$.}
\label{fig:ogle_completeness}
\end{figure*}

In Fig.~\ref{fig:ogle_completeness} we compare the properties of our RR Lyrae stars (both RRab and RRcd) to the OGLE sample. The on-sky distribution of OGLE RRab and our (completeness-corrected, see later) VVV RRab sample with $p>0.9$ and $K_s<16$ are shown in the bottom right. The overall morphology of the samples is similar with the VVV sample reaching further into the high-extinction plane. The RRab related to the Sagittarius stream are visible in the OGLE data around $(\ell,b)=(5,-8)$ but are too faint ($K_s>17$) for VVV.

Due to `dead' regions in the OGLE camera, the absolute completeness of any OGLE dataset is reduced by $6-7\,\percent$. Furthermore, by inspecting duplicate sources, \cite{soszynskilyr} evaluate the completeness of the remaining RRab and RRcd as $96$ and $91\,\percent$ respectively.
We can thus treat the completeness with respect to OGLE as being indicative of the overall completeness. In the top left and middle panels of Fig.~\ref{fig:ogle_completeness}, we show the fraction of VIRAC-2 cross-matches in the full OGLE RR Lyrae star sample (using a cross-match radius of $0.4\,\mathrm{arcsec}$ and accounting for the epoch difference using the VIRAC-2 proper motions) which enter our final variable star catalogue. We see that the completeness is around $90\,\percent$ for RRab with more than $\sim150$ epochs and $14<K_s<15$. A similar conclusion was reached by \cite{Contreras2018} who found $\sim10\percent$ of the OGLE RRab were low-amplitude and difficult to detect with VVV. Restricting to only those with $p>0.9$ only weakly changes the completeness suggesting many objects with $p<0.9$ are not genuine RRab. The tail at faint magnitudes arises due to increased uncertainties in VVV. The bright magnitude tail is a reflection of the higher misclassification rate seen in Fig.~\ref{fig:misclass}. For RRcd, the completeness is lower and decays at brighter magnitudes due to their lower amplitude. Furthermore, limiting to $p>0.9$ sources more significantly changes the contamination suggesting this cut is too strict and removes some genuine RRc. In the top right of Fig.~\ref{fig:ogle_completeness} we show the cumulative misclassifications by us of the OGLE RR Lyrae stars as a function of classification probability. We also show the RRab misclassification rate determined by eye from Fig.~\ref{fig:eyeball_misclass} which agrees very well with the OGLE result. This gives us confidence that for $p>0.9$ the misclassification rate (contamination) is below $1\,\percent$.

In the lower left of Fig.~\ref{fig:ogle_completeness} we show the density profile of OGLE and VVV RRab with $K_s<16$ as a function of Galactic latitude. Here we have corrected the VVV number counts for the incompleteness of the VVV RRab sample with respect to OGLE (top left). For $b\lesssim-3\,\mathrm{deg}$, we see the very good correspondence between the two density profiles, which is well fitted by an exponential profile with scalelength $4.5\,\mathrm{deg}$. This test is not trivial as it requires low contamination in our VVV sample. Note at positive latitudes the OGLE density is lower than symmetry would suggest, whilst our sample well matches the symmetric prediction, suggesting OGLE is incomplete above the plane. Finally, we see that the VVV density profile continues as predicted by the simple model to slightly lower latitudes than OGLE, reflecting VVV's ability to probe further through the midplane dust.

\section{Science potential of the catalogue}\label{sect:science}
A full investigation of the contents of the catalogue is beyond the scope of this work, and we intend on producing more detailed science studies resulting from this catalogue in future works. However, to close the presentation of this new catalogue, we will present some preliminary results highlighting the scientific potential of the catalogue.

\subsection{Spatial and kinematic distributions of RR Lyrae stars}
We briefly investigate the spatial and kinematic distributions of the RR Lyrae stars in our catalogue. As highlighted in the introduction, these stars are useful Galactic tracers due to their tight period-luminosity relation and bias towards old, metal-poor systems. Our investigation here is brief and we refer readers to the more thorough work of \cite{Du2020} using OGLE RR Lyrae stars, which we corroborate here.

\begin{figure*}
\includegraphics[width=\linewidth]{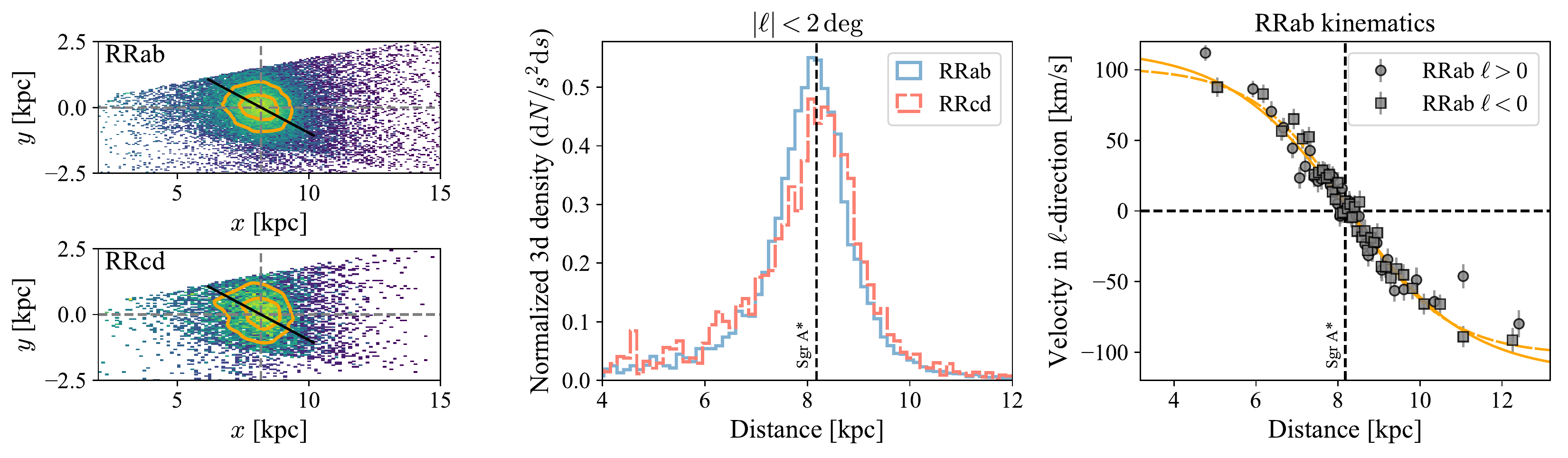}
\caption{Properties of the RR Lyrae star sample: the left two panels show the top-down view of RRab (top) and RRcd (bottom) samples in Galactocentric coordinates. Contours show the $20$ ($33$) and $50$ ($66$) per cent of peak density for the top (bottom) panel. The black line corresponds to a bar angle of $28\,\mathrm{deg}$. The middle panel shows the volume density of sources within $2\,\mathrm{deg}$ of the bar-bulge minor axis using the period-luminosity-metallicity calibrations from \protect\cite{Cusano2021}. The right panel shows the mean velocities of RRab in the Galactic longitude ($\ell$) direction, separated into $\ell>0$ (circles) and $\ell<0$ (squares). Simple $\tanh$ models are fitted in orange (dashed is $\ell<0$).}
\label{fig::rrl_illustration}
\end{figure*}

We use the period-luminosity-metallicity relations for RR Lyrae stars in the Large Magellanic Cloud (LMC) from \cite{Cusano2021}. These relations are in the VISTA bands. For RRab type the relations are
\begin{equation}
\begin{split}
	J = &(17.888\pm 0.008) - (2.45\pm 0.02) \log_{10} P \\&+ (0.121\pm0.004)[\mathrm{Fe/H}],\\
	K_s = &(17.547\pm 0.008) - (2.80\pm 0.02) \log_{10} P\\&+ (0.114\pm0.004)[\mathrm{Fe/H}],
\end{split}
\end{equation}
whilst for RRc type they give
\begin{equation}
\begin{split}
	J = &(17.225\pm 0.03) - (2.53\pm 0.05) \log_{10} P\\&+ (-0.010\pm0.012)[\mathrm{Fe/H}],\\
	K_s = &(16.850\pm 0.026) - (2.99\pm 0.05) \log_{10} P\\&+ (0.011\pm0.012)[\mathrm{Fe/H}].
\end{split}
\end{equation}
We adopt an LMC distance modulus of $18.477\,\mathrm{mag}$ from \cite{Pietrzynski2019}. \cite{Dekany2021} has shown bar-bulge OGLE RRL have a metallicity mode of $-1.38\,\mathrm{dex}$ corresponding approximately to $-0.94\,\mathrm{dex}$ on the \cite{Skowron2016} scale used by \cite{Cusano2021}. For simplicity, we adopt this metallicity for all RRL.
To avoid dealing directly with extinction, we utilise the Wesenheit magnitude \citep{Madore1982}
\begin{equation}
    W_{JK_s} = K_s - R_{JK_s}(J-K_s).
\label{eq:weis}
\end{equation}
\cite{AlonsoGarcia2018} report $R_{JK_s}=(0.428\pm0.04)$ in the VISTA bands based on the observed colour-magnitude slope of bulge red clump in VVV, and \cite{WangChen2019} report $R_{JK_s}=(0.443\pm0.036)$ for 2MASS bands using red clump stars from the APOGEE spectroscopic survey. We use $R_{JK_s}=0.428$ which also matches very nicely the slope of the colour-magnitude distribution of the RRab sample.

In Fig.~\ref{fig::rrl_illustration} we display the top-down view in Galactocentric coordinates of the high-confidence ($p>0.9$) RR Lyrae ab and c/d stars. Both populations show a clear elongation in the same sense as the more metal-rich red clump stars \citep[e.g.][]{Wegg2013} with a major axis very close to $28\,\mathrm{deg}$ away from the Galactic Centre-Sun line. \cite{Du2020} reached similar conclusions based on the OGLE-4 RR Lyrae ab star sample, showing that the metal-rich RRab are aligned with the bar whilst the metal-poor RRab have a smaller bar angle. We also find that the volume density for both populations of RR Lyrae stars peaks very near the distance of Sgr A* \citep[$8.18\,\mathrm{kpc}$,][]{GravCollab}.
Some small discrepancies are perhaps expected due to the lack of more metal-rich RR Lyrae stars in the LMC sample used by \cite{Cusano2021} for the period-luminosity-metallicity calibration.
Finally, we show the velocity in the Galactic longitudinal direction corrected for the motion of Sgr A* \citep{Reid} for RRab either side of the bulge minor axis. We have fitted simple models of the form $v_\ell = (A/b) \tanh(b(s-s_0))$ to both populations finding a rotation rate of $A=(39.0\pm 2.9)\,\mathrm{km\,s}^{-1}\mathrm{kpc}^{-1}$ and $A=(43.2\pm 2.2)\,\mathrm{km\,s}^{-1}\mathrm{kpc}^{-1}$ for $\ell>0$ and $\ell<0$ respectively. As noted by \cite{Du2020}, this rotation rate is slower than that of the metal-rich bar-bulge stars indicating their greater pressure support. It remains to be seen whether the kinematics indicates the pattern speed of the RRab stars is similar to that of the metal-rich stars as suggested by their spatial alignment.

\subsection{Period-luminosity relation of contact Eclipsing Binaries (EW) and the Gaia EDR3 parallax zeropoint}
The two components of a contact eclipsing binary (EW) fill their Roche lobes and their envelopes are in contact. This typically leads to equilibrium of the surface temperatures and hence a highly symmetric light curve where the minima from each eclipse are difficult to distinguish. In this regime, the enclosed mean density is directly related to the orbital period of the binary, leading to tight period-luminosity relations. \cite{Chen2018} calibrated the EW period-luminosity relations in the 2MASS bands using Gaia DR1 data, finding
\begin{equation}
\begin{split}
	J_2 &= (-0.04\pm 0.11) - (6.87\pm 0.25) \log_{10} P,\\
	K_{s2} &= (0.00\pm 0.09) - (5.95\pm 0.21) \log_{10} P,
\end{split}
\end{equation}
with dispersions of $0.23$ and $0.19$ respectively.
Using the VISTA-2MASS calibrations as reported by \cite{GonzalezFernandez2018}, we find
\begin{equation}
\begin{split}
	J &= -0.039 - 6.84\log_{10} P,\\
	K_{s} &= - 5.94\log_{10} P.
\end{split}
\label{eqn::ew_pl}
\end{equation}
The difference in the magnitude systems produces smaller differences than the uncertainties in the calibration.

\begin{figure*}
\includegraphics[width=\linewidth]{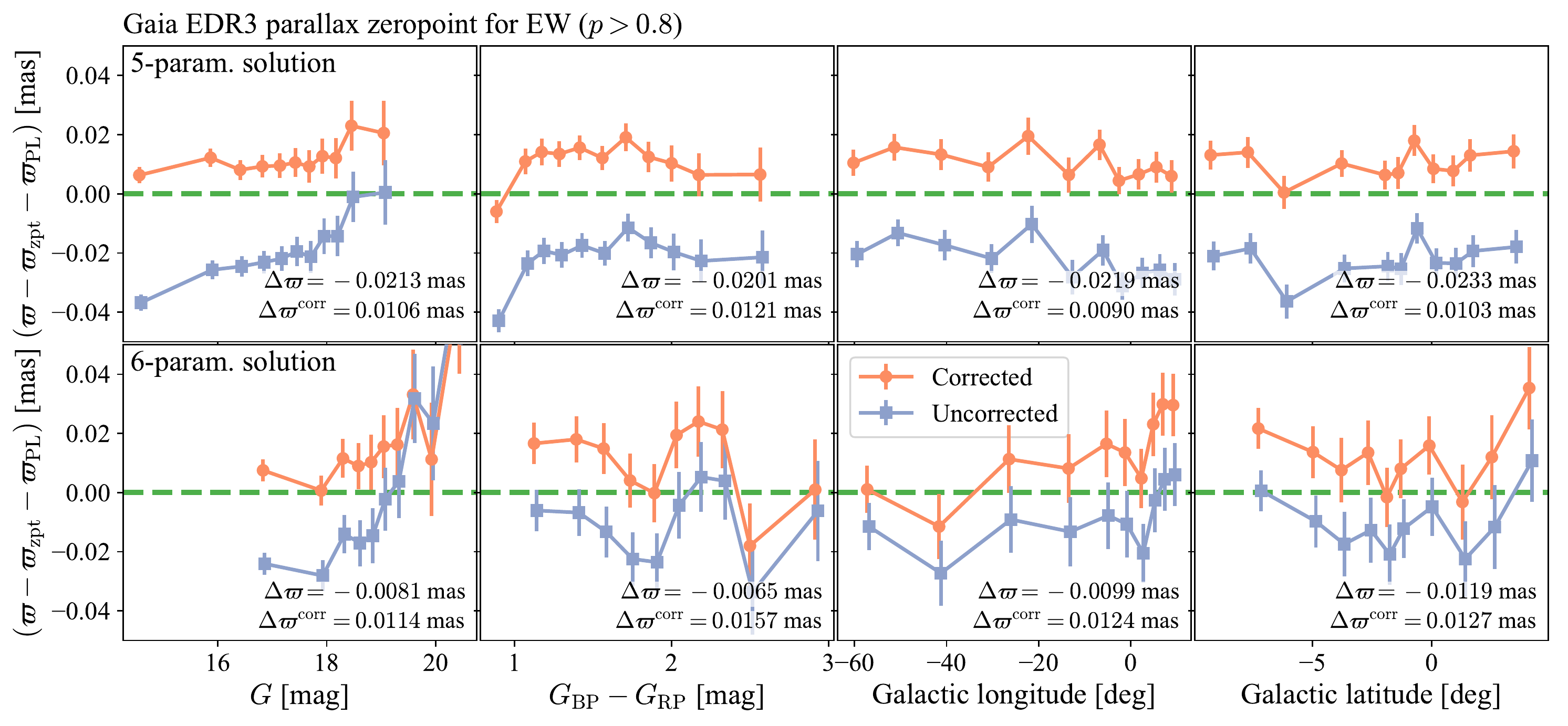}
\caption{Comparison of parallaxes computed from contact eclipsing binary (EW) period-luminosity relations and those from Gaia EDR3 for EW identified by us with classification probability $p>0.8$. Orange circles show differences with the suggested corrections from \protect\cite{Lindegren2021} and purple squares without. The top (bottom) row shows the results for five(six)-parameter astrometric solutions. The numbers in each subpanel give the median difference over the displayed bins.}
\label{fig::edr3_zpt}
\end{figure*}

As highlighted in the introduction, eclipsing binaries are fairly unbiased tracers of the entire Galactic population so well represented across the sky. Furthermore, they are predominantly dwarf stars so relatively nearby. For these reasons, they make ideal calibrators for the Gaia parallax zeropoint. \cite{Lindegren2021b} describes how astrometric solutions were computed for around $1.5$ billion stars in the early third data release of Gaia \citep{Gaia,GaiaEDR3}. \cite{Lindegren2021} provided a magnitude-, colour- and ecliptic-latitude-dependent correction for both $5$ and $6$ parameter solutions based on a sample of quasars. We take all of our high-confidence ($p>0.8$) EW candidates with periods $-0.55<\log_{10}P/\,\mathrm{day}<-0.25$ and with Gaia EDR3 cross-matches within $0.5\,\mathrm{arcsec}$ accounting for epoch difference using Gaia proper motions). We further restrict to stars with RUWE$<1.4$ and which have Gaia EDR3 parameters within the recommended bounds for applying the \cite{Lindegren2021} parallax zeropoint corrections\footnote{
\href{https://gitlab.com/icc-ub/public/gaiadr3_zeropoint/}{https://gitlab.com/icc-ub/public/gaiadr3\_zeropoint/}}. This results in $9872$ stars with five-parameter astrometric solutions and $23030$ with six.

In Fig.~\ref{fig::edr3_zpt} we show the median difference between the parallax computed via the period-luminosity relations from equation~\eqref{eqn::ew_pl} (again using the Wesenheit magnitude from equation~\eqref{eq:weis} and adopting $R_{JK_s}=0.428$) and the Gaia EDR3 parallax, both corrected and uncorrected. We split the sample by stars with $5$ and $6$ parameter astrometric solutions in Gaia EDR3, and also by magnitude, colour and on-sky location of the stars. We observe that the raw Gaia EDR3 parallaxes are significantly underestimated. This is somewhat fixed by applying the \cite{Lindegren2021} parallax zeropoint corrections although in general results in an overestimate of the parallaxes relative to those computed from the period-luminosity relations. Indeed, without any correction the median difference (Gaia EDR3 minus EW period-luminosity relation) for $5$ ($6$) parameter astrometric solutions is
$(-0.022\pm0.002)\,\mathrm{mas}$ ($(-0.010\pm0.003)\,\mathrm{mas}$), whilst after applying the recommended correction we find $0.011\pm0.002$ ($0.013\pm0.003$).
This agrees well with the results of \cite{Ren2021} who performed a similar analysis using $110,000$ contact binaries from ASAS-SN \citep{jayasinghe2018}, ATLAS \citep{atlas}, WISE \citep{Chen2018WISE} and ZTF \citep{Chen2020}, finding the recommended parallax corrections overcorrect the parallaxes by $\sim4\,\mu\mathrm{as}$ with some dependence on on-sky position. They find in particular a tendency for larger overcorrections near the Galactic bulge region. They also find a significant trend of the parallax shift with magnitude as shown in the upper left panel of Fig.~\ref{fig::edr3_zpt}. Additionally, \cite{Huang2021} reach very similar conclusions using a sample of red clump stars from LAMOST with $10\lesssim G\lesssim15$, finding after correction the Gaia EDR3 parallaxes are overestimated by $4.0(1.1)\,\mu\mathrm{as}$ for $5$-($6$-)parameter solutions. We find for our sample there is very little colour or Galactic latitude dependence to the mean shift (although note the sample only covers a narrow range in Galactic $b$). However, we do find there is a weak Galactic longitude dependence in the $5$-parameter solution results with the bulge region having a smaller (corrected) parallax overestimate than for the disc. This could be related to variations in the extinction coefficients across the Galaxy \citep{Schlafly2016}. Finally, we note that this analysis rests on an accurate period-luminosity calibration for the contact binaries. We also looked at the period-luminosity relations provided by \cite{Jayasinghe2020CB} which are $\sim0.1\,\mathrm{mag}$ brighter than those reported by \cite{Chen2018}. However, these produce a corrected parallax zero-point of $\sim0.04\,\mathrm{mas}$. This more suggests a bias in the \cite{Jayasinghe2020CB} relations possibly caused by the quite strict parallax signal-to-noise cut of $20$. A fuller analysis would simultaneously measure the Gaia EDR3 parallax zeropoint \emph{and} the period-luminosity relation \citep[e.g.][]{Sesar2017}.

\section{Conclusion}\label{sect:conclusion}

We have applied an automated variability classification algorithm to the Vista Variables in the Vi\'a La\'ctea (VVV) survey data to generate VIVACE (the VIrac VAriable Classification Ensemble), a catalogue of variable star classifications. We use a new astrometric and photometric reduction of VVV, VIRAC-2, which provides point-spread function photometry, includes a new zeropoint calibration with respect to 2MASS and combines common detections into an iterative astrometric fit. We ran our algorithm on an initial sample of $\sim490$ million sources. The automated selection process was performed hierarchically using a two-stage classifier. The first stage utilised a random forest to separate constant and likely variable sources using a set of simple variability summary statistics. The constant training set was constructed using low-amplitude stars selected from the Gaia 3rd Early Data Release, whilst the variable set was formed primarily from OGLE classifications and the VSX compilation and included RR Lyrae stars, eclipsing and ellipsoidal binaries, Cepheids and long-period variables. A unique classifier was trained for each tile in VVV ($1.5 \times 1.1\,\mathrm{deg}^2$) to accommodate variations in observing strategy and quality. The binary classifier was on average able to correctly recall $\sim86\,\percent$ of the variable training set with a precision of $\sim92\,\percent$. In particular, the incorrect classification of known RRab as constant sources was below $1\,\percent$.

The second stage used gradient boosting trees to perform a detailed classification of the likely variable sources. The feature set was complemented with periodic features computed from Fourier fits to the candidate light curves. We also introduced features to accentuate the differences between different binary classes and multi-band photometric features to distinguish between intrinsic and extrinsic variability. The resulting macro averaged precision of our classification was $81\,\percent$ with $86\,\percent$ recall. Our most successful classifications are for RR Lyrae ab stars for which we achieve a $94.5\,\percent$ precision and $97.5\,\percent$ recall.

The algorithm was applied to the entire VVV survey yielding $\sim1.4$ million classified variable sources. Alongside this publication, we have provided the properties of these sources, their classification and the set of light curve summary statistics. Proper motions (and potentially a subset of the light curves) will be published along with the full VIRAC-2 catalogue (Smith et al., in prep.). Through a visual inspection, we have found restricting to classification probabilities $>0.9$ yields samples which are $<5\,\percent$ contaminated for all classes. This drops to $<1\,\percent$ for RR Lyrae ab stars which is corroborated by comparison to the overlapping OGLE RR Lyrae star sample. This comparison also suggests the presented RR Lyrae ab star sample is approximately $90\,\percent$ complete whilst the RR Lyrae c/d star sample is approximately $50\,\percent$ complete. The final high-confidence (prob.$>0.9$) subset of our catalogue consists of $\sim38,600$ RR Lyrae ab stars, $\sim7900$ RR Lyrae c/d stars, $\sim17,900$ contact eclipsing binaries, $\sim186,900$ detached/semi-detached eclipsing binaries, $\sim1400$ classical Cepheids and $\sim2200$ Type II Cepheids of which we find $\sim205,000$ of these noted in VSX (accessed July 2021).

As an illustration of the potential of the VIVACE catalogue, we investigated the properties of RR Lyrae and contact eclipstars sing binaries using the tight empirical period-luminosity relations for these sources. We demonstrated how the spatial alignment of the bar-bulge RRab and RRcd stars is very similar to that of the more metal-rich populations, whilst exhibiting a lower rotational velocity consistent with increased pressure support. Using the contact binaries we have shown that, assuming the period-luminosity relations of \cite{Chen2018}, the Gaia EDR3 parallax zeropoint corrections from \cite{Lindegren2021b} overcorrect the parallaxes by $\sim10\,\mu\mathrm{as}$ with evidence of a weak magnitude dependence and variation with Galactic longitude (Gaia EDR3 parallaxes typically more overestimated in the southern disc).

Our catalogue is the first covering the entirety of the VVV footprint, and represents a significant advance in indexing variable stars within the high extinction regions of the Milky Way. We look forward to this catalogue being used to further investigate and understand the structure and formation of the Milky Way.

\section*{Data Availability}
The VIVACE table of classified sources along with their light curve statistics will be made available through Vizier. We have made the table available at \href{https://people.ast.cam.ac.uk/~jls/data/vproject/vivace_catalogue.fits}{https://people.ast.cam.ac.uk/\textasciitilde jls/data/vproject/vivace\_catalogue.fits} temporarily. We have provided the basic identifying properties, light curve statistics used in our classification, mean $ZYJHK_s$ photometry and a cross-match to Gaia EDR3 \citep{GaiaEDR3}. Further properties of the sources will be published along with the full VIRAC v2 catalogue (Smith et al., in prep.). All other data used in the training steps (OGLE and VSX) are in the public domain. The code used in this project is available at \href{https://github.com/thomasmolnar/virac_classifier}{https://github.com/thomasmolnar/virac\_classifier}.

\section*{Acknowledgements}
We thank the referee, Andrej Pr{\v{s}}a, for a thorough report that helped to clarify the paper.
J.L.S. acknowledges support from the Royal Society (URF\textbackslash R1\textbackslash191555). Based on data products from observations made with ESO Telescopes at the La Silla or Paranal Observatories under ESO programme ID 179.B-2002. This work has made use of data from the European Space Agency (ESA) mission
{\it Gaia} (\url{https://www.cosmos.esa.int/gaia}), processed by the {\it Gaia}
Data Processing and Analysis Consortium (DPAC,
\url{https://www.cosmos.esa.int/web/gaia/dpac/consortium}). Funding for the DPAC
has been provided by national institutions, in particular the institutions
participating in the {\it Gaia} Multilateral Agreement. This paper made use of the Whole Sky Database (wsdb) created by Sergey Koposov and maintained at the Institute of Astronomy, Cambridge by Sergey Koposov, Vasily Belokurov and Wyn Evans with financial support from the Science \& Technology Facilities Council (STFC) and the European Research Council (ERC). This study made use of the q3c software \citep{koposov}. This paper made use of
\textsc{numpy} \citep{numpy},
\textsc{scipy} \citep{scipy},
\textsc{matplotlib} \citep{matplotlib},
\textsc{seaborn} \citep{seaborn},
\textsc{astropy} \citep{astropy:2013,astropy:2018} and
\textsc{phoebe} \citep{Phoebe1,Phoebe2,Phoebe3}.

\bibliographystyle{mnras}
\bibliography{bibliography.bib}

\appendix
\section{Optimal regularization for Fourier model fits}\label{appendix::optimal_regularization}
When fitting Fourier models described by equation~\eqref{eq:model2} with a large number of terms to relatively sparse light curves, higher likelihood fits can be obtained by increasing the power in the higher frequency terms at the expense of large oscillations in regions with no data. Such unphysical behaviour is a priori unlikely so we introduce a prior (also known as regularization) that biases the Fourier amplitudes $a_i,b_i$ towards zero, increasingly so for the high frequency terms. Following \cite{Vanderplas2015}, a regularization term $\bs{\theta}^T\Lambda\bs{\theta}$ was included in $\chi^2$ which is equivalent to putting a normal prior on $\bs{\theta}$ with mean zero and covariance matrix $\Lambda^{-1}$. We choose $\Lambda$ as a diagonal matrix with diagonal $\lambda\mathsf{Tr}\Sigma^{-1}\bs{n}^2=\lambda\mathsf{Tr}\Sigma^{-1}(0,0,0,1,1,4,4,\cdots)$, which minimises the curvature of the Fourier fits by increasingly penalising the higher frequency terms. $\Sigma$ is the covariance matrix of the magnitude measurements and gives an appropriate scaling for the prior. The choice of $\lambda$ is somewhat arbitrary: too high and we fail to fit genuine periodic signal; too low and we fail to penalise highly oscillatory solutions. The optimal $\lambda$ for a given light curve can be selected by minimising the linear model cross-validation score
\begin{equation}
    \mathrm{CV} = \frac{1}{N}\sum_{i=1}^N \frac{\chi^2}{(1-\mathsf{H}_{ii})^2},
\end{equation}
where the projection matrix $\mathsf{H}$ is given by
\begin{equation}
    \mathsf{H} = \mathsf{X}_\omega(\mathsf{X}_\omega^T\Sigma^{-1}\mathsf{X}_\omega + \Lambda)^{-1}\mathsf{X}^T_\omega\Sigma^{-1}.
\end{equation}
The CV is the sum of the magnitude residuals for a set of leave-one-out models evaluated at the left-out times \citep{Seber2003}.
In the notation of \cite{Vanderplas2018} $\mathsf{X}_\omega$ is the matrix of Fourier (and polynomial) terms at each time.

\begin{figure}
    \centering
    \includegraphics[width=\columnwidth]{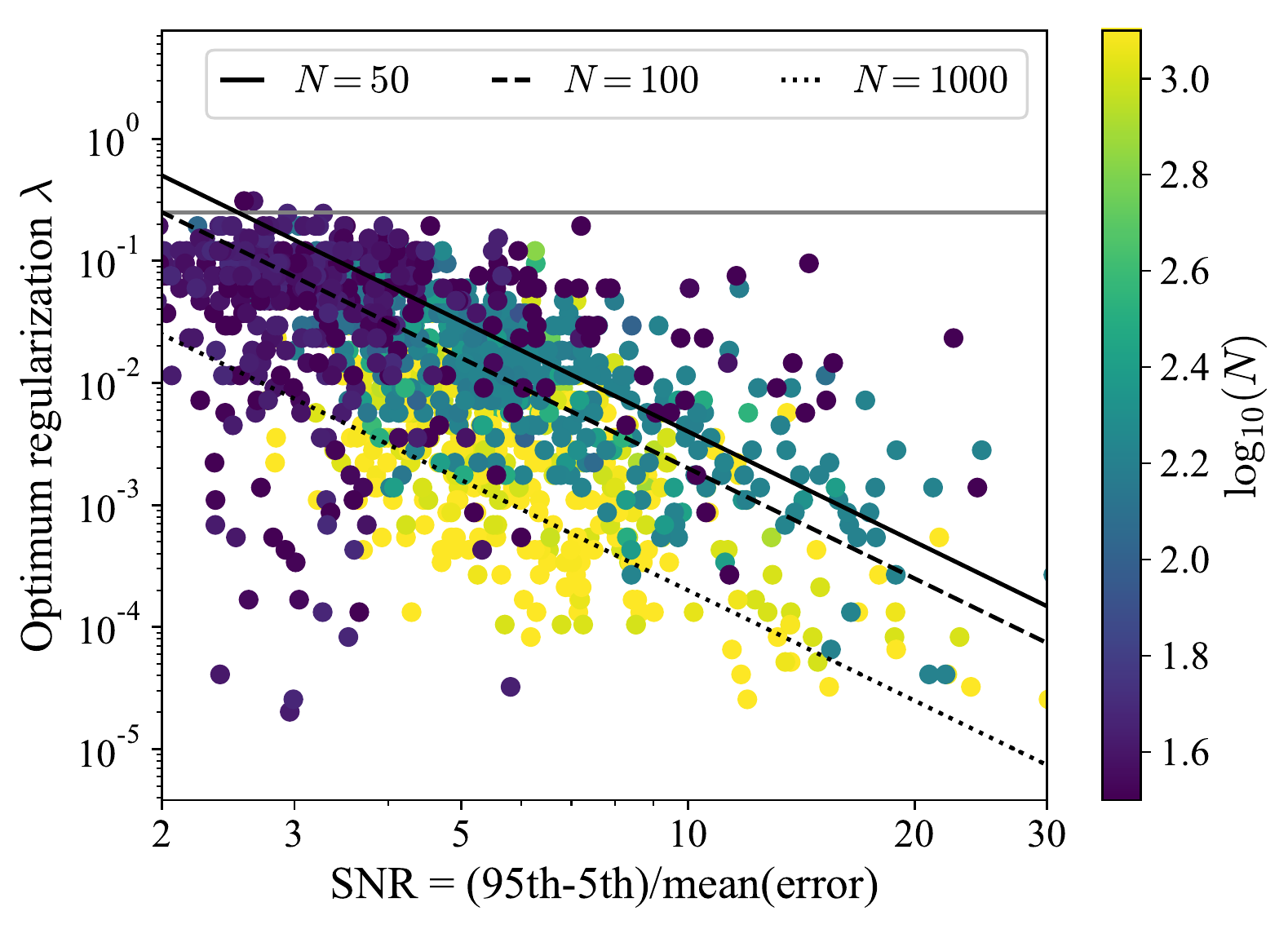}
    \caption{Optimal light curve regularization parameter, $\lambda$, as a function of light curve signal-to-noise ratio (SNR, $S$) characterised as the difference in the $95$th and $5$th percentile divided by the mean error. The points are coloured by the number of epochs. The proposed approximation is shown by the black lines for $N=(50,100,1000)$. The horizontal grey line is the absolute maximum.}
    \label{fig:opt_reg}
\end{figure}

For a sample of VIRAC v2 light curves of OGLE non-contact eclipsing binaries \citep{soszynskibin}, we find the optimal regularization $\lambda=\lambda^*$ by numerically minimising CV at fixed period using $N_f=10$. The results are shown in Fig.~\ref{fig:opt_reg}. We have found that the optimal regularization correlates best with the amplitude of the light curve compared to the error (characterised as the difference in the $95$th and $5$th percentile divided by the mean error, $S$). There is also a weaker dependence on the number of epochs $N$ and a maximum $\lambda^*\approx0.25$. We therefore employ the fitting function
\begin{equation}
\lambda^* = \mathrm{min}\Big[0.25,\frac{200}{NS^3}\Big].
\end{equation}
We find very similar results using $N_f=4$ or $N_f=6$ terms.

\section{Summary of features}\label{appendix::feature_table}
\newlength{\storetabcolsepa}
\setlength{\storetabcolsepa}{\tabcolsep}
\setlength{\tabcolsep}{2pt}
\begin{table*}
    \begin{threeparttable}
    \caption{A list and definition of the various features used in our classification algorithm. The features are defined for a light curve consisting of $N$ magnitudes $m_i$ with errors $\sigma_i$. We define $\hat\delta_i=(m_i-\bar{m})/\sigma_i$ and $\delta_i=\sqrt{N/(N-1)}\hat\delta_i$ where $\bar{m}$ is the mean magnitude. The indices are separated into non-periodic (above the line) used in the first and second classification steps and periodic (below) only used in the second step.}
    \centering
    \begin{tabular}{lllcc}
         Name & Description & Definition &  \makecell{Feature\\Importance\\Stage 1} & \makecell{Feature\\Importance\\Stage 2} \\
         \hline
         MAD\tnote{1} & Median absolute deviation &
    $\text{median} \left(|m_i - \text{median}\left(m_i \right)|  \right)$&0.035 (\textbf{0.073})& 0.017 (0.021)\\
         mags\_p$j$\_p$k$\tnote{1} & \makecell[l]{$j,k$ percentile difference where\\$(j,k) \in \{ \left(100,0 \right),\left(99,1\right),$\\$ \left(95,5\right),\left(84,16\right), \left(75,25\right)\}$}& $\mathrm{percent}(m_i,j)$-$\mathrm{percent}(m_i,k)$&\makecell{
         0.010,0.016,0.037,0.056,0.052\\ (0.004,0.006,0.017,0.043,0.063)}&\makecell{0.003,0.006,0.020,0.011,0.006\\ (0.004,0.003,0.006,0.005,0.035)}\\
         SD\tnote{1} & Standard deviation &
         $
         \sqrt{\frac{1}{N-1}\sum\limits_{i=1}^N {(m_i-\bar{m})}^2}
         $&0.043 (0.011)& 0.003 (0.006)\\
         Skewness&-&
         $
         \frac{N}{(N-1)(N-2)}\sum\limits_{i=1}^N {(m_i-\bar{m})}^3\Big/\mathrm{SD}^3
         $
         &0.022&0.003\\
         Kurtosis&
        \multicolumn{2}{r}{
        $
         \frac{1}{(N-2)(N-3)}\Big(\frac{N(N+1)}{(N-1)}\sum\limits_{i=1}^N {(m_i-\bar{m})}^4\Big/\mathrm{SD}^4
         -3(N-1)^2\Big)$
         }
         &0.004&0.004\\
         $\eta$&\makecell[l]{von Neumann ratio of the\\ consecutive to total variance}&
         $
         \sum\limits_{i=1}^{N-1}\left(m_{i+1} - m_i \right)^2\Big/\sum\limits_{i=1}^N \left(m_i - \bar{m} \right)^2
         $
         &0.069&0.007\\
         Stetson $I$&\makecell[l]{Correlation between $N_\mathrm{p}$ \\pairs of observations}&
        $\sqrt{\frac{1}{N_\mathrm{p}\left(N_\mathrm{p}-1\right)}}\sum\limits_{i=1}^{N_\mathrm{p}}\hat\delta_{1i}\hat\delta_{2i}$; $|t_{1i}-t_{2i}|<1\mathrm{h}$
         &\textbf{0.242}&0.004\\
         Stetson $J$&\makecell[l]{Correlation between\\consecutive observations}&$\frac{1}{N-1}\sum\limits_{i=1}^{N-1}\mathrm{sgn}(\delta_i\delta_{i+1})|\delta_i\delta_{i+1}|$&\textbf{0.119}&0.010\\
         Stetson $K$&\makecell[l]{Robust kurtosis measure}&$\frac{1}{N}\sum\limits_i^N|\delta_i|\Big/\sqrt{\frac{1}{N}\sum\limits_i^N\delta_i^2}$&0.006&0.002\\
         \hline
         $f_{>\mathrm{SD}}$& \makecell[l]{Fraction of magnitudes outside\\ one standard deviation of mean}&$\frac{1}{N}\sum_{i,|m_i-\bar{m}|>\mathrm{SD}}^N 1$&&0.003\\
         $\Delta\phi_\mathrm{max}$&\makecell[l]{Max. difference between\\consecutive folded phases}&$\mathrm{max}_i\,\Delta\phi_i;\,\Delta\phi_i=\omega(t_{i+1}-t_i) \,\mathrm{mod}\,2\pi $&&0.003\\
         Norm. $\Delta\phi_\mathrm{max}$&\makecell[l]{$\Delta\phi_\mathrm{max}$ relative to mean\\normalized by dispersion}&$(\mathrm{max}_i\,\Delta\phi_i-\mathrm{mean_i}\,\Delta\phi_i)/\mathrm{std}_i\,\Delta\phi_i$&&0.003\\
         Period ($P$)&\makecell[l]{Best-fitting Fourier period}&$2\pi/\omega$&&\textbf{0.081}\\
         $\mathrm{LS}_{\mathrm{max}}$&\makecell[l]{Maximum power in \\Lomb-Scargle periodogram}&$\mathrm{max}_\omega\mathcal{P}_\omega;\,\mathcal{P}_\omega=\Big(1-\chi_\mathrm{LS}^2(\omega)/\chi_\mathrm{LS,ref}^2\Big)$&&0.023\\
         $\mathrm{LS}_\mathrm{disp}$&\makecell[l]{Dispersion of maximum\\ Lomb-Scargle power from mean}&$(\mathrm{max}_\omega\,\mathcal{P}_\omega-\mathrm{mean}_\omega\,\mathcal{P}_\omega)/\mathrm{std}_\omega\,\mathcal{P}_\omega$&&0.061\\
         $A_\mathrm{model}$&Model amplitude&$\mathrm{max}_tm(t|\omega,\bs{\theta})-\mathrm{min}_tm(t|\omega,\bs{\theta})$&&\textbf{0.062}\\
         $A_\mathrm{data}$&Data amplitude&$\mathrm{max}_i\,m_i-\mathrm{min}_i\,m_i$&&0.049\\
         $A_j$ &Amplitude of $j$th  harmonic&$\sqrt{a_j^2+b_j^2}$&&0.017,0.010,0.005,0.003\\
         $R_{ij}$&Amplitude ratio&$A_j/A_i$&&\makecell{0.012,0.010,0.005,\\0.005,0.003,0.002$^2$}\\
         $\Phi_{ij}$&Phase difference&$j\Phi_i-i\Phi_j$ where $\Phi_i=\arctan(-b_i/a_i)$&&\makecell{0.010,0.005,0.003,\\0.005,0.003,0.003$^2$}\\
         $A^*_j$&Double-period amplitude&$\sqrt{{a^*}_j^2+{b^*}_j^2}$&&0.016,0.051,0.005,0.055\\
         $R^*_{ij}$&Double-period amplitude ratio&$A^*_j/A^*_i$&&\makecell{0.006,0.004,0.009,\\0.007,0.010,0.006$^2$}\\
         $\Phi^*_{ij}$&Double-period phase difference&$j\Phi^*_i-i\Phi^*_j$ where $\Phi^*_i=\arctan(-b^*_n/a^*_n)$&&\makecell{0.003,0.002,0.003,\\0.002,0.018,0.002$^2$}\\
         $(\Delta\log\mathcal{L})/N$&\makecell[l]{Log-likelihood diff. between\\Fourier and constant fit}&$\frac{1}{N}\sum\limits_{i=1}^N\Big[-\frac{(m_i-m(t_i|\omega,\bs{\theta}))^2}{2\sigma_i^2}+\frac{(m_i-\bar{m})^2}{2\sigma_i^2}\Big]$&&0.013\\
         FAP&\makecell[l]{False-alarm probability of \\highest peak, $\mathrm{LS}_{\mathrm{max}}$\tnote{3}}&$1-(1-\mathrm{e}^{-\mathrm{LS}_\mathrm{max}})\mathrm{e}^{-\tau(\mathrm{LS}_\mathrm{max})}$&&\textbf{0.073}\\
         $r_\mathrm{model}$&\makecell[l]{Ratio between consecutive\\minima depths of model}&$
         \frac{m(t_\mathrm{min}|\omega,\bs{\theta})-s}{m(t_\mathrm{2nd\,min}|\omega,\bs{\theta})-s}$; $s=\mathrm{max}_i\,m_i$
         &&0.011\\
         $r_\mathrm{data}$&\makecell[l]{Ratio between consecutive\\minima depths of data}&$\frac{\sum_{i\in\mathrm{min}}(m_i-s)}{\sum_{i\in\mathrm{2nd\,min}}(m_i-s)}$; $s=\mathrm{mags\_p1}$&&0.003\\
         $(J-K_s)_0$&Unextincted $(J-K_s)$ colour&$(J-K_s)-E(J-K_s)_\mathrm{RC}$&&0.028\\
         $(H-K_s)_0$&Unextincted $(H-K_s)$ colour&$(H-K_s)-E(H-K_s)_\mathrm{RC}$&&0.006\\
         $X_\mathrm{RMS}/K_{s,\mathrm{RMS}}$&\makecell[l]{RMS of $X=\{Z,Y,J,H\}$ \\relative to RMS of Fourier model} &$\mathrm{RMS}(X_i)\Big/\mathrm{RMS}(m(t_i|\omega,\bs{\theta}))$&&0.008,0.004,0.006,0.003\\
         $X_\mathrm{scale}$&\makecell[l]{Scaling of $K_s$ Fourier model\\to match band $X=\{Z,Y,J,H\}$}&$\argmin_{X_\mathrm{scale}}\sum\limits_i^{}(X_i-X_\mathrm{scale}m(t_i|\omega,\bs{\theta}))^2$&&0.004,0.006,0.005,0.002\\

         \hline

    \end{tabular}
    \label{tab:var_indices}
    \begin{tablenotes}
    \item[1] Median error-weighted versions of these indices were also included (feature importances given in brackets).
    \item[2] Ratios and phase-differences are given in the order ((2,1), (3,1), (4,1), (3,2), (4,2), (4,3)). Phase-differences are the geometric mean of the sine and cosine components.
    \item[3] See \cite{Baluev2008} for the full definition.
    \end{tablenotes}
    \end{threeparttable}
\end{table*}
\setlength{\tabcolsep}{\storetabcolsepa}

In Table~\ref{tab:var_indices} we summarise all of the features used in the two classification stages and give their relative importances.

\bsp
\label{lastpage}

\end{document}